\pdfoutput=1
\documentclass[amsmath,amssymb,preprintnumbers,nofootinbib,a4paper,11pt]{article}

\usepackage{graphicx}
\usepackage{epstopdf}
\usepackage{jcappub}
\usepackage{amsmath,amsthm,latexsym,amssymb,amsfonts,epsfig}
\usepackage{fontenc,layout}
\usepackage{hyperref}
\usepackage{psfrag}
\usepackage[cp1250]{inputenc}
\usepackage[usenames,dvipsnames]{xcolor}

\newcommand{\be}{\begin{equation}}
\newcommand{\ee}{\end{equation}}
\newcommand{\bea}{\begin{eqnarray}}
\newcommand{\eea}{\end{eqnarray}}
\newcommand{\beq}{\begin{eqnarray}}
\newcommand{\eeq}{\end{eqnarray}}

\numberwithin{equation}{section}

\title{On the Maximal Strength of a First-Order Electroweak Phase Transition and its Gravitational Wave Signal}

\author[1,2,3]{John ~Ellis}
\author[1]{, Marek ~Lewicki}
\author[1,4]{, Jos\'e ~Miguel ~No}
\affiliation[1]{Department of Physics, King's College London, Strand, WC2R 2LS London, UK}
\affiliation[2]{Theoretical Physics Department, CERN, Geneva, Switzerland}
\affiliation[3]{National Institute of Chemical Physics \& Biophysics, R\"avala 10, 10143 Tallinn, Estonia}
\affiliation[4]{Departamento de F\'isica Te\'orica and Instituto de F\'isica Te\'orica, IFT-UAM/CSIC,
Cantoblanco, 28049, Madrid, Spain}
\emailAdd{John.Ellis@cern.ch}
\emailAdd{Marek.Lewicki@kcl.ac.uk}
\emailAdd{Josemiguel.no@uam.es}

\abstract{
What is the maximum possible strength of a first-order electroweak phase transition and the resulting gravitational wave (GW) signal? 
While naively one might expect that supercooling could increase the strength of the transition to very high values,
for strong supercooling the Universe is no longer radiation-dominated and the vacuum energy of the unstable minimum of the potential dominates the expansion,
which can jeopardize the successful completion of the phase transition. After providing a general treatment for the nucleation, growth and 
percolation of broken phase bubbles during a first-order phase transition that encompasses the case of significant 
supercooling, we study the conditions for successful bubble percolation and completion of the electroweak phase transition in theories 
beyond the Standard Model featuring polynominal potentials. For such theories, these conditions set a lower bound on the temperature of the transition.
Since the plasma cannot be significantly diluted, the resulting GW signal originates mostly from sound waves and turbulence 
in the plasma, rather than bubble collisions. We find the peak frequency of the GW signal from the phase transition to 
be generically $f \gtrsim 10^{-4}$ Hz.
We also study the condition for GW production by sound waves to 
be {\it long-lasting} (GW source active for approximately a Hubble time), showing it is generally not fulfilled in concrete scenarios. 
Because of this the sound wave GW signal could be weakened, with turbulence setting in earlier, resulting in a smaller overall GW signal
as compared to current literature predictions.
\\
\\
\\
\\
\\
KCL-PH-TH/2018-46, CERN-TH/2018-197, IFT-UAM/CSIC-18-94}

\notoc

\begin{document}

\maketitle

\newpage

\tableofcontents

\section{Introduction}

There are many scenarios for physics beyond the Standard Model (BSM) that predict a first-order electroweak phase transition, 
a number of them motivated by the hope of realising electroweak baryogenesis~\cite{Kuzmin:1985mm,Cohen:1993nk,Riotto:1999yt,Morrissey:2012db}.
Recently these models have enjoyed renewed attention because a strong phase transition 
is also a potential source of observable gravitational wave (GW) 
signals~\cite{Kamionkowski:1993fg,Apreda:2001us,Grojean:2006bp,Huber:2008hg,Espinosa:2008kw,Ashoorioon:2009nf,Dorsch:2014qpa,Enqvist:2014zqa,Kakizaki:2015wua,Jaeckel:2016jlh,
Dev:2016feu,Hashino:2016rvx,Chala:2016ykx,Tenkanen:2016idg,Kobakhidze:2016mch,Huang:2016cjm,Artymowski:2016tme,Hashino:2016xoj,
Vaskonen:2016yiu,Dorsch:2016nrg,Beniwal:2017eik,
Marzola:2017jzl,Kang:2017mkl,Chala:2018ari,Huang:2015izx,Huang:2016odd,Huang:2017kzu,Hashino:2018zsi,Vieu:2018zze,Huang:2018aja,Bruggisser:2018mrt,Megias:2018sxv,Croon:2018erz}. 

It is widely thought that a thermally-induced phase transition at the electroweak scale results in a GW signal in the frequency window 
where space-based interferometers such as LISA~\cite{Caprini:2015zlo,Audley:2017drz,Caprini:2018mtu} offer the best hope of detection. 
It has also been suggested that strong supercooling could have resulted in the 
electroweak phase transition lasting much longer and ending at a significantly lower 
temperature \footnote{It has been wrongly hypothesized that such a scenario could yield a 
GW signal from the electroweak phase transition peaking at a lower frequency than in the standard case and
extending into the range of frequencies accessible to pulsar timing array (PTA) studies, 
much below the space-based interferometer band (see e.g.~\cite{Kobakhidze:2017mru,Cai:2017tmh}).}.
%
However, for long-lasting supercooled first-order phase transitions
the field remaining in the initial vacuum generates an effective cosmological constant 
term, due to the non-zero energy of this false vacuum. If the transition lasts too long, the cosmological constant term eventually dominates over the
red-shifting radiation background, and the horizons still occupied by undecayed false vacuum begin to inflate. In this case it {may not be} 
possible for the transition to complete successfully, echoing the well-known {\it graceful-exit} problem of old inflation~\cite{Guth:1982pn}.

In this work we provide a detailed treatment of supercooled cosmological first-order phase 
transitions (see~\cite{Turner:1992tz,Megevand:2016lpr} for related work), 
taking into account the impact of the vacuum energy on the expansion of the Universe and discussing in detail the various stages of the transition: 
bubble nucleation, growth and percolation/reheating. We study
the extent to which a supercooled electroweak phase transition is possible, 
investigating the maximal possible strength of an electroweak phase transition 
that completes successfully, and constraining the corresponding maximal GW signal. 

We illustrate this detailed treatment in two specific BSM scenarios with polynomial potentials, namely
the Higgs effective field theory (EFT) with a dimension-6 term $\propto |H|^6$ and the SM with an extra real singlet scalar field.
In both scenarios we obtain the GW spectrum produced by sound waves~\cite{Hindmarsh:2013xza,Hindmarsh:2015qta,Hindmarsh:2017gnf} and plasma 
turbulence~\cite{Caprini:2006jb,Caprini:2009yp,Kosowsky:2001xp,Gogoberidze:2007an,Niksa:2018ofa}. We 
study the conditions for sound waves to be a long-lasting (active for more than a Hubble time) source of 
GW production, showing these are generically not fulfilled in these concrete BSM scenarios.
This indicates that the sound wave GW signal could be weakened, with turbulence setting in relatively early. 
This would result in a smaller than predicted amplitude of the GW spectrum, 
revising previous estimates of the GW signal from the electroweak 
phase transition in BSM models~\cite{Caprini:2015zlo}.

However, we note that the above results may not hold in the particular case of the electroweak phase transition 
triggered by conformal dynamics~\cite{Creminelli:2001th,Randall:2006py,Nardini:2007me,Konstandin:2010cd,Konstandin:2011dr,Bruggisser:2018mrt,
Megias:2018sxv,Bunk:2017fic,Dillon:2017ctw,vonHarling:2017yew,Jaeckel:2016jlh,Baldes:2018emh}. 
After discussing general features of such a scenario, we leave a detailed analysis for future work.

\vspace{2mm}

This paper is organized as follows: In Section~\ref{sec:FOPT} we
review the formalism used to describe the various stages of phase transition dynamics, and introduce the
new elements needed when the vacuum energy of the unstable minimum becomes cosmologically relevant.
Then, in Section~\ref{sec:GWsignals} we describe the method of extracting parameters relevant for a 
GW signal from a {cosmological first-order phase transition that is also appropriate for supercooled scenarios, and 
discuss the computation of the GW spectrum from the electroweak phase transition in such a case.
In Section~\ref{sec:models} we apply the above formalism to two BSM scenarios: the SM supplemented with a 
dimension-six $|H|^6$ EFT operator and the SM with an extra real singlet scalar field, and also discuss briefly 
the case of conformal scenarios. Finally, we present our conclusions in Section~\ref{sec:conx}.
The Appendix contains some technical details of our calculations.

\section{First-order phase transitions with strong supercooling}\label{sec:FOPT}

\subsection{Nucleation and bubble growth}\label{sec:Nucleation}

We begin by reviewing the formalism governing the nucleation of bubbles in first-order phase transitions, and discuss in detail the specific case of 
a supercooled transition. Our starting point is the decay rate of the false vacuum~\cite{Coleman:1977py,Linde:1980tt,Linde:1981zj}:
\be
\Gamma (T) \simeq {\rm max} \left[ T^4 \left(\frac{S_3}{2\pi T}\right)^{\frac{3}{2}} \exp \left(-S_3/T\right) , \
R_0^{-4} \left(\frac{S_4}{2\pi}\right)^{2} \exp \left(-S_4\right)
 \right] \, ,
 \label{GammaT}
\ee
where the first term corresponds to the thermally-induced decay rate, and the second to quantum tunnelling, which can dominate 
at very low temperatures when there is a potential barrier between vacua at $T = 0$. In (\ref{GammaT}), $S_3$ and $S_4$ are the 
$3-$ and $4-$dimensional Euclidean actions for the $O(3)$- and $O(4)$-symmetric tunnelling (``bounce") solutions, respectively, and  
$R_0$ is the size of the nucleating bubble in the latter case.
 
Once the decay rate is obtained in some specific model, of which we discuss examples
in Section~\ref{sec:models}, one can compute the nucleation temperature $T_n$ at which one bubble is nucleated per horizon
on average, given by
\be
\label{eq:T_n}
N(T_n)=\int_{t_c}^{t_n} dt \frac{\Gamma(t)}{H(t)^3} = \int_{T_n}^{T_c} \frac{d T}{T} \frac{\Gamma(T)}{ H(T)^4} =1\,,
\ee
where in the second step we have used the adiabatic {time-temperature} relation
\be 
\label{eq:entropyconserved}
\frac{dt}{dT} = - \frac{1}{T\,H(T)} \, ,
\ee
and $T_c$ is the temperature at which the two minima are degenerate, below which the decay of the false vacuum becomes possible.

In the case of a fast phase transition it is customary to assume also that the transition finishes at a temperature $\simeq T_n$.
However, a more accurate prescription, valid in a {general} case, is obtained in terms of the 
probability of finding a point still in the false vacuum, given by~\cite{Guth:1979bh,Guth:1981uk}
\be
\label{eq:prob_false_vacuum}
P(t)=e^{-I(t)}, \quad
I(t)=\frac{4\pi}{3}\int_{t_c}^{t} dt' \,\Gamma(t')\, a(t')^3\, r(t,t')^3\,,
\ee
where $a(t')$ is the Friedmann-Robertson-Walker scale factor and $r(t,t')$ is the comoving size of a bubble nucleated 
at $t'$ after growing until $t$:
\be \label{eq:bubblesizedef}
r(t,t') = \int_{t'}^{t}\frac{v_w\,d\tilde{t}}{a(\tilde{t}\hspace{0.4mm})} \,,
\ee
where $v_w$ is the wall velocity of the expanding bubble, which we discuss in more detail in Section~\ref{sec:hydro}. 
The exponent $I(t)$ in~\eqref{eq:prob_false_vacuum} 
yields the amount of true vacuum volume per unit comoving volume, where $\Gamma(t')$ and $a(t')^3$ are the 
nucleation rate per unit time and the unit comoving volume, respectively. 

A common way to proceed has been to assume radiation 
domination and compute the above expressions in terms of the temperature. However, for scenarios with strong 
supercooling due to the existence of a potential barrier between minima that persists down to $T=0$, 
it is possible for the energy associated with the non-zero value of the potential in the false vacuum -
which acts as a ``cosmological constant" - to become important and even dominate at low temperatures~\cite{Guth:1982pn}.
In this case the Friedmann equation becomes, in terms of the radiation and vacuum energy densities $\rho_{\rm R}$ and $\rho_{\rm V}$:
\be \label{eq:Hubble}
H^2=\frac{1}{3 M_{\mathrm{pl}}^2}\left(\rho_{\rm R}+\rho_{\rm V} \right) = \frac{1}{3 M_{\mathrm{pl}}^2} \left(\frac{T^4}{\xi_g^2} + \Delta V \right) =
H_{\rm V}^2 \left( \chi^{-1}+1 \right) \,,
\ee
with $\xi_g=\sqrt{30/(\pi^2 g_*)}$, where $g_*=106.75$ is the number of degrees of freedom in the
plasma (which we assume to be constant, for simplicity), $M_{\mathrm{pl}} = 2.435 \times 10^{18}$ GeV
and we have defined $\chi \equiv \rho_{\rm V}/\rho_{\rm R}$ 
and $H_{\rm V}^2 \equiv \Delta V / (3 M_{\mathrm{pl}}^2)$. 

Using~\eqref{eq:bubblesizedef} and~\eqref{eq:Hubble}, and assuming
$v_w \simeq 1$, the comoving size of bubbles $r(T,T')$ (with $T' > T$) is given by
\be   \label{eq:RV}
a(T') \, r(T,T') = a(T')\,\int_{T}^{T'}\frac{d\tilde{T}}{\tilde{T}\,H(\tilde{T})\, a(\tilde{T})}
= \frac{1}{T'}\,\int_{T}^{T'}\frac{d\tilde{T}}{H_{\rm V}\,\sqrt{1 + \chi(\tilde{T})^{-1}}} \, ,
\ee
which can readily be computed in terms of elliptic functions. From~\eqref{eq:prob_false_vacuum}, the volume fraction converted to the true vacuum $I(T)$ is then
\be \label{eq:prob_false_vacuum_2}
I(T)  = \frac{4\pi}{3} \int^{T_c}_{T} 
\frac{dT'\,\Gamma(T')}{H_{\rm V}\,T'^4\,\sqrt{1 + \chi(T')^{-1}}}\,\left(\int_{T}^{T'}\frac{d\tilde{T}}{H_{\rm V}\,\sqrt{1 + \chi(\tilde{T})^{-1}}} \right)^3 \, .
\ee
It is possible to obtain approximate analytic solutions to~\eqref{eq:RV} and~\eqref{eq:prob_false_vacuum_2} 
by assuming that one of the components dominates the r.h.s. of~\eqref{eq:Hubble}, 
and defining the temperature $T_V$ (with $\chi(T_V) = 1$) 
below which the vacuum energy $\Delta V$ dominates:
\be \label{eq:Hubbleapprox}
\frac{T_V^4}{\xi_g^2} = \Delta V \Longrightarrow 
H(T)= \begin{cases}
 H_{\rm R}(T)= \frac{T^2}{\sqrt{3}\, M_{\mathrm{pl}}\, \xi_g}, \ \ T > T_V \, , \\ H_{\rm V} = \frac{T_V^2}{\sqrt{3}\, M_{\mathrm{pl}} \,\xi_g} ,\quad \quad  T < T_V \, .
 \end{cases}
\ee 
We assume that the potential difference between the false and true vacua $\Delta V$ (and thus $H_{\mathrm{V}}$) is
temperature-independent, which is a 
good approximation at sufficiently low temperature, as in the strong supercooling case of interest to us.
There is then a contribution to~\eqref{eq:RV} from bubble growth during vacuum domination ($T \leq T' \leq T_V$):
\be 
\label{eq:r_VD}
a(T') r_{\rm V}(T,T')= \frac{1}{T'}\int_{T}^{T'}\frac{d\tilde{T}}{ H_{\rm V}} =
\frac{1}{H_{\rm V}} \frac{T'-T}{T'} \,,
\ee
as well as from bubbles nucleated during radiation domination and evolving down to vacuum domination ($T \leq T_V \leq T'$), given by  
\be
\begin{split}
 \label{eq:r_RVD}
a(T') r_{\rm RV}(T,T') = & \frac{1}{T'} \left( \int^{T'}_{T_V}\frac{d\tilde{T}}{H_{\rm R}(\tilde{T})}  +
\int^{T_V}_{T}\frac{d\tilde{T}}{H_{\rm V}} 
\right) \\
= & \frac{1}{T'\,H_{\rm V}}\left(2\, T_V - T - \frac{T_V^2}{T'} \right)\,. 
\end{split}
\ee
The probability for a point in space to remain in the false vacuum at a temperature 
$T < T_V$ is then $P_{\rm RV}(T) = e^{-I_{\rm RV}(T)}$, with
\be
\begin{split} \label{eq:I_RVD}
I_{\rm RV}(T) & =\frac{4\pi}{3}\left( \int^{T_c}_{T_V}\frac{dT' \Gamma(T')}{T'H_{\rm R}(T')}a(T')^3 r^3_{\rm RV}(T,T')+\int^{T_V}_{T}
\frac{dT' \Gamma(T')}{T'H_{\rm V}}a(T')^3 r^3_{\rm V}(T,T') \right) 
\\ &=
\frac{4\pi}{3\,H_{\rm V}^4} \left(
 \int^{T_c}_{T_V}\frac{dT' \Gamma(T')}{T'^6} T_V^2 \left(2\, T_V - T - \frac{T_V^2}{T'} \right)^3 
 +   \int^{T_V}_{T}
 \frac{dT' \Gamma(T')}{T'}\left(1-\frac{T}{T'}\right)^3 \right)
.
\end{split}
\ee
We can compare these expressions with the analogous ones assuming radiation domination (i.e. neglecting $\Delta V$ in~\eqref{eq:Hubble}):
\be
\label{eq:r_RD}
a(T') r_{\rm R}(T,T') = \frac{1}{T'} \int_{T}^{T'}\frac{d\tilde{T}}{H_{\rm R}(\tilde{T})}
= \frac{\sqrt{3}\, M_{\mathrm{pl}}\,\xi_g}{T'}\int_{T}^{T'}\frac{d\tilde{T}}{\tilde{T}^2} = 
\frac{\sqrt{3}\, M_{\mathrm{pl}}\,\xi_g}{T'}\left( \frac{1}{T}-\frac{1}{T'} \right) \, ,
\ee
\be \label{eq:I_RD}
I_{\rm R}(T)  = \frac{4\pi}{3} \int^{T_c}_{T}\frac{dT'\, \Gamma(T')}{T'H_{\rm R}(T')}a(T')^3 r^3_{\rm R}(T,T') 
= 12 \pi (M_{\mathrm{pl}}\,\xi_g)^4 \int^{T_c}_{T}\frac{dT'\, \Gamma(T')}{T'^6}\left( \frac{1}{T}-\frac{1}{T'} \right)^3  \, .
\ee
%

\begin{figure}[h!]
\centering
\includegraphics[width=0.68\textwidth]{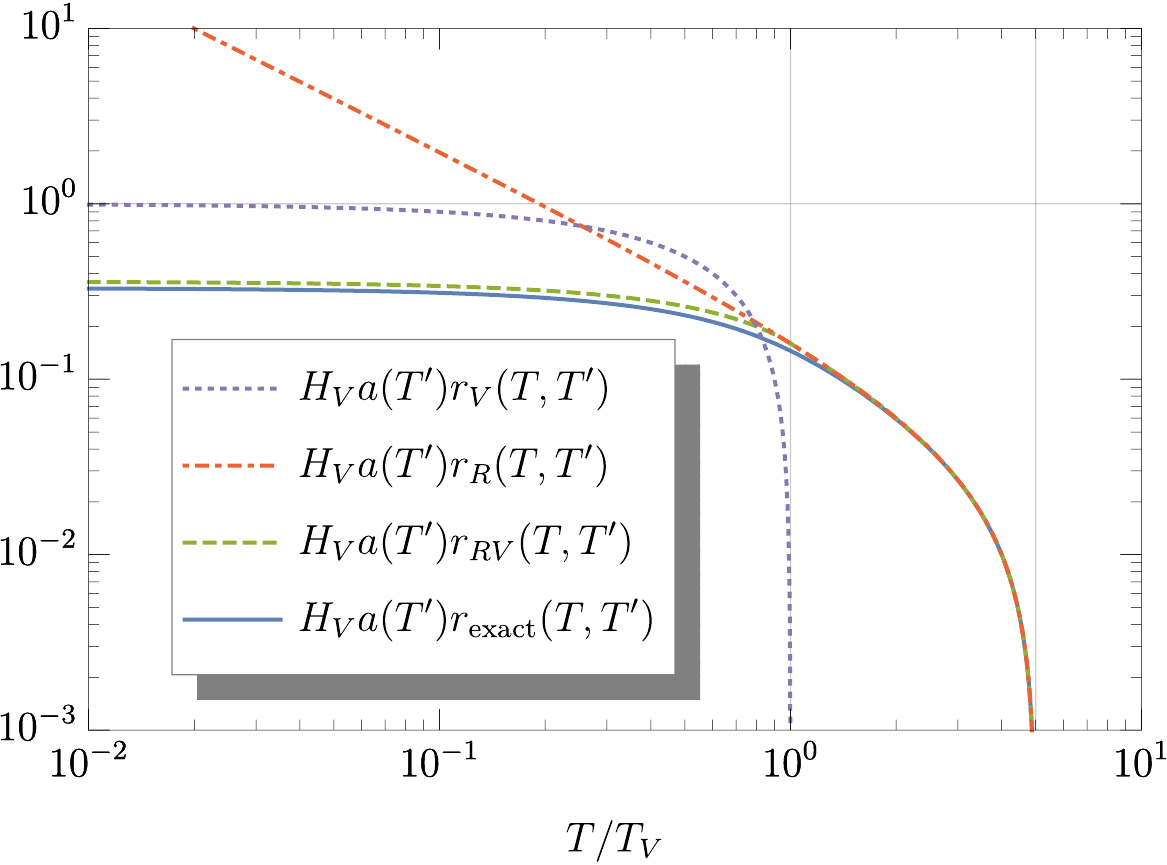}
\caption{\it
The normalized size $H_{\mathrm{V}} \,a(T')\, r(T,\, T')$ of a bubble nucleated at $T' = 5\, T_V$, as a function of $T/T_V$, 
for pure radiation domination~\eqref{eq:RV} (red dot-dashed line), the approximate vacuum domination solution~\eqref{eq:r_RVD}
(green dashed line), and the exact~\eqref{eq:RV} (solid blue line) evolution from radiation 
to vacuum domination. We also show the size of a bubble nucleated at $T = T_{V}$ and growing in vacuum~\eqref{eq:r_VD} (purple dashed line).
We have verified these behaviours in all four cases using the specific models from Section~\ref{sec:models}.
\label{fig:Rplot}}
\end{figure}

\noindent
 The difference in bubble growth between a purely radiation-dominated scenario and that of~\eqref{eq:Hubble}, featuring a transition from 
radiation to vacuum domination, is shown in Figure~\ref{fig:Rplot}. 
It demonstrates that the analytic approximation~\eqref{eq:r_RVD} to the exact growth~\eqref{eq:RV} works very well, and 
shows the change in the bubble expansion for $T < T_V$. Specifically, it highlights that once the vacuum energy dominates the 
bubbles only grow to a finite size in comoving coordinates~\cite{Guth:1982pn} (recall~\eqref{eq:bubblesizedef} and~\eqref{eq:r_VD}):
\be
\label{rcomoving_Vacuum}
r_{\mathrm V}(t\to \infty ,t') = v_w \int_{t'}^{\infty}\frac{\,d\tilde{t}}{e^{H_{\mathrm V}\tilde{t}}} = \frac{v_w}{H_{\mathrm V}} e^{ - H_{\mathrm V}t'}\, .
\ee
In this way, bubbles separated by a distance  larger than twice \eqref{rcomoving_Vacuum} (in comoving coordinates) 
when they nucleate will never meet. 
This has a crucial impact on percolation and the completion of the transition, which we discuss in Section~\ref{sec:percolation}. 

\subsection{More on bubble growth: hydrodynamics}\label{sec:hydro}

In this Section we delve into the details of the bubble expansion, which will be of importance when discussing the generation of GWs from the phase 
transition in Section~\ref{sec:GWsignals}. 

\vspace{1mm}

We start by discussing briefly why, 
in the case of a very strong cosmological first-order phase transition, the bubbles are expected 
to expand with a velocity $v_w \to 1$. A leading-order evaluation of the friction on a phase transition bubble
expanding at ultrarelativistic speeds $\gamma \gg 1$~\cite{Bodeker:2009qy} showed that the 
relativistic flux of particles crossing the bubble wall exert only a 
bounded, i.e., independent of $\gamma$ in the $\gamma \to \infty$ limit, pressure on the wall: $\Delta {\cal P}_{\mathrm{LO}} \sim \Delta m^2\, T^2$,
where $\Delta m$ denotes the particle's change of mass across the phase boundary.  
Hence, if the pressure difference produced by the difference in vacuum free energy $\Delta V$ 
exceeds the friction $\Delta {\cal P}$, 
the bubble wall keeps on accelerating to highly relativistic velocities with no upper bound on $\gamma$, 
a situation known as ``runaway"~\cite{Bodeker:2009qy}.   

Higher-order contributions to the friction have recently been evaluated~\cite{Bodeker:2017cim}, and shown 
to scale roughly as $\Delta {\cal P}_{\mathrm{NLO}} \sim \gamma\, g^2\, \Delta m \,T^3$ for the electroweak phase transition, 
where $g$ is the electroweak gauge coupling. This implies that
the bubbles reach a state expanding with $\gamma \sim \Delta V/(g^2 \,\Delta m \,T^3) \gg 1$ shortly after nucleation,
even if $\Delta V$ exceeds the leading-order friction $\Delta {\cal P}_{\mathrm{LO}}$, 
as expected for very strong phase transitions. Still, the leading-order runaway criterion allows us
to assess the expansion of bubbles at ultrarelativistic velocities $v_w \to 1$~\cite{Bodeker:2009qy}, and we employ it below when discussing 
explicit models in Section~\ref{sec:models}.   

\vspace{1mm} 

When the bubble reaches a terminal expansion velocity, the subsequent bubble growth in the 
presence of a thermal plasma can be described via a hydrodynamical 
treatment~\cite{Steinhardt:1981ct,Laine:1993ey,Ignatius:1993qn,KurkiSuonio:1995pp,Espinosa:2010hh,Konstandin:2010dm}, 
whose most relevant aspects we review 
here (see~\cite{Espinosa:2010hh} for a detailed discussion). This hydrodynamic description assumes that the plasma is in local thermal equilibrium 
and can be described by a perfect fluid, with an energy-momentun tensor $T_{\mu\nu} = w U_{\mu} U_{\nu} - g_{\mu\nu} p$. 
Here $ p $ is the pressure, $w = T \, (\partial p / \partial T)$ is the enthalpy and $U_{\mu}$ the four-velocity field of the plasma:
\be 
U_{\mu} = \frac{(1,\,\vec{v})}{\sqrt{1 - \left|\vec{v}\right|^2  }} = (\gamma,\,\gamma \,\vec{v}) \, .
\ee
The behaviour of the plasma can then be obtained from the conservation of energy and momentum,
$\partial_{\mu} T^{\mu\nu} = 0$, with the appropriate boundary conditions 
on the bubble wall, where conservation of energy-momentum becomes non-trivial
due to a non-zero change in pressure across the phase boundary:
$\Delta p = - \Delta V$. The equations that match energy-momentum across the bubble wall 
(with ``$+$" denoting the symmetric phase and ``$-$" the broken phase) read 
\be
w_{+} \gamma^2_{+} v^2_{+} + p_{+} = w_{-} \gamma^2_{-} v^2_{-} + p_{-} \quad\quad, \quad \quad w_{+} \gamma^2_{+} v_{+} = w_{-} \gamma^2_{-} v_{-} \, \,,
\ee
with $v = \left|\vec{v}\right|$. From these equations, and assuming an appropriate
equation of state (EoS) for the fluid, one arrives at the relation~\footnote{The relation~\eqref{matchingEQsHydro} is 
usually derived in the context of the bag-model EoS, which follows from a relativistic gas approximation. It has nevertheless been shown in~\cite{Espinosa:2010hh} to hold in 
a more general context.}
\be
v_{+} = \frac{1}{1 + \alpha} \left[ \left(\frac{v_{-}}{2} + \frac{1}{6 v_{-}}   \right) \pm \sqrt{\left(\frac{v_{-}}{2} + \frac{1}{6 v_{-}}   \right)^2 + 
\alpha^2 + \frac{2}{3} \alpha - \frac{1}{3}}\right] \, .
\label{matchingEQsHydro}
\ee
Here $\alpha(T)$ is the latent heat of the transition~\footnote{For the case of strong phase transitions considered in this work, 
the definition of $\alpha$ in terms of the latent heat or 
in terms of the free-energy difference $ \Delta V(T)$, which would yield $\alpha(T) = \chi(T)$ 
(recall eq.~\eqref{eq:Hubble}), are approximately equivalent.  
However, for the case of weak phase transitions it is not completely settled 
how $\alpha$ should be defined precisely~\cite{ThomasK_private}.} 
normalized to the 
radiation energy of the plasma $\rho_{\mathrm{R}}$ in the symmetric phase (outside the bubbles):
\be
\alpha(T) = \frac{\Delta V(T) - T \frac{\partial \Delta V(T)}{\partial T}}{\rho_{\mathrm{R}}} \, .
\label{alpha_DEF}
\ee
%
Away from the bubble wall, energy-momentum conservation of the plasma yields
\be
\label{Tmunuplasma_conservation}
\partial_{\mu} T^{\mu\nu} = U^{\nu} \partial_{\mu} (U^{\mu} w) + U^{\mu} w \partial_{\mu} U^{\nu} - \partial^{\nu}p = 0 \, .
\ee
We consider a spherically-symmetric bubble configuration. In addition, as there are no characteristic length scales 
in the system apart from the microphysical (EW) scale and the Hubble radius, the velocity and temperature profiles of the 
plasma show a self-similar behaviour, depending only on the combination $\xi = r/t$,
where $r$ is the distance from the centre of the bubble and 
$t$ is the time since nucleation. Then, from~\eqref{Tmunuplasma_conservation} we obtain the equation describing the plasma velocity 
profile $v(\xi)$:
\be
\label{vfluid_1}
\frac{2\, v}{\xi} = \frac{1 - \xi\,v}{1- v^2} \left[\frac{1}{c_s^2} \frac{(\xi-v)}{(1-\xi\, v)} - 1 \right] \partial_\xi v \, ,
\ee
where $c_s$ is the speed of sound in the plasma, which is given by $c_s^2 = 1/3$ in a relativistic fluid.
For $v_w \to 1$, the solution to~\eqref{vfluid_1} with the appropriate boundary conditions on the phase boundary
(the $+$ branch of~\eqref{matchingEQsHydro} with $v_w = v_+$ and $v_- \geq c_s$,
see~\cite{Espinosa:2010hh} for details) 
is called a ``detonation" (see Figure~\ref{fig:detonation}), with the velocity of the fluid just behind the phase boundary given by
\be
v(\xi = v_w \to 1) = \frac{3 \alpha}{2 + 3 \alpha} \, .
\ee
The above discussion can easily be extended to $v_w < 1$, with detonation solutions being realized for a given value of $\alpha$ down to 
a minimum value of $v_w$ (given by the condition $v_{-} = c_s$), below which other fluid solutions (``deflagrations" and 
``hybrids", see~\cite{Espinosa:2010hh}) are instead realized, a regime we do not consider in this work. 

\begin{figure}[h!]
\centering
\includegraphics[width=0.68\textwidth]{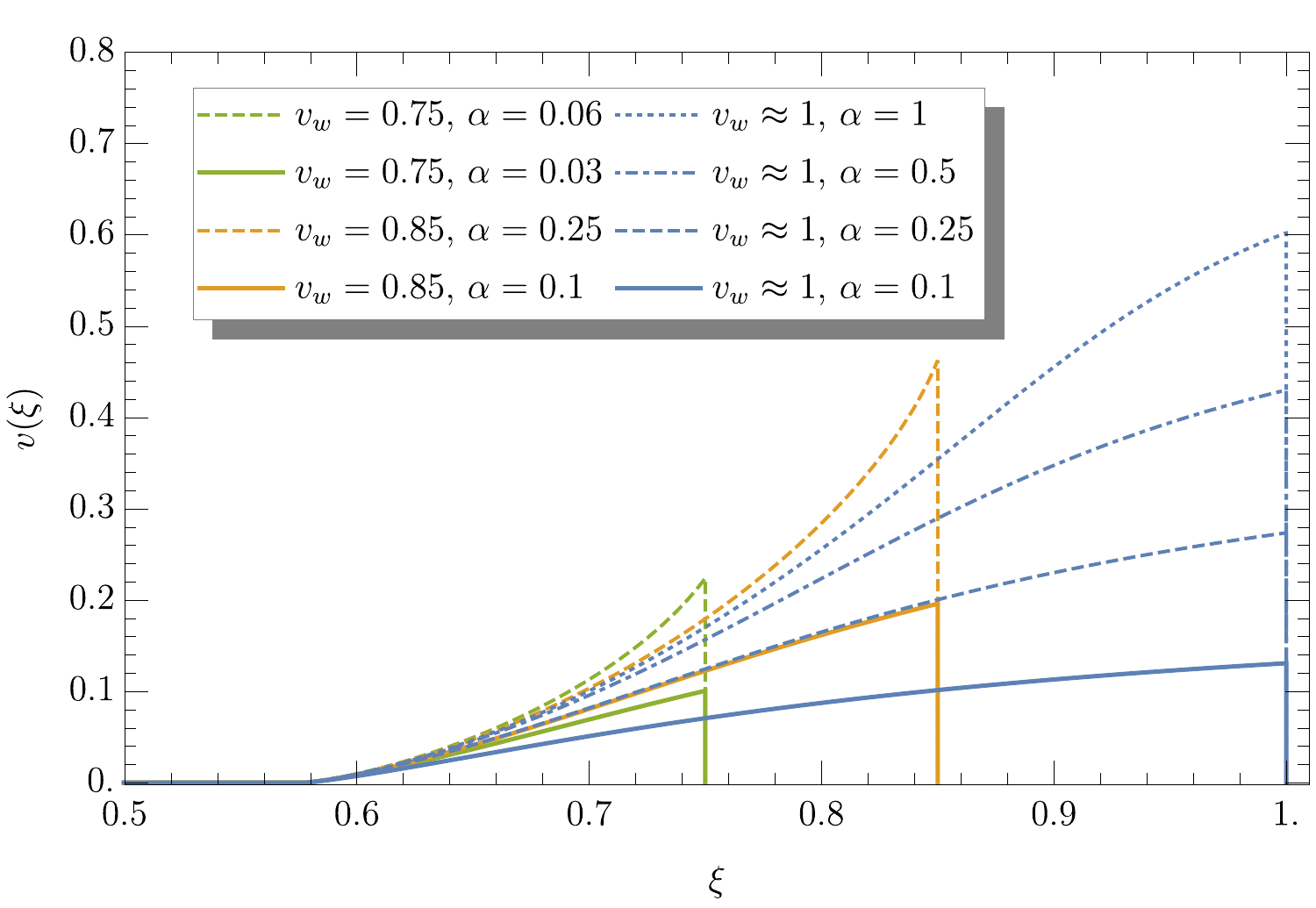}

\vspace{-2mm}

\caption{\it Detonation plasma velocity profile $v(\xi)$ for various values of $\alpha$ and $v_w$.
\label{fig:detonation}}
\end{figure}

Before continuing, we define two quantities (related to each other) that will appear in the analysis of GW signatures later on. 
The first is the root-mean-square four-velocity of the plasma $\bar{U}_f$, which for a single expanding bubble as discussed above 
reads~\cite{Hindmarsh:2017gnf}
\be \label{eq:plasma_rms_velocity}
\bar{U}^2_f = \frac{3}{v_w^3} \int^{v_w}_{c_s} \xi^2 \frac{v^2}{1-v^2} d\xi \, .
\ee
The second is the ``efficiency" ratio of the plasma kinetic energy to the available energy of the transition (the latent heat), given by
\be
\kappa = \frac{3}{\alpha \, \rho_{\mathrm{R}} \,v_w^3} \int^{v_w}_{c_s} w\, \xi^2 \frac{v^2}{1-v^2} d\xi \, ,
\label{eq:kappa_eff}
\ee
where $w(\xi)$ is the enthalpy profile of the plasma~\cite{Espinosa:2010hh}.


\subsection{Bubble percolation and reheating: completing the phase transition}\label{sec:percolation}

In the standard radiation-dominated scenario, bubbles are assumed to percolate when 
$I(T) \gtrsim n_c = 0.34$. This is the ratio of the volume in 
equal-size and randomly-distributed spheres (including overlapping regions) to the total volume of space for which percolation occurs  
in three-dimensional Euclidean space~\cite{1971Shante},
and implies that at least $34\%$ of the (comoving) volume has been converted to the true minimum, leading then to 
the completion of the phase transition for $P(T) \lesssim 0.7$. The percolation temperature can then be approximately defined 
as~\footnote{Some early works in the context of cosmological first-order phase transitions~\cite{Enqvist:1991xw} defined the 
onset of bubble coalescence as $P(t_0)=1/e \longrightarrow I(t_0)=1$, yielding a slightly lower percolation temperature.} 
$I(T_p) = 0.34$.

However, this simple percolation criterion is known to be misleading in a vacuum-dominated scenario, as the false vacuum is inflating 
and percolation may never be achieved in this case~\cite{Guth:1982pn}, even though $P(t)$ decreases with time and can reach the required value, since
\bea
I_{\mathrm{V}}(t) &=& \frac{4\pi}{3}\int_{t_c}^{t} dt' \,\Gamma(t')\, a_{\mathrm{V}}(t')^3\, r_{\mathrm{V}}(t,t')^3  =
\frac{4\pi}{3} \left(\frac{v_w}{H_{\mathrm V}}\right)^3 \int_{t_c}^{t} dt' \,\Gamma(t')\, 
\left(1 - e^{-H_{\mathrm V}(t-t')} \right)^3 \nonumber \\
&\underset{t \to \infty}{\longrightarrow} & \,\,\, \frac{4\pi}{3} \left(\frac{v_w}{H_{\mathrm V}}\right)^3 \Gamma \times t \, ,
\eea
where in  the last step we have assumed a constant $\Gamma$ in the $t \to \infty$ limit.
Instead, a necessary requirement for successful completion of the phase transition is that the physical 
volume of the false vacuum $\mathcal{V}_{\rm false}\propto a(t)^3P(t)$ decreases~\cite{Turner:1992tz} around/after percolation.
This is a strong requirement, since not only does the probability $P(t)$ need to decrease, it has to drop faster than 
the increase in the volume of the space, which is inflating in the case of vacuum domination. This condition reads
\be \label{eq:falsevacuumvol}
\frac{1}{\mathcal{V}_{\rm false}}\frac{d \mathcal{V}_{\rm false}}{d t}=3 H(t)- \frac{d I(t)}{d t}=H(T)\left( 3+T\,\frac{d I(T)}{dT} \right) < 0 \, .
\ee
Intuitively, one would evaluate this condition at the percolation temperature $T_p$ and conclude
that the transition ends successfully with percolation provided it is fulfilled. 
However, as we discuss below in specific models, there usually exists a region in the 
parameter space with an even stronger transition for which this criterion is 
not satisfied at $T_p$ but becomes true at some lower temperature. 
We will indicate when such situations occur, keeping in mind that successful percolation is then 
not guaranteed as in transitions for which the above criterion is never fulfilled,  though it cannot be immediately disproved.    
We will see in the models studied in the following Sections that condition~\eqref{eq:falsevacuumvol} 
severely limits the duration of the vacuum-domination period, and leads in general 
to a lower bound on $T_p$, constraining the possible amount of supercooling.

The criterion~\eqref{eq:falsevacuumvol} may also be compared to the one originally used in~\cite{Guth:1982pn}
for strongly supercooled phase transitions. Expanding $I(t)$ around some instant $t_0$, 
the percolation criterion from~\cite{Guth:1982pn} reads
\be
\epsilon \equiv \frac{3}{4\pi\,H(t_0)}\left.\frac{d I(t)}{d t}\right|_{t_0} > \frac{9\, n_c}{4 \pi} \simeq 0.243 \, ,
\ee
which is automatically fulfilled (a weaker condition) if~\eqref{eq:falsevacuumvol} is satisfied at time $t_0$.

\vspace{2mm}

A reheating process takes place after successful percolation, and for strongly supercooled phase transitions this brings
the system back to a state of radiation domination.
If~\eqref{eq:falsevacuumvol} is satisfied at $T = T_p$, we may assume reheating to be a relatively fast process that occurs as percolation 
happens (in contrast, when~\eqref{eq:falsevacuumvol} is only satisfied at some temperature lower than $T_p$, percolation, if successful, is bound to 
be a fairly slow process, and the same should be assumed for reheating).
Then, immediately after percolation the Universe would evolve into a state of radiation domination with a
temperature $T_{\rm rh}$ given by~\footnote{We note the lack of an efficiency factor $\kappa$ accompanying $\alpha$ in~\eqref{Trh},
in contrast to the result in~\cite{Cai:2017tmh}. This is because~\cite{Cai:2017tmh} includes only the reheating contribution 
from the energy conversion of plasma bulk motion, and does not take into
account the plasma heating occurring already during the expansion of the bubbles~\cite{Espinosa:2010hh}. In addition, the above expression 
assumes the number of relativistic degrees of freedom to be similar in the two phases, which is in any case well-justified for 
strong supercooling in the symmetric, unstable phase.}:
\be
\label{Trh}
T_{\rm rh} \simeq T_p \left[ 1 + \alpha(T_p) \right]^{1/4}\, .
\ee
A more quantitative picture of the reheating process could in principle be obtained in the following way: 
{\it(i)} For small plasma velocities $v(\xi) \ll 1$, the superposition of fluid shells is approximately a linear process and  
gives rise to sound waves~\cite{Hindmarsh:2013xza}.
{\it(ii)} In this case, the average temperature of the plasma during/after percolation can be obtained
from the linear, random superposition of plasma shells. The temperature profile for each plasma shell 
can be obtained using the hydrodynamical treatment
discussed in Section~\ref{sec:hydro}. {\it(iii)} When the sound waves in the plasma cease to be active due to the increase of non-linearities,
the bulk-motion kinetic energy stored in the fluid is used to heat the plasma up further and the reheating process is completed.

We note that since both the thermal and kinetic energy of the fluid contribute to the radiation energy density, under the assumption of relatively 
fast reheating we have~\footnote{We thank Mark Hindmarsh for reminding us of this.} $H(T_p) \simeq H(T_{\rm{rh}})$
as a result of energy conservation. As such, the details of the reheating process are not very crucial.
However, we stress that reheating poses a fundamental challenge to shifting the GW frequency from the electroweak phase transition 
below the space-based interferometer range of frequencies and towards the PTA band (e.g. invalidating the claim 
made in~\cite{Kobakhidze:2017mru,Cai:2017tmh}).

\section{Gravitational wave signals}\label{sec:GWsignals}

We now turn to the discussion of GWs from supercooled phase transitions. 
Throughout this work, we assume that the production of GWs from the phase transition occurs around $T = T_p$.
The key parameters 
from the phase transition used conventionally to determine the GW spectrum are $v_w$, $\alpha$ (defined in~\eqref{alpha_DEF}) and the 
parameter $\beta/H$~\cite{Kamionkowski:1993fg,Apreda:2001us,Grojean:2006bp}, 
which yields the approximate timescale of the transition in terms of the Hubble parameter $H$. 
(Alternatively, one may consider the relevant length scale for GW generation instead of $\beta$, as we discuss below).

As has been discussed in Section~\ref{sec:hydro}, $\alpha$ describes the available energy from the 
transition, normalized to the radiation energy of the plasma. 
For strong, supercooled transitions 
\begin{equation}\label{eq:alpha}
\alpha(T_p) \simeq \chi(T_p) = \frac{\Delta V(T_p)}{\rho_{\rm R}(T_p)} \, .
\end{equation}
For a fast cosmological first-order 
phase transition occurring during radiation domination, the parameter $\beta$ is conventionally defined in terms of the false vacuum decay 
rate~\eqref{GammaT} as 
\begin{equation}
\Gamma\propto e^{-\frac{S_3(T)}{T}} = e^{\,\beta\,(t-t_{0})\, +\, ...} \, .
\label{eq:Gamma_beta}
\end{equation}
This means the nucleation rate rises exponentially, and the timescale of the transition is given by (minus) 
the first derivative of the thermal bounce action
$S_3(T)/T$ (the action of the bounce field solution driving the thermal tunnelling~\cite{Linde:1981zj, Linde:1980tt}) 
in a Taylor expansion around some time $t_0$, yielding
\begin{equation}
\beta = \beta_{\rm R} \equiv -\frac{d}{dt}\left.\left(\frac{S_3(T)}{T}\right)\right|_{t=t_0}= 
H(T) \left. T\frac{d}{dT}\left(\frac{S_3(T)}{T}\right)\right|_{T=T_0} \, .
\label{eq:beta}
\end{equation}
For $t_0$ corresponding to $P(t_0)=1/e$ ($I(t_0)=1$), the bubble number density is~\cite{Enqvist:1991xw}
\begin{equation}\label{eq:bubble_separation_RD}
n_B = (R_{*\rm R})^{-3}= \frac{1}{8\pi}\,\left(\frac{\beta}{v_w}\right)^3 \, ,
\end{equation}
yielding a direct relation between $\beta$ and the mean bubble centre separation 
(in the case of a fast phase transition) $R_{*\rm R}$. This relation is approximately maintained at~$T= T_p$.

\vspace{2mm}

For very strong phase transitions, a potential barrier between the symmetric and broken 
phases may still be present at $T = 0$, which results in $S_3(T)/T$ having a minimum at some finite $T$. The linear 
approximation~\eqref{eq:Gamma_beta} {may then break} down~\cite{Megevand:2016lpr,Jinno:2017ixd} 
(see also~\cite{Cutting:2018tjt}), as the first derivative of the bounce 
action {can} vanish (the timescale of the transition defined by~\eqref{eq:beta} then
yields $\beta_{\rm R} \to 0$ and even turns negative). In this case, going to the next order in the Taylor expansion of the 
bounce action, we obtain a Gaussian approximation
\begin{equation}
\label{eq:Gamma_beta_2}
\Gamma\propto e^{-\frac{S_3(T)}{T}} = e^{-\frac{1}{2}\beta_{\rm V}^2 (t-t_{m})^2\, +\, ...} \, , 
\end{equation}
where $t_m$ corresponds to $(d/dt)(S_3(T)/T)|_{t=t_m} = 0$, and $\beta_{\rm V}$ is given by
%
\begin{equation}
\beta_{\rm V} \equiv \left.\sqrt{\frac{d^2}{dt^2}\left(\frac{S_3(T)}{T}\right)}\right|_{t=t_m} = 
H(T)  T \left. \sqrt{\frac{d^2}{dT^2}\left(\frac{S_3(T)}{T}\right)}\right|_{T=T_m} \, .
\label{eq:beta_V}
\end{equation}
In this scenario the majority of the bubbles are nucleated around time $t_m$, and the bubble number density at this time is 
given by~\cite{Cutting:2018tjt} (we set $v_w \to 1$, as expected for a very strong phase transition):
\begin{equation}
\label{number_density_Gaussian}
n_B = (R_{*\rm V})^{-3} \simeq \sqrt{2\pi}\,\frac{\Gamma (T_m)}{\beta_{\rm V}} \,,
\end{equation}
with $R_{*\rm V}$ the mean bubble separation in this limit. It is nevertheless clear that the regime in which the approximation~\eqref{eq:Gamma_beta_2}
is valid will only be relevant for GW generation if $T_p \lesssim T_m$, since otherwise~\eqref{eq:Gamma_beta} holds down to $T_p$.   

\vspace{2mm}

In the general case, the bubble number density at time $t$ can be written in terms of $P(t)$ 
and $\Gamma(t)$ as~\cite{Enqvist:1991xw,Turner:1992tz}:
\begin{equation}
\label{bubble_number_density}
n_B = (R_*)^{-3} = \int^{t}_{t_c} dt'\, \frac{a(t')^3}{a(t)^3} \, \Gamma(t') P(t') \,.
\end{equation}
The mean bubble separation $R_*$ may be regarded as the relevant length scale for the generation of GWs from the 
phase transition (see e.g. the discussion in~\cite{Hindmarsh:2017gnf}). 
In this work we argue for a related, 
but we believe quantitatively more appropriate, choice for the GW length scale as follows:  
\begin{itemize}
\item At time $t$ the physical size of bubbles nucleated at some earlier time $t'$ is given by
\be
\label{physical_R}
R(t,t')=a(t)\,r(t,t')\,.
\ee 
\item From~\eqref{bubble_number_density} the distribution of bubble sizes at temperature $T$ is
\be
\label{bubble_distribution}
\frac{dn}{dR}(t,R) = -\frac{dt'}{dR} \frac{a(t'(R))^3}{a(t)^3} \Gamma(t'(R)) P(t'(R))\,.
\ee
where both $t'(R)$ and $dt'/dR$ may be obtained by inverting~\eqref{physical_R}.
\item As the dominant contribution to the GW generation comes from the bubbles that contain the largest fraction of the energy budget 
of the phase transition (see~\cite{Nicolis:2003tg} and Appendix A of~\cite{Huber:2007vva} for a related discussion), 
the relevant GW length scale should be of the order of the bubble size for which the energy 
distribution is maximized~\footnote{Since the energy budget of a bubble scales with its volume $R^3$.} as a function of $R$: 
\begin{equation}
\label{bubble_energy_fraction}
\mathcal{E}_B(t,R) \equiv R^3\,\frac{dn}{dR}(t,R) \, .
\end{equation}
\end{itemize}
In obtaining~$dt'/dR = (-1 / H(T') T')\, dT'/dR$ to maximize $\mathcal{E}_B(t,R)$,
we can use the approximation~\eqref{eq:Hubbleapprox} and consider separately 
the cases $T_p > T_{V}$ and $T_p < T_{V}$. In the former case, bubbles expand during radiation domination 
(recall~\eqref{eq:r_RD}), and we have 
\be
T' =  T_p\,\frac{1}{1-\frac{T_p^2}{T_V^2} (H_{\rm V} R)}\, .
\ee
In the latter case, bubbles expand during both vacuum and radiation domination (recall~\eqref{eq:r_VD} and~\eqref{eq:r_RVD}), and we have
\be
T' =
\begin{cases}
 T_p\, (1 + H_{\rm V} R),
  \quad \quad \quad \  H_{\rm V} R < \frac{T_V}{T_p} - 1 \, ,
  \\
T_V\, \frac{1}{2 - \frac{T_p}{T_V}(1 + H_{\rm V} R)},
 \quad\quad  H_{\rm V} R > \frac{T_V}{T_p} - 1 \, .
\end{cases}
\ee
We show in Figure~\ref{fig:distplot} the distribution $\mathcal{E}_B(t,R)$ (normalized to the total energy stored in the bubbles at $T_p$) 
for two representative scenarios from 
Section~\ref{subsec:H6} featuring $T_p > T_{V}$ and $T_p < T_{V}$, respectively. 
In each case we highlight the {value of $R$ for which the energy distribution is maximal, $R_{\rm MAX}$,} 
together with the mean bubble separation $R_{*}$ for the purpose of comparison.
{Figure~\ref{fig:distplot} illustrates that in specific scenarios the difference between $R_{\rm MAX}$ and $R_{*}$ can sometimes be sizeable.}
We further note that the length scale fixing the peak of the GW spectrum does not correspond to 
the size of the bubble $R_{\rm MAX}$ but rather to the thickness of its fluid shell~\cite{Hindmarsh:2016lnk}, which for $v_w \to 1$ is given 
by $(1 - c_s) R_{\rm MAX} \simeq 0.422\, R_{\rm MAX}$. The relevant scale for GW generation (yielding the peak of the GW power spectrum)
is then given by~\footnote{The impact of the continuous increase of $R_{\rm MAX}$ during the completion 
of the transition and the generation of GWs, which may
yield an extra source of GW power on scales larger than $\bar{R}$~\cite{Jinno:2017fby}, is currently under 
debate~\cite{Konstandin:2017sat}.} $\bar{R} \sim (v_w -c_s) R_{\rm MAX}$.


\vspace{2mm}

As already argued in Section~\ref{sec:hydro}, in the presence of the surrounding thermal plasma 
the growing bubbles can only accelerate for a short amount of time before they reach a time-independent expansion rate. 
From that moment on, the energy is mostly pumped into fluid shells around the bubbles (e.g.~corresponding to the detonation profiles discussed in 
Section~\ref{sec:hydro}), leading to only a negligible fraction of the energy being carried by the Higgs bubble walls. 
Only an extreme dilution of the plasma due to supercooling could allow the bubbles to keep accelerating until the transition completes. 
However, as we show in the following Section via explicit examples, a significant amount of supercooling 
is not possible in polynominal potentials  as it spoils percolation.   
Such plasma dilution is then excluded and the energy stored in the bubble walls at the end of the transition is negligible.
This is however not necessarily true for (nearly-)conformal scalar potentials, which we comment on in Section~\ref{sec:dilaton}.

\begin{figure}[t]
\centering
\includegraphics[width=0.46\textwidth]{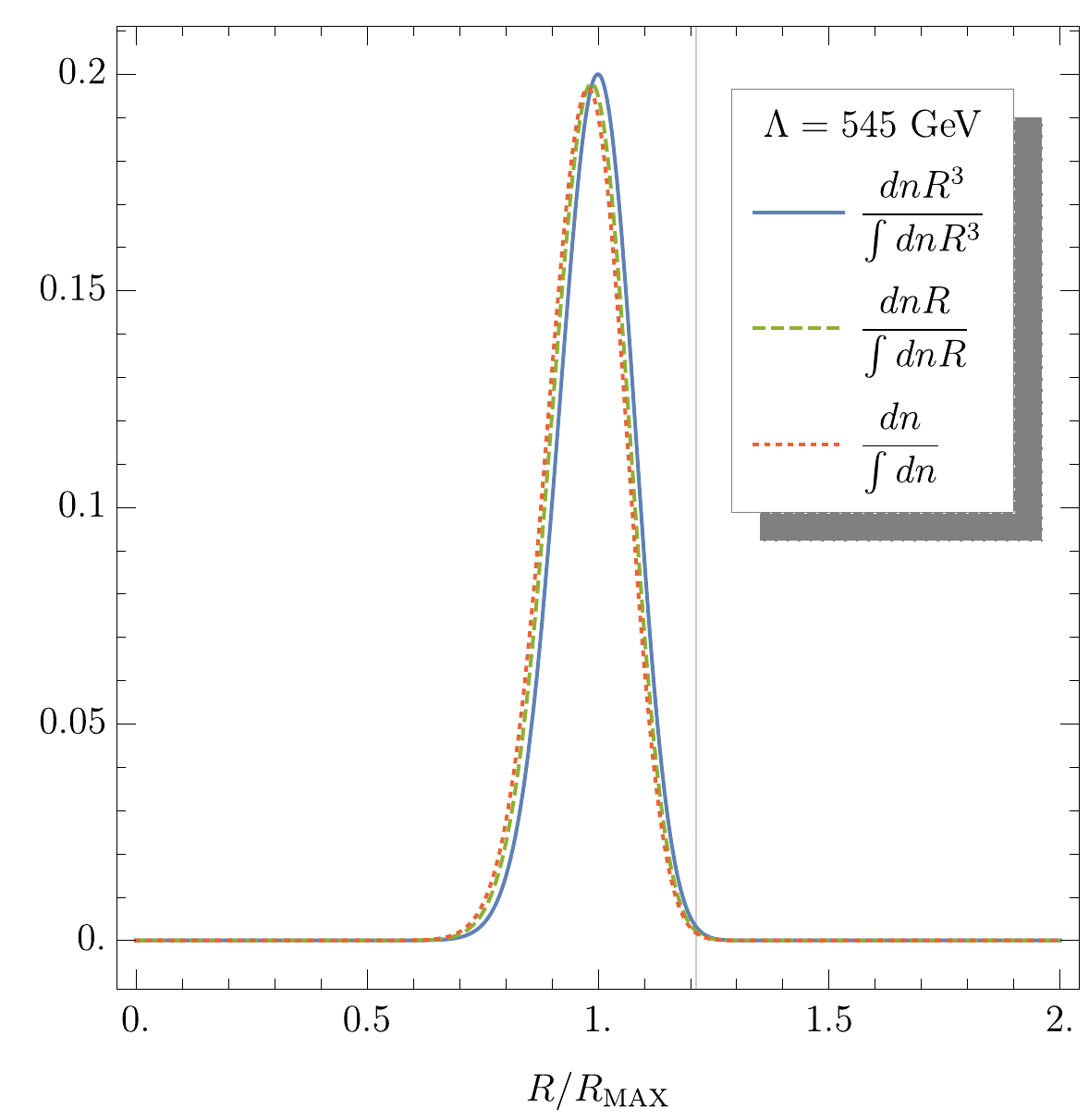}
\hspace{2mm}
\includegraphics[width=0.46\textwidth]{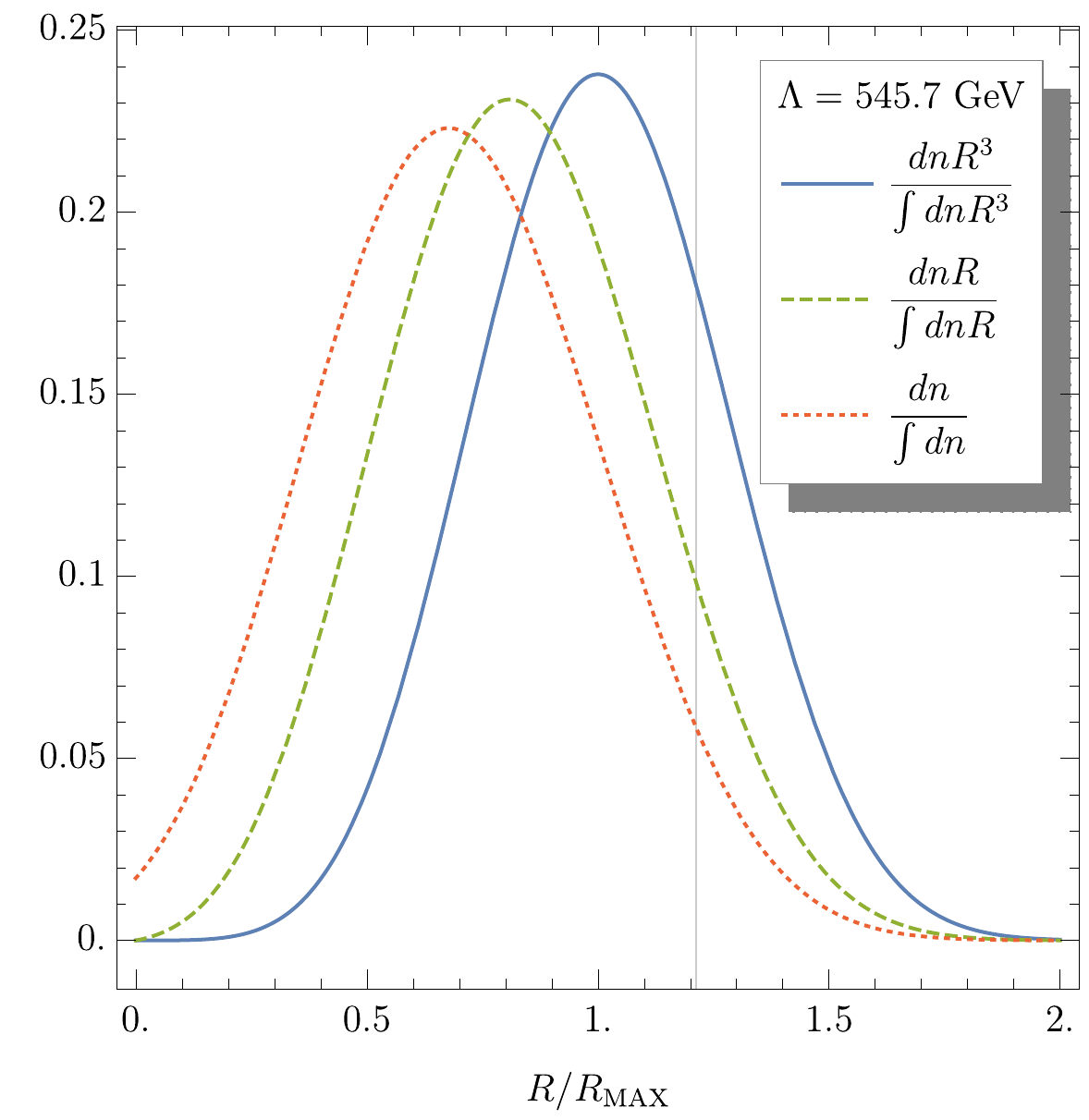}
\vspace{-2mm}
\caption{\it
The energy distribution $\mathcal{E}_B(t_p,R)=dnR^3/\int dn R^3$ (blue solid), bubble size distribution $dnR/\int dn R$ (green dashed) 
and number density $dn/\int dn$ (red dotted) as functions of the bubble size $R$ normalised to $R_{\rm MAX}$ (the maximum of the energy distribution), 
all computed at the nucleation temperature. 
The thin {vertical} line indicates {$R_*/R_{\rm MAX}$, where $R_*$is the mean bubble separation.}
The left panel shows the example with the strongest transition where percolation can still be possible, namely
$\Lambda = 545 \ {\rm GeV}$ and $T_p < T_{V}$, while the right panel shows the strongest transition for which percolation is assured,
{corresponding to} $\Lambda = 545.7 \ {\rm GeV}$ and $T_p > T_{V}$ for the EFT model from Section~\ref{subsec:H6}.
\label{fig:distplot}}
\vspace{-2mm}
\end{figure}

As a result of the above, there is no observable GW signal from the collisions of bubbles~\cite{Kamionkowski:1993fg,Huber:2008hg} 
at the end of the phase transition in the BSM scenarios we consider.
The dominant GW sources from a first-order electroweak phase transition are then expected to be sound waves propagating in the 
plasma~\cite{Hindmarsh:2013xza,Hindmarsh:2015qta,Hindmarsh:2016lnk,Hindmarsh:2017gnf} after percolation happens and the transition completes. 
The peak frequency of the sound wave GW spectrum (as would be observed today) is (see~\cite{Caprini:2015zlo})
\begin{equation}
f_{\rm sw}=1.9 \times 10^{-5}\, \frac{\left(8\pi\right)^\frac{1}{3}}{H \bar{R}} \frac{1}{v_w} 
\frac{T_*}{100}\left({\frac{g_*}{100}}\right)^{\frac{1}{6}} {\rm Hz } \, ,
\label{eq:frequency_soundwaves}
\end{equation}
where $T_*$ is the temperature after the transition has completed and reheating has taken place, from which 
the GW signal is redshifted up until today. We identify it with the reheating temperature $T_{\rm rh}$ (see~\eqref{Trh}).
The amplitude of the GW spectrum is given by
\begin{equation}
\Omega h^2_{\rm sw}(f)=2.65 \times 10^{-6}\frac{H \bar{R}}{\left(8\pi\right)^\frac{1}{3}}
\left(\frac{\kappa \,\alpha }{1+\alpha }\right)^2 
\left(\frac{g_*}{100}\right)^{-\frac{1}{3}}
v_w
\left(\frac{f}{f_{\rm sw}}\right)^3 \left(\frac{7}{4+3 \left(f/f_{\rm sw}\right)^2}\right)^{7/2} \, ,
\label{eq:amplitude_soundwaves}
\end{equation}
with $\kappa$ given by~\eqref{eq:kappa_eff}. 
It is important to stress that the numerical lattice simulations from which 
this result (\eqref{eq:frequency_soundwaves} and \eqref{eq:amplitude_soundwaves}) 
is obtained are reliable only if $H R_*/\bar{U}_f > 1$~\cite{Hindmarsh:2017gnf} (with $\bar{U}_f$ given by~\eqref{eq:plasma_rms_velocity}).
If this is not the case, the flow in the plasma becomes turbulent in less than a Hubble time, cutting short
the sound wave period. The result~\eqref{eq:amplitude_soundwaves} would then overestimate the GW signal \footnote{The amplitude of the GW signal 
will probably be reduced by a factor $\sim H R_*/\bar{U}_f$ in this case. However, dedicated numerical lattice simulations would be needed to test this simple intuition.}.
As we will see in the following Section, the criterion $H R_*/\bar{U}_f > 1$ for sound waves to be {\it long-lasting} 
(active for more than a Hubble time) is challenging to satisfy in explicit models,
putting into question whether the prediction in the literature for GW from sound waves is reliable in these BSM models. 

The final important 
source of a GW signal from the electroweak phase transition, usually subdominant relative to sound waves, is associated 
with magneto-hydrodynamical (MHD) turbulence in the plasma. 
The peak frequency and amplitude of this contribution is~\cite{Caprini:2009yp}
\begin{equation}f_{\rm turb}=2.7  \times 10^{-5}\,
\frac{\left(8\pi\right)^\frac{1}{3}}{H \bar{R}} \frac{1}{v_w} \frac{T_*}{100}\left({\frac{g_*}{100}}\right)^{\frac{1}{6}} {\rm Hz } \, ,
\end{equation}
\begin{equation}
\Omega h^2_{\rm turb}(f)=3.35 \times 10^{-4}\frac{H \bar{R}}{\left(8\pi\right)^\frac{1}{3}}
\left(\frac{\epsilon\, \kappa\, \alpha }{1+\alpha }\right)^{\frac{3}{2}} 
\left(\frac{g_*}{100}\right)^{-\frac{1}{3}}
v_w\,
\frac{\left(f/f_{\rm turb}\right)^3\left(1+f/f_{\rm turb}\right)^{-\frac{11}{3}}}{\left[1+8\pi f a_0/(a_* H_*)\right]} \, ,
\end{equation}
{where $\epsilon$ is an efficiency factor for vorticity to develop in the plasma after bubble percolation, which is estimated 
to be~$\epsilon \approx 0.05$ (see e.g.~\cite{Caprini:2015zlo}).}
We note that the above formula for the GW spectrum from turbulence is a subject of 
ongoing debate (see~\cite{Kosowsky:2001xp,Gogoberidze:2007an,Niksa:2018ofa}). 
Nevertheless, we continue using it as an estimate. 

We also stress that the shortening of the sound wave period in the specific 
BSM models discussed here would probably be accompanied by an enhancement 
of the GW spectrum from turbulence, through an increase in the efficiency factor $\epsilon$. While this would tend to reduce the hierarchy
between the GW contributions from sound 
waves and turbulence, we expect the overall effect to still be a significant reduction of the GW amplitude as compared to the current predictions for BSM scenarios, 
as we discuss in the following Section.

\vspace{-2mm}

\section{Supercooled electroweak phase transitions in specific models} \label{sec:models}

We turn now to the application of the formalism developed in the previous Sections to specific BSM scenarios that yield a first-order electroweak phase transition.

\subsection{Standard Model with an $|H|^6/\Lambda^2$ interaction}
\label{subsec:H6}

We start with a simple example, namely the SM supplemented by a single non-renormalisable dimension-6
operator $|H|^6/\Lambda^2$ (see e.g.~\cite{Bodeker:2004ws,Delaunay:2007wb,Chala:2018ari,Huang:2016odd,Huang:2015izx}), to show the impact of taking 
vacuum domination correctly into account.
Thus, we consider the following potential 
\begin{equation}\label{eqn:classpot}
V(H)=-m^2|H|^2+\lambda|H|^4+\frac{1}{\Lambda^2} |H|^6 \, ,
\end{equation}
with
$H^T=\left(\chi_1+i\chi_2,\varphi+i\chi_3\right)/\sqrt{2}$.
As usual, only the real part of the neutral component obtains a vev: $\varphi=h+v$. 
We identify $h$ as the physical Higgs boson, 
which leads to the tree-level potential
\begin{equation}\label{eqn:treepot}
V(h)^{\textrm tree}=-\frac{m^2}{2}h^2+\frac{\lambda}{4}h^4+\frac{1}{8}\frac{h^6}{ \Lambda^2} \, .
\end{equation} 
We use the observed mass of the Higgs boson $m_h=125 \ {\rm GeV}$, and the measured Higgs vev 
$v  = 246 \ {\rm GeV}$ in the renormalisation conditions 
\begin{equation}
V'(h=v)=0, \quad V''(h=v)=m_h^2 \, ,
\end{equation} 
to express the parameters in the potential (\ref{eqn:treepot}) as
\begin{equation}
m^2=\frac{m_h^2}{2}-\frac{3 v^4}{4 \Lambda^2}, \quad \lambda=\frac{m_h^2}{2v^2}-\frac{3 v^2}{2 \Lambda^2} \, .
\end{equation}
We also include one-loop corrections to the zero-temperature potential and thermal corrections 
as described in Appendix~\ref{sec:effpotappendixLambda}.

\begin{figure}[h!]
\centering
\includegraphics[width=0.88\textwidth]{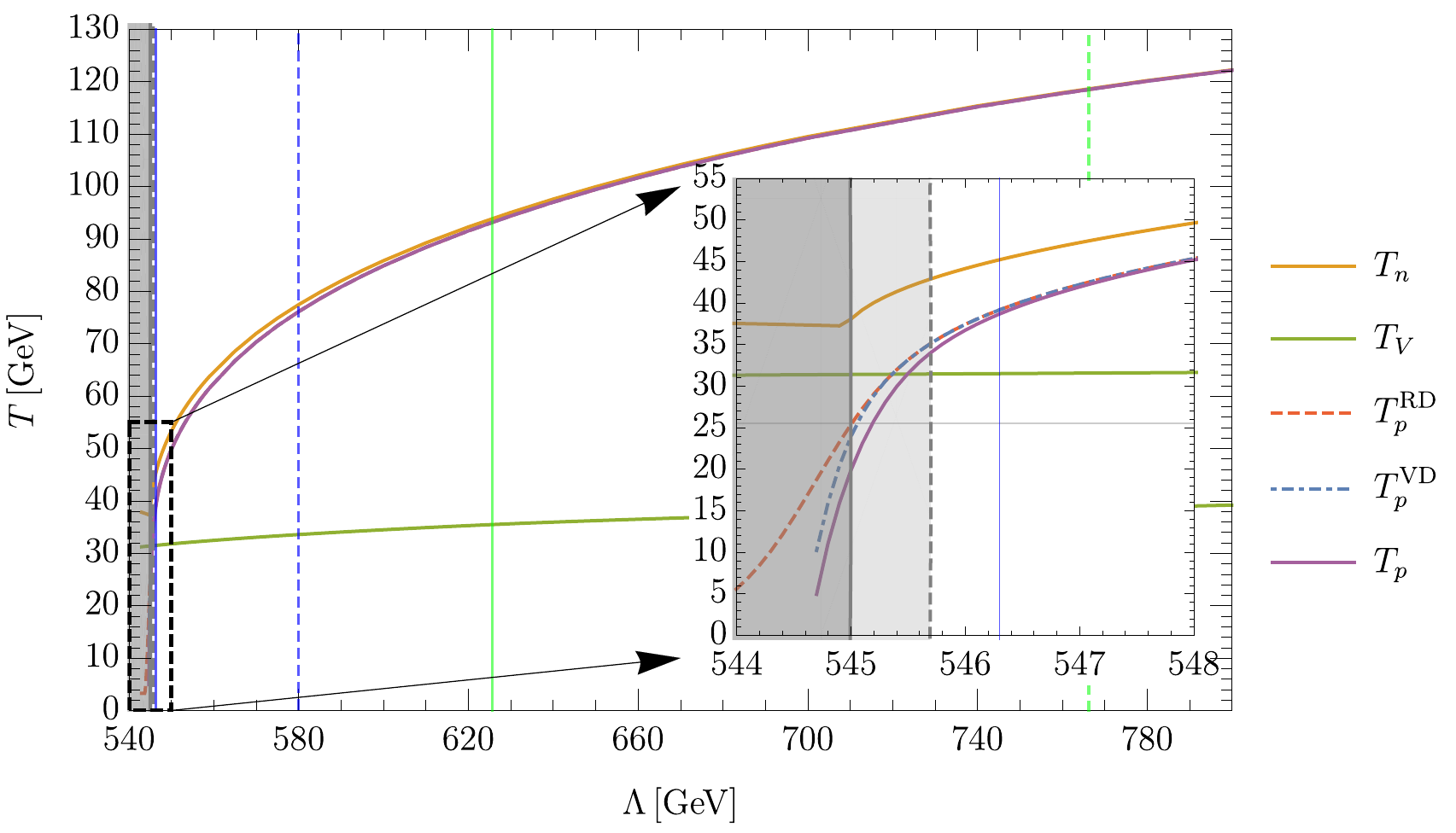}
\caption{\it
The nucleation temperature $T_n$ (orange solid line), the percolation temperature $T_p$ (purple solid line)
and the temperature $T_V$ (green solid line) below which vacuum energy dominates the expansion of the Universe.
The temperature $T_p^{\rm RD}$ (red dashed line) and $T_p^{\rm VD}$ (blue dash-dot line) respectively show 
the percolation temperature obtained neglecting the vacuum energy contribution to the expansion, and using the 
approximation~\eqref{eq:Hubbleapprox}.
The dark grey area is excluded by our percolation criterion~\eqref{eq:falsevacuumvol}, while in the light gray area percolation is questionable, 
as the criterion is only satisfied below $T_p$.
Vertical lines show the projected reach of various detection methods: 
The green lines indicate values of $\Lambda$ to which the HL-LHC will be
sensitive at the $3$- (solid) and 2-$\sigma$ (dashed) level respectively, and the  
blue lines show the reach of LISA from sound waves (dashed line) and turbulence (solid line).
\label{fig:Lambda6Tplot}}
\end{figure}

In Figure~\ref{fig:Lambda6Tplot} we show the relevant temperatures discussed in Section~\ref{sec:FOPT} 
as functions of $\Lambda$, the only free parameter in the $|H|^6$ model. 
The dark grey shaded area given by $\Lambda < 545 \, {\rm GeV}$ is excluded 
by the absence of percolation, while in the light gray area percolation is questionable,
as our percolation criterion~\eqref{eq:falsevacuumvol} is only fulfilled at temperatures below $T_p$. 
The lowest possible percolation temperature is $T_p \geq 26.13 \, {\rm GeV}$, with 
the corresponding lower limit on the model parameter being $\Lambda = 545 \, {\rm GeV}$. 
The lines in Figure~\ref{fig:Lambda6Tplot} show the nucleation temperature $T_n$~\eqref{eq:T_n},
the temperature $T_V$ below which vacuum energy dominates the expansion, 
and the percolation temperature $T_p$ at which $I = 0.34$ 
in~\eqref{eq:prob_false_vacuum_2}. The dashed-red and dot-dashed-blue lines show approximations to 
the percolation temperature obtained, respectively, neglecting the vacuum energy contribution to the expansion, $T_p^{\rm RD}$, 
and using the simple approximation of switching to vacuum domination below $T_V$ as in~\eqref{eq:Hubbleapprox}).

Figure~\ref{fig:Lambda6NucleationExample} shows instead various quantities used in calculating the percolation temperature 
for $\Lambda=545 \, {\rm GeV}$ (left panel),
which is the strongest transition not excluded by our percolation criterion~\eqref{eq:falsevacuumvol}, and $\Lambda=545.7 \, {\rm GeV}$ (right panel), 
which is the strongest transition for which percolation is assured.
We note that even for the most supercooled transition possible, with $\Lambda=545 \, {\rm GeV}$, 
the quantity $\Gamma/H^4$ exceeds one. While this is commonly used as a criterion 
to approximate the nucleation temperature~\eqref{eq:T_n}, in this case it would lead to 
a value for $T_n$ almost twice larger than the correct value, namely $T_n \simeq 37$ GeV.
The number density of bubbles $N$ from~\eqref{eq:T_n} also exceeds one, 
so we have about one bubble per horizon on average. The final percolation temperature defined by
 $I(T_p) = 0.34$ (see~\eqref{eq:prob_false_vacuum_2}) is significantly lower than the nucleation temperature and given by $T_p \simeq 23.09$ GeV. 
For values $\Lambda < 545$ GeV, despite having nearly one bubble per horizon, 
the horizons without bubbles would start inflating and the expansion of bubbles nucleated at 
higher temperatures would be too slow ever to complete the transition.
Our results also imply that the quantum tunnelling contribution to the decay rate~\eqref{GammaT} is never relevant 
if percolation is to be achieved~\cite{Espinosa:2008kw}.

\begin{figure}[h!]
\centering
\includegraphics[width=0.48\textwidth]{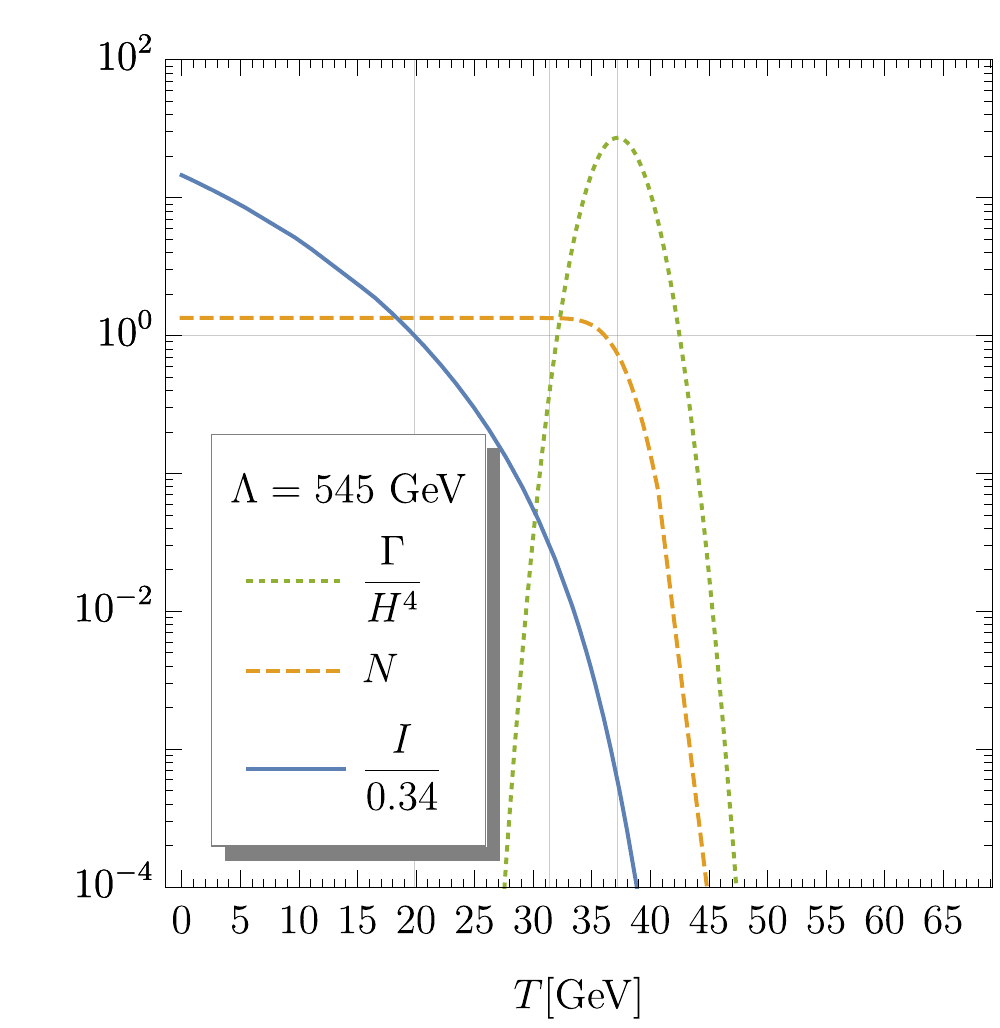}
\includegraphics[width=0.48\textwidth]{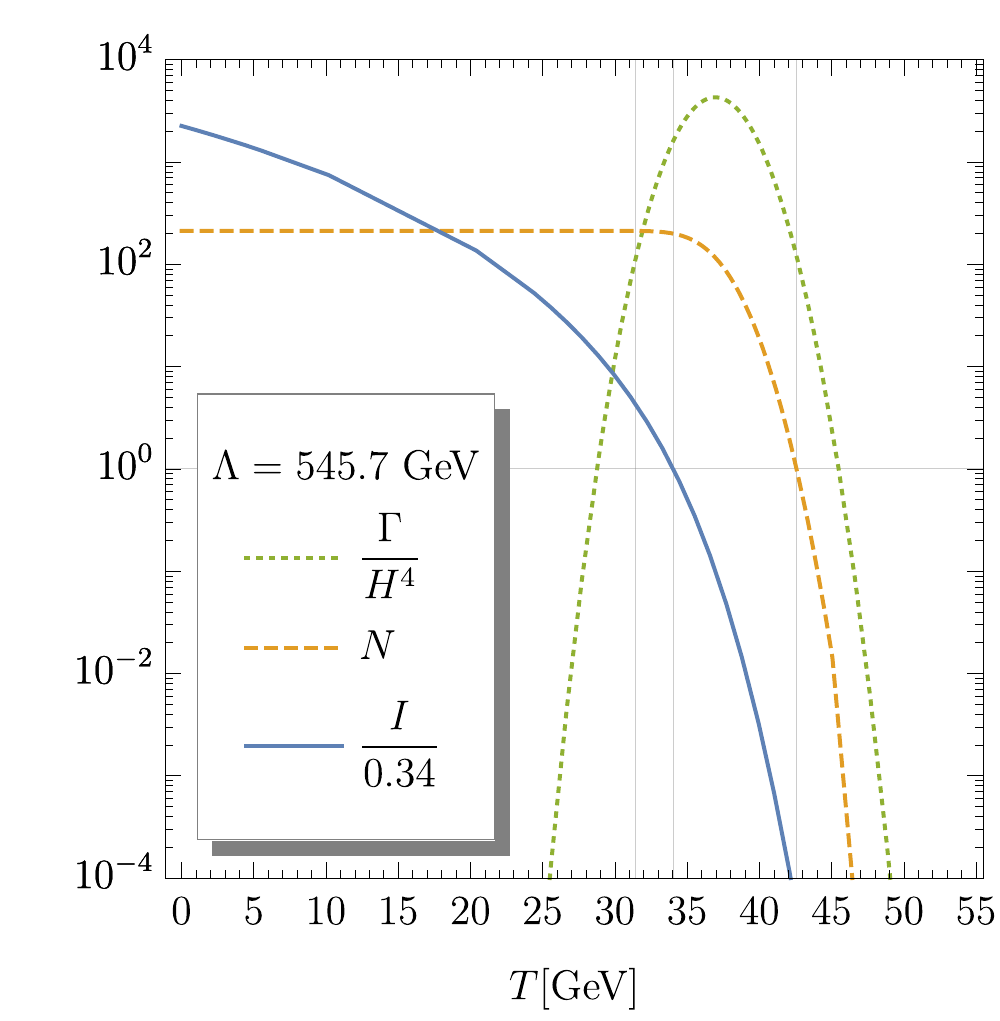}
\caption{\it
Values of $\Gamma/H^4$ (green dashed line), the number $N$ of bubbles per horizon~\eqref{eq:T_n} 
(red dashed line) and $I(T)$ (see~\eqref{eq:prob_false_vacuum} and~\eqref{eq:prob_false_vacuum_2}), as a function of $T$ for 
$\Lambda = 545 \ {\rm GeV}$, corresponding to the strongest transition where percolation can still be possible (left panel) 
and $\Lambda = 545.7 \ {\rm GeV}$, corresponding to the strongest transition for which percolation is assured (right panel).
The vertical lines show (from left to right) the temperatures $T_p$, $T_V$ and $T_n$.
\label{fig:Lambda6NucleationExample}}
\end{figure}

We turn now to the discussion of the quantities relevant for obtaining the GW spectrum from the phase transition. 
Figure~\ref{fig:H6alphaplot} shows the values of $\alpha$ as a function of $\Lambda$ obtained from~\eqref{eq:alpha}.
We find that percolation is assured only if $\alpha<1$, i.e.~when vacuum energy does not yet dominate the expansion of the Universe. 
However, in scenarios where percolation is not assured but could still be possible, this value goes up to a factor of a few,
corresponding to a brief period of vacuum domination.
In Figure~\ref{fig:H6HRplot} we plot $H R$ for the various scales $R$ relevant for GW generation: $R_{\rm MAX}$ (solid blue),
the size of the bubbles carrying the largest fraction of energy on completion of the transition, and $R_*$ (dashed green), the mean bubble separation.
We also show the approximation to the latter assuming 
radiation domination $R_{*R}$~\eqref{eq:bubble_separation_RD} (dashed red) and vacuum domination $R_{*V}$~\eqref{number_density_Gaussian} (dash-dot yellow).
We see that the three first quantities agree very well in this model as long as the vacuum contribution to the total energy 
density is negligible. 
When this is no longer the case (for $\Lambda \lesssim 550 \, {\rm GeV}$) the approximation $R_{*R}$ fails (as expected) 
and has to be replaced with the full result. The approximation $R_{*V}$ works when 
vacuum energy dominates the expansion, but the full result $R_{*}$ is still needed to interpolate between this case and 
that of radiation domination. Finally, we note that in this model the values of $R_{\rm MAX}$ and $R_*$ are very similar,
with only a very mild mismatch for the strongest phase transitions.

\begin{figure}[h]
\centering
\includegraphics[width=0.78\textwidth]{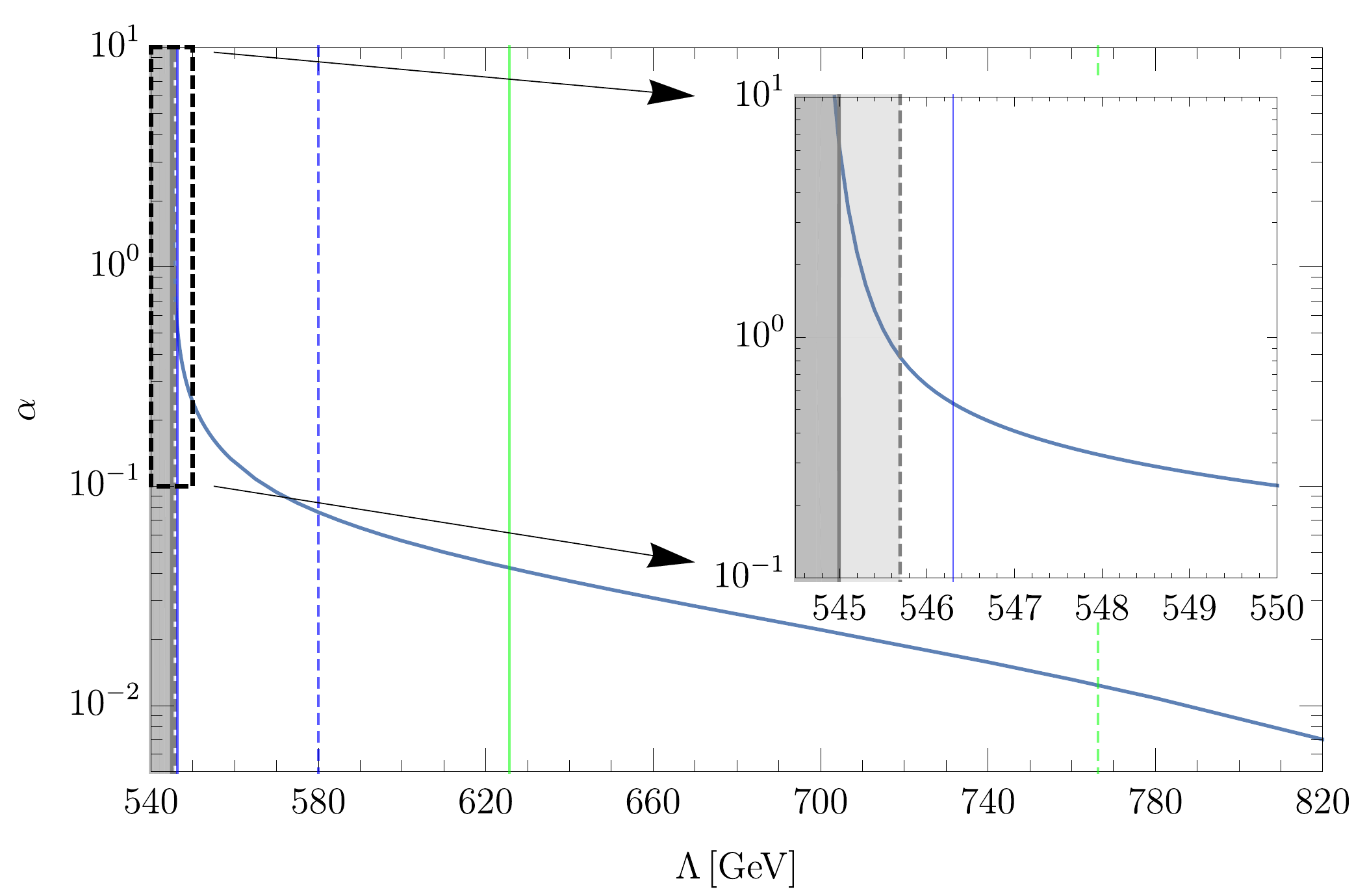}
\caption{ \it
Value of $\alpha$ as a function of $\Lambda$.
The dark grey area is excluded by our percolation criterion \eqref{eq:falsevacuumvol},
while in the light grey area percolation is questionable as the criterion is only satisfied at a temperature below $T_p$.
Vertical lines show the projected reaches of various detection methods, as in Figure~\ref{fig:Lambda6Tplot}.
\label{fig:H6alphaplot}}
\end{figure}

We find that for values $\Lambda < 580 \, {\rm GeV}$ the expanding broken phase bubbles 
satisfy the leading-order runaway criterion~\cite{Bodeker:2009qy} (recall the discussion in Section~\ref{sec:hydro}), 
which can be expressed as~\cite{Espinosa:2010hh,Caprini:2015zlo}
\begin{equation}
\alpha > \alpha_{\infty} = 4.9 \times 10^{-3}\, \left(\frac{\left\langle h \right\rangle_p}{T_p} \right)^2 \, .
\end{equation}
Thus, for $\Lambda < 580 \, {\rm GeV}$ we expect ultrarelativistic bubble expansion velocities $v_w \to 1$.
At the same time, we note that for the maximum dilution of the plasma compatible with percolation, corresponding to $T_p \simeq 29$ GeV 
the bubbles continue to reach a stationary state shortly after nucleation~\cite{Bodeker:2017cim} 
since there is just a mild hierarchy between $T_c$ and $T_p$.

\vspace{2mm}

We then check the applicability of the sound wave GW lattice simulations leading to~\eqref{eq:amplitude_soundwaves}, which we recall 
requires $H R_*/\bar{U}_f > 1$. 
In Figure~\ref{fig:SWapplicableplot} (top) we show the relation between $H R_*$ and 
$\bar{U}_f$ through the entire range of $\Lambda$ predicting a first-order phase transition.  
For very strong transitions we expect $v_w \to 1$. Nevertheless
we plot the results for several different bubble wall velocities $v_w$ so as to check the applicability more 
thoroughly and for weaker transitions. 
Our results show that the formula~\eqref{eq:amplitude_soundwaves} for the GW amplitude from sound waves cannot reliably 
be applied and will overestimate the GW 
signal~\footnote{We note we can only conclude this for bubbles expanding as detonations (recall 
the discussion in~\ref{sec:hydro}), as we have not analyzed the case of deflagrations. We note,
however, that these occur for rather slow bubbles ($v_w < c_s$), and thus 
will generally not be relevant for GW generation.} 
(we note that the region that seemingly satisfies $H R_*/\bar{U}_f > 1$ does not guarantee percolation).
The bottom panel of Figure~\ref{fig:SWapplicableplot} shows the same result in terms of 
the more commonly used variables $\alpha$ and $\beta/H$. 
The dotted part of the line indicates where the standard definition of $\beta/H$ 
as the transition timescale~\eqref{eq:beta} ceases to be accurate, 
and a more thorough treatment should be used. 
The blue regions at the bottom of the Figure indicate parameter values for which the sound wave GW spectrum 
prediction should be robust, with the conclusion being again clearly that the GW signal prediction from sound waves is 
not reliable anywhere in the parameter space.

\begin{figure}[h]
\centering
\includegraphics[width=0.88\textwidth]{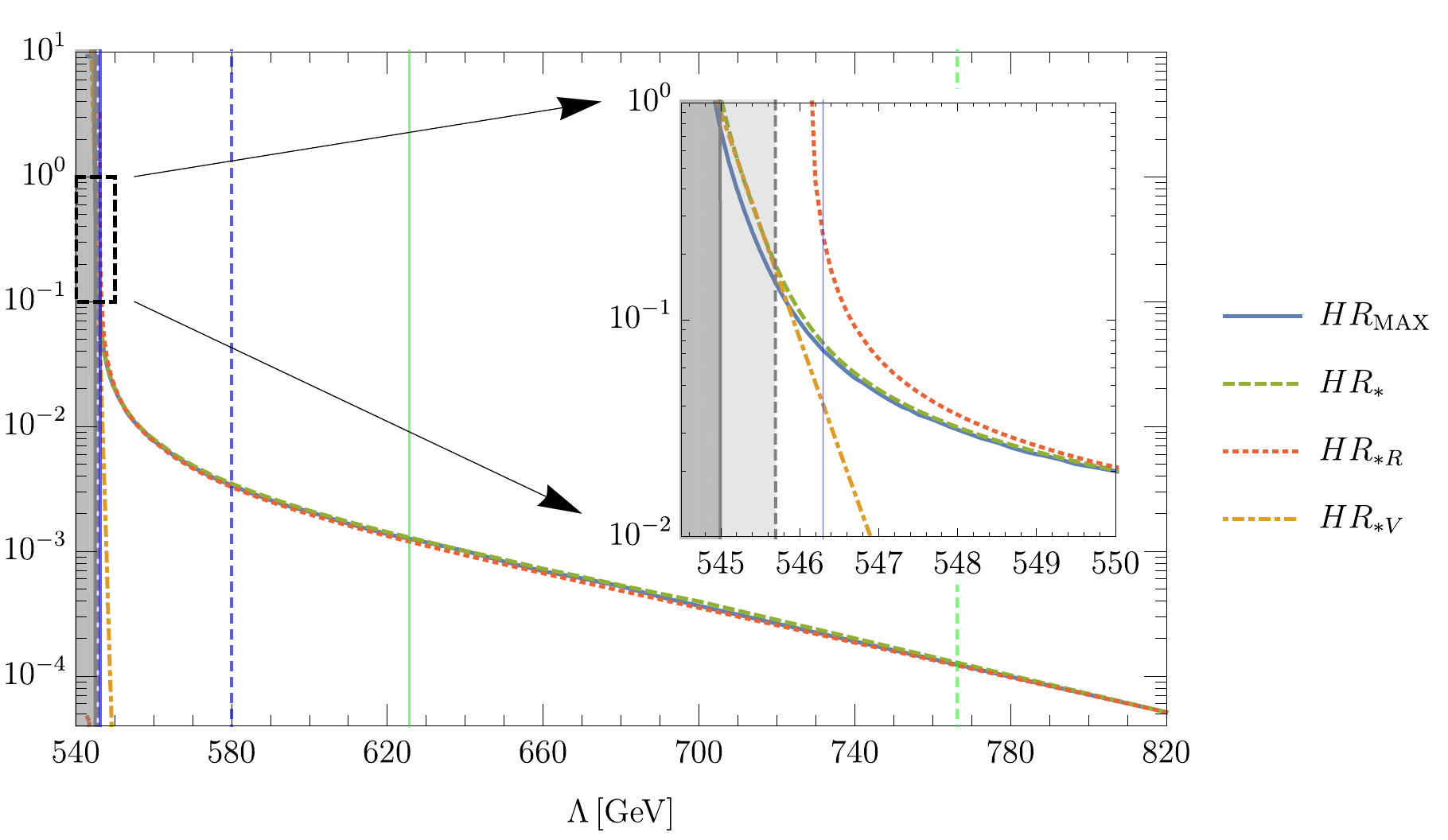}
\caption{\it
Relevant scales for GW generation: $R_{\rm MAX}$ (solid blue), $R_*$ (dashed green) and two approximations to the latter
valid respectively in radiation domination, $R_{*R}$~\eqref{eq:bubble_separation_RD} (dashed red) and 
vacuum domination, $R_{*V}$~\eqref{number_density_Gaussian} (dash-dot yellow).
Vertical lines show the projected reach of various detection methods, as in Figure~\ref{fig:Lambda6Tplot}.
\label{fig:H6HRplot}}
\end{figure}

Bearing in mind that these yield an overestimate of the final GW amplitude from sound waves 
(and a corresponding underestimate of the GW amplitude from turbulence), we
compute the GW signals using the {formulae from} Section~\ref{sec:GWsignals},
and show the results in Figure~\ref{fig:GWresTp}. The upper panel corresponds to the sound wave GW spectrum,
while the lower panel corresponds to that from MHD turbulence.
As also seen in Figure~\ref{fig:GWresTp}, the value of $\Lambda$ in the $|H|^6$ extension of the Standard Model
that could be probed by LISA is $\Lambda < 580$ GeV according to the sound wave GW spectrum prediction,
and $\Lambda < 546.3$ GeV for the GW spectrum from turbulence.
In the Figure we show explicitly the 
current and future sensitivities (see also~\cite{Figueroa:2018xtu}) of the European Pulsar Timing Array~(EPTA)~\cite{vanHaasteren:2011ni} 
and LIGO~\cite{TheLIGOScientific:2014jea,Thrane:2013oya,TheLIGOScientific:2016wyq}, as well as the
projected sensitivities of LISA~\cite{Bartolo:2016ami}, the
Einstein Telescope~(ET)~\cite{Punturo:2010zz,Hild:2010id}, the
 Cosmic Explorer~(CE)~\cite{Evans:2016mbw}, and
the Square Kilometre 
Array~(SKA)~\cite{Janssen:2014dka}.
Finally, we also show the prospective
sensitivities of the DECIGO~\cite{Kawamura:2006up} and Big Bang Observer (BBO)~\cite{Yagi:2011wg,Crowder:2005nr} projects in the frequency 
range between LISA and LIGO (another project aimed at probing
a similar frequency range is MAGIS-100~\cite{Graham:2017pmn}).
{From the results shown in Figure~\ref{fig:GWresTp} it is also clear that there exists a lower bound 
on the GW peak frequency from successful percolation, which} leads to
the conclusion that PTA experiments will not be sensitive to GWs from a first-order electroweak phase transition,
contrary to some earlier claims~\cite{Kobakhidze:2017mru,Cai:2017tmh}.

\begin{figure}[ht]
\centering
\includegraphics[width=0.655\textwidth]{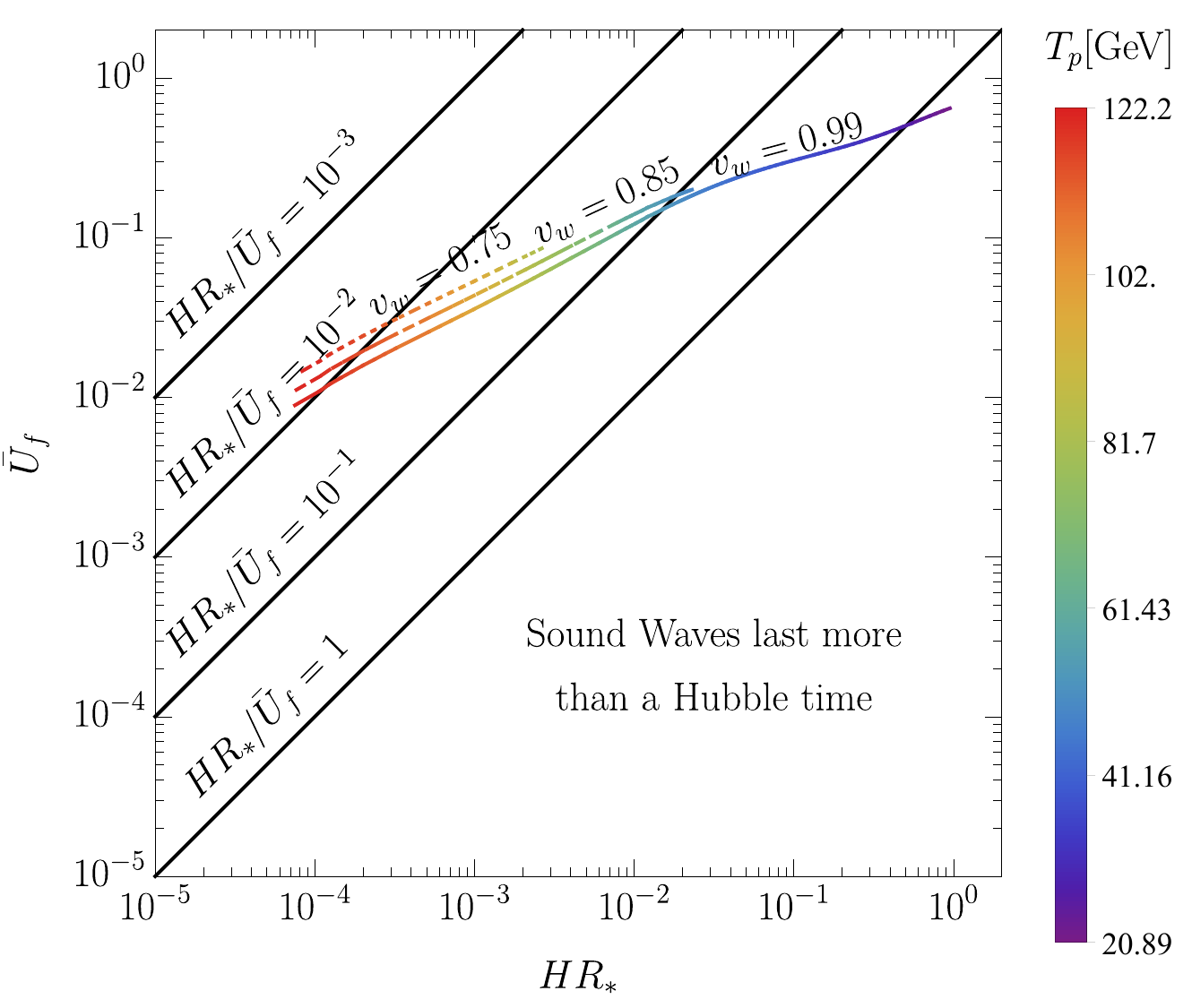}
\includegraphics[width=0.655\textwidth]{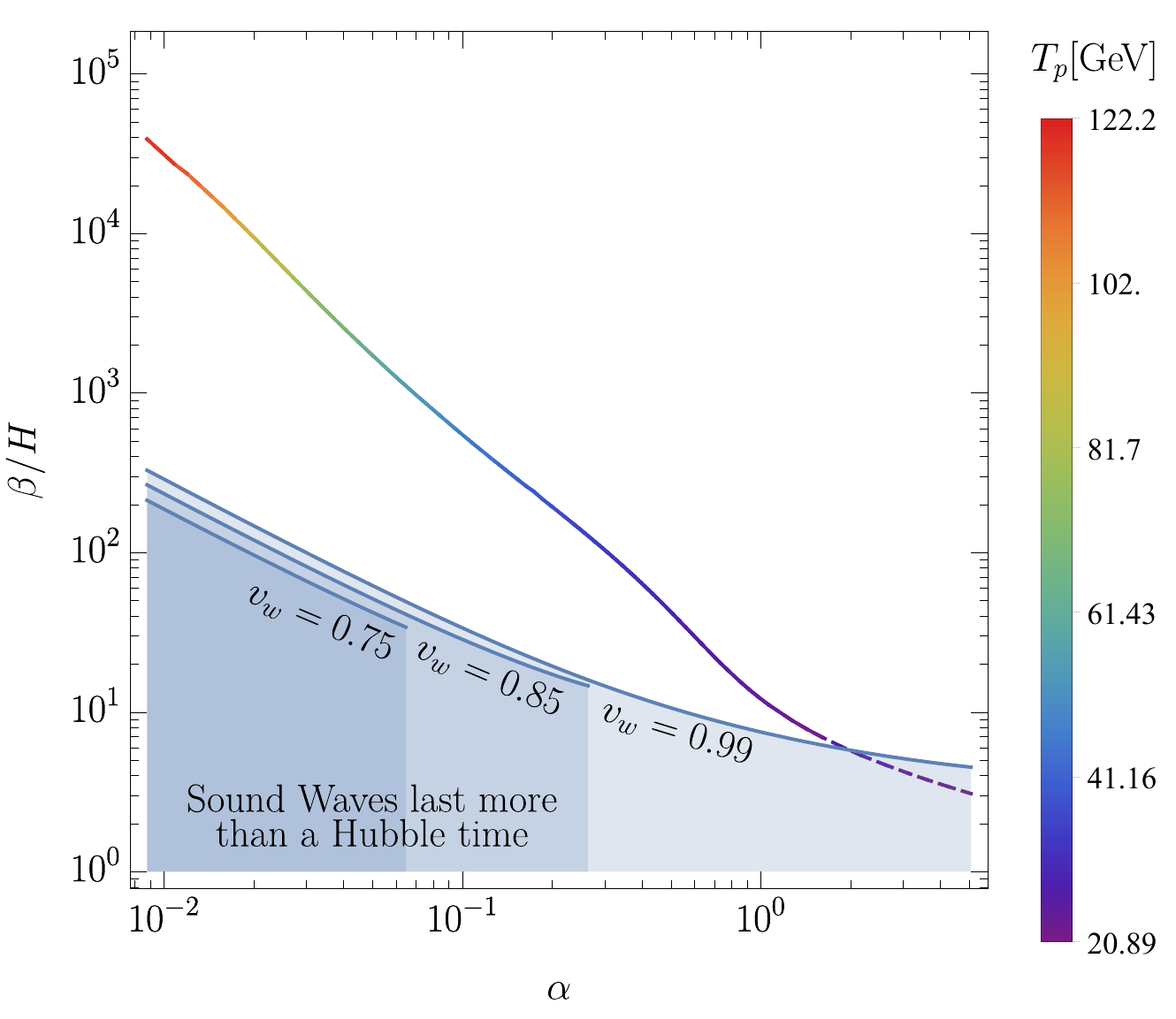}
\caption{\it
Top panel: Plasma RMS velocity $\bar{U}_f$~\eqref{eq:plasma_rms_velocity} as a function of the relevant scale for GW generation $R_*$ (normalized to 
$H^{-1}$), for the range of $\Lambda$ predicting a strong first-order electroweak phase transition
(only if $H R_*/\bar{U}_f > 1$ can~\eqref{eq:amplitude_soundwaves} be used to reliably predict the GW spectrum
from sound waves). We show the results for several different bubble wall velocities $v_w$ as indicated in the plot. 
Bottom panel: Same result in terms of the more commonly used variables $\alpha$ and $\beta/H$. 
(The dotted part of the line indicates where the standard approximation~\eqref{eq:beta} used to define $\beta/H$ breaks down.)
Blue regions at the bottom of the Figure indicate parameter values for which the sound wave GW spectrum prediction should be robust.} 
\label{fig:SWapplicableplot}
\end{figure}

\begin{figure}
\centering
\includegraphics[width=0.93\textwidth]{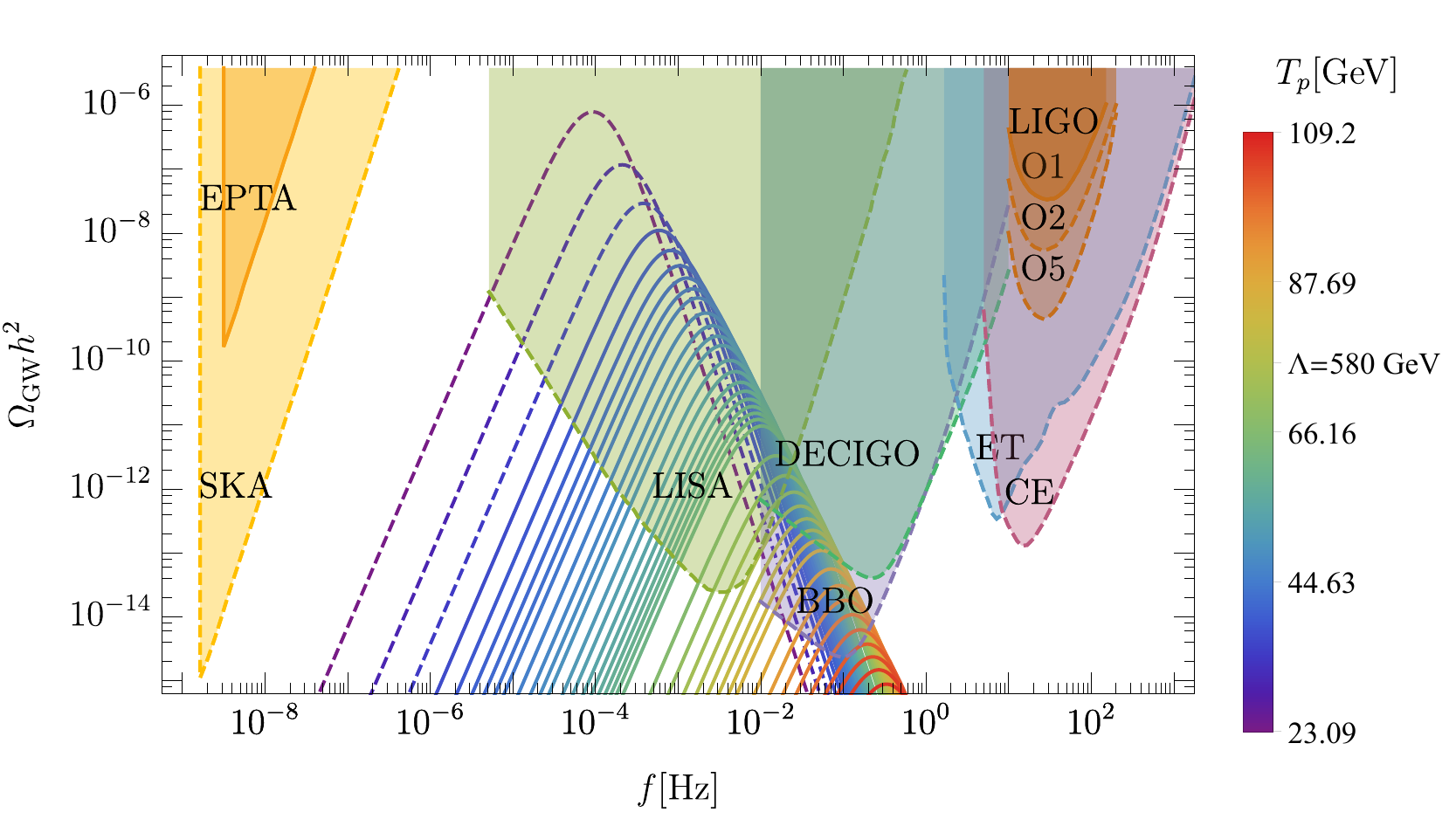}\\
\includegraphics[width=0.93\textwidth]{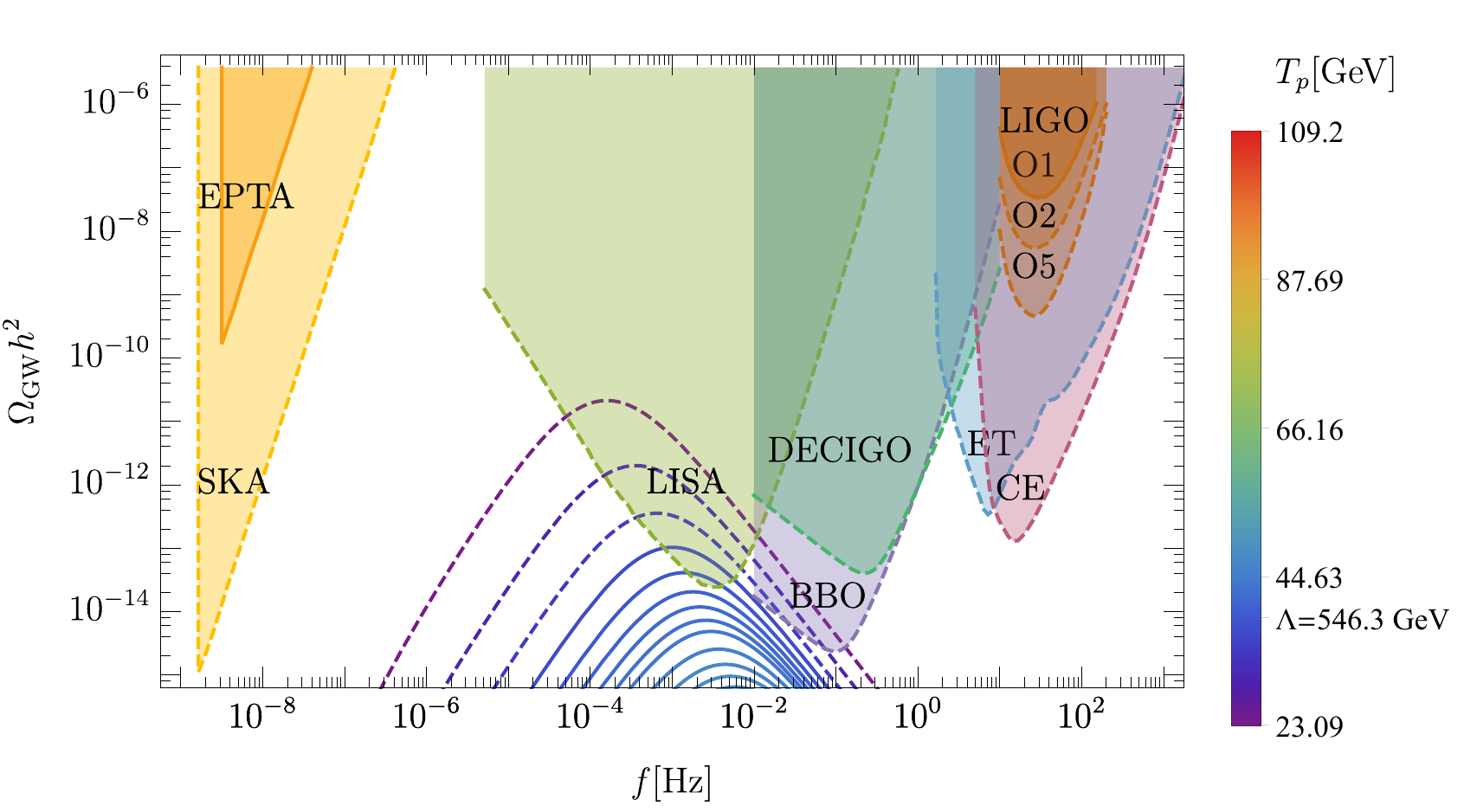}
\caption{\it
The GW signal as a function of frequency for different values of the percolation temperature $T_p$. 
The upper panel shows the sound wave contribution to the GWs, which is questionable, and the lower panel
shows the contribution from turbulence, which we regard as a reliable baseline estimate.
Dashed lines correspond to situations where the percolation criterion~\eqref{eq:falsevacuumvol} is satisfied 
only at temperatures lower than $T_p$, and successful percolation is not assured.
In the legends we also highlight the temperature corresponding to the highest value of $\Lambda$ observable by LISA.
\label{fig:GWresTp}}
\end{figure}

Finally, we show in Figure~\ref{fig:GW_Uncertainty} the corresponding GW spectra from the combination of sound waves and turbulence 
(for a few selected scenarios from those shown in Figure~\ref{fig:GWresTp}), showing in addition the possible effect of including 
a reduction in the GW amplitude from sound waves by a factor $H R_*/\bar{U}_f$, taking into account the shortening of the sound 
wave period as active GW source w.r.t. the naive {\it long-lasting}  
(active for more than a Hubble time) estimate. This effect is in any case milder for stronger transitions, as Figure~\ref{fig:GW_Uncertainty} shows.

\begin{figure}
\centering
\includegraphics[width=0.8\textwidth]{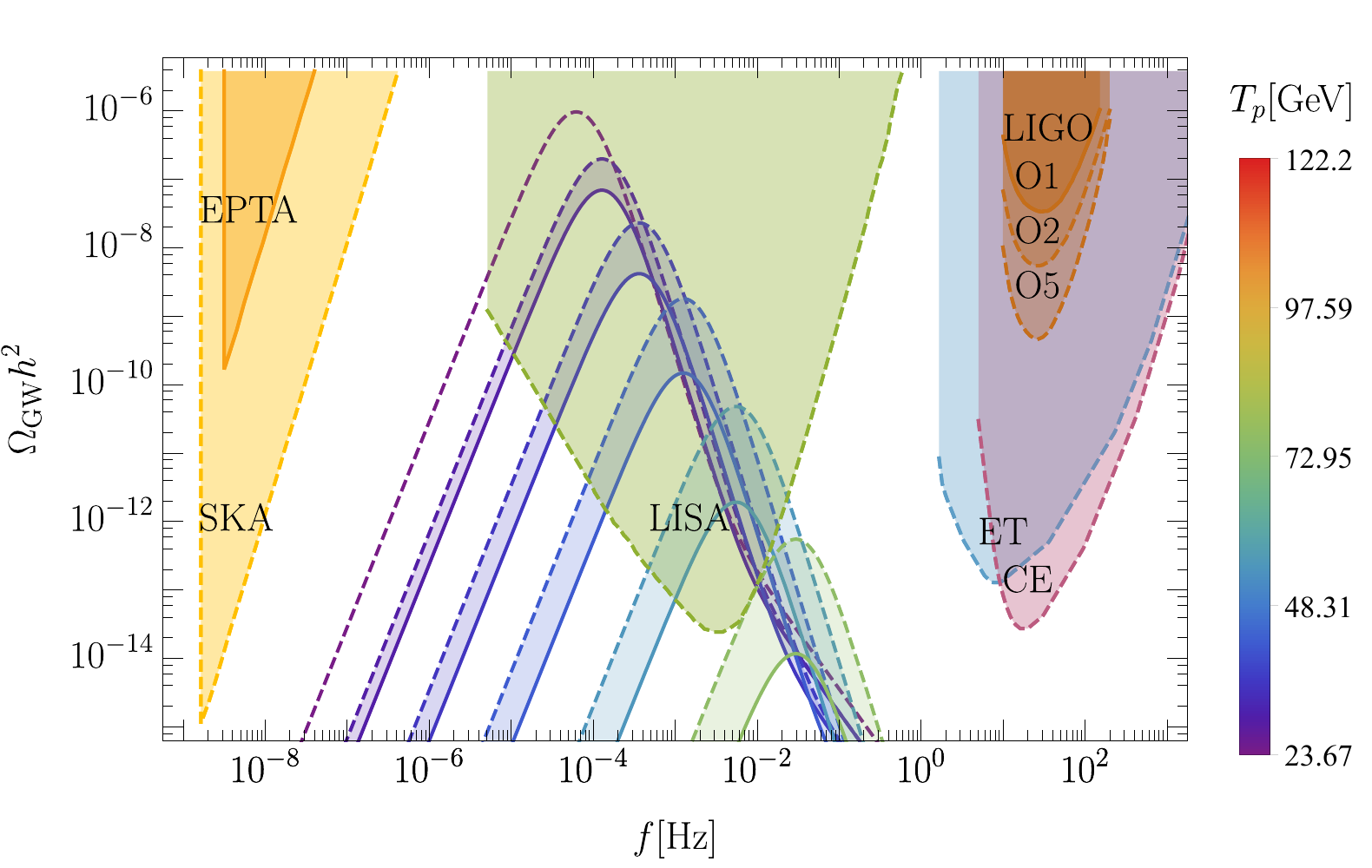}
\caption{\it
The combined GW signal (sound waves and turbulence) as a function of frequency for different values of the percolation temperature $T_p$. 
The uncertainty bands correspond to the GW amplitude range from including / not including a reduction factor in the sound wave GW amplitude by a factor 
$H R_*/\bar{U}_f$ (corresponding to the shortening of the sound wave period as active GW source w.r.t. the {\it long-lasting}  
estimate).
\label{fig:GW_Uncertainty}}
\end{figure}

\begin{figure}[h!]
\centering
\includegraphics[width=0.7\textwidth]{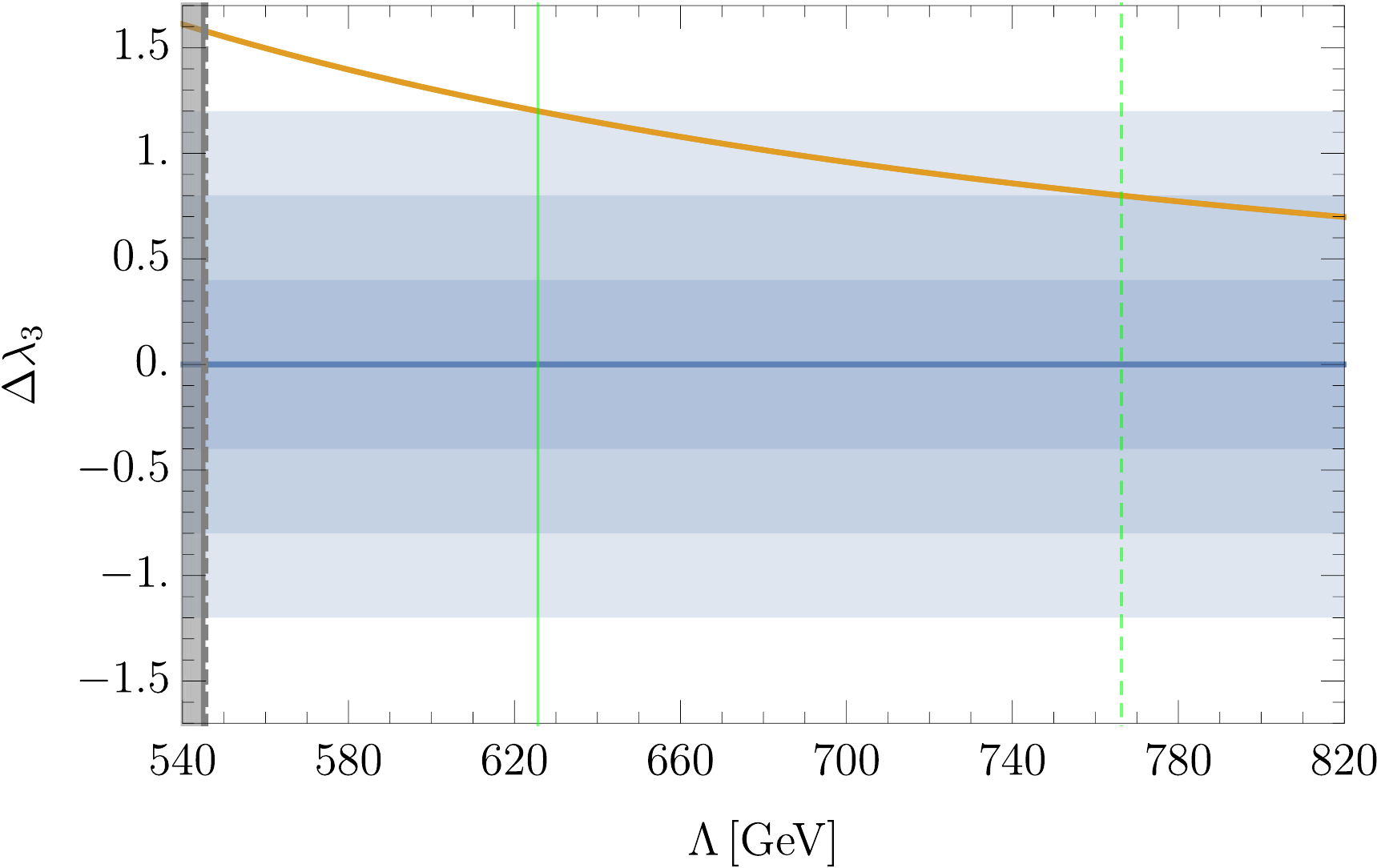}
\caption{\it
The modification of the triple Higgs coupling with respect to the SM value induced by the $|\phi|^6$
 nonrenormalisable operator: $\Delta \lambda_3=(\lambda_3-\lambda_3^{\rm SM})/\lambda_3^{\rm SM}$. The coloured areas correspond to 
 the $1$-, $2$- and 3-$\sigma$ experimental reach of the HL-LHC, and the green vertical lines indicate values of $\Lambda$ to which HL-LHC
will be sensitive at these significance levels.
\label{fig:lambda3plot}}
\end{figure}

{The $|H|^6/\Lambda^2$ extension} of the Standard Model can also
be probed indirectly at colliders {through the modification of the Higgs self-coupling
induced by the $|H|^6/\Lambda^2$ term}
\begin{equation}\label{mod}
\lambda_3 = \left. \frac{1}{6} \frac{\partial^3 V}{\partial h^3} \right|_{h=v} = \frac{m_h^2}{2v} + \frac{ v^3}{\Lambda^2} 
 = {\lambda_3^{\rm SM} + \frac{ v^3}{\Lambda^2}}\, .
\end{equation} 
{Currently there is no significant experimental constraint on $\lambda_3$,}
but the HL-LHC will be able to constrain this coupling to within about $40 \%$ of the SM 
result at 68\% C.L.~\cite{Goertz:2013kp,Barger:2013jfa,Barr:2014sga}. 
Figure~\ref{fig:lambda3plot} shows the modification with respect to the SM value induced by the 
non-renormalisable operator $\Delta \lambda_3=(\lambda_3-\lambda_3^{\rm SM})/\lambda_3^{\rm SM} = {(2\,v^4)/(m_h^2\,\Lambda^2)}$. 
The coloured areas correspond to the $1$-, $2$- and 3-$\sigma$ experimental reach 
of the HL-LHC, and the thin vertical lines indicate values of $\Lambda$ to which HL-LHC
will be sensitive at these significance levels
The modification diminishes as the cutoff scale grows, but HL-LHC would still observe a 
2-$ \sigma$ deviation up to $\Lambda\approx 766\, {\rm GeV}$ and 3-$ \sigma$ up to $\Lambda\approx 625\, {\rm GeV}$,
covering a significant part of the parameter space of interest.

\subsection{Singlet scalar extension of the Standard Model}
\label{sec:singlet}

We now consider the addition to the SM of a real singlet scalar field $s$, assuming that
the scalar potential is invariant under a $\mathbb{Z}_2$ symmetry, which is arguably the simplest renormalisable extension 
of the SM. The tree-level scalar potential reads
\begin{eqnarray}
\label{ScalarPotential_singlet}
V^{\rm tree}(H,s) = -\mu^2_h \left|H\right|^2 + \lambda_h \left|H\right|^4 + \frac{\mu^2_s}{2} s^2 + \frac{\lambda_s}{4} s^4
+ \frac{\lambda_{hs}}{2} \left|H\right|^2 s^2 \, ,
\end{eqnarray}
with the mass of the scalar singlet after electroweak symmetry breaking (and assuming that the $\mathbb{Z}_2$ symmetry remains unbroken) given by
$m_s^2 = \mu^2_s + \lambda_{hs} \, v^2/2$.

The real singlet scalar extension of the SM has long been recognized as a scenario which can yield a strongly first-order electroweak phase 
transition (see e.g.~\cite{Anderson:1991zb,Choi:1993cv,Espinosa:1993bs,Profumo:2007wc,Espinosa:2011ax}), with the presence of the singlet field direction rendering
possible very strong phase transitions resulting from tree-level potential barriers between vacua. In order to illustrate this, we first consider 
the tree-level potential~\eqref{ScalarPotential_singlet} with the addition of the leading thermal corrections, corresponding to terms scaling as 
$\phi^2 T^2$ (with $\phi = h,\, s$) in a high-temperature expansion of the 1-loop thermal potential (see Appendix~A from~\cite{Beniwal:2017eik} for details). 
Such corrections read
\be
\Delta V_T(h,s) = D_h \,T^2\, h^2  + D_s \,T^2 \, s^2 \, 
\ee 
with $D_h$ and $D_s$ given by (see e.g.~\cite{Espinosa:2011ax, Chen:2017qcz})
\begin{equation}
D_h = \frac{1}{96}\left( 24 \lambda_h + 9 g^2 + 3 g'^2 + 12 y^2_t + 2 \lambda_{hs} \right) \quad,\quad
D_s = \frac{1}{24}\left(2 \lambda_{hs} + 3 \lambda_{s} \right) \, .
\end{equation}
The resulting scalar potential is simply given by
\bea
\label{ScalarPotential_singlet_total}
V(h,s) &=&  V^{\rm tree}(h,s) + \Delta V_T(h,s) \nonumber \\
&=& \left(-\frac{\mu^2_h}{2} + D_h \,T^2\right) h^2 + \frac{\lambda_h}{4} h^4 + 
\left( \frac{\mu^2_s}{2} + D_s \,T^2 \right) s^2 + \frac{\lambda_s}{4} s^4 + \frac{\lambda_{hs}}{4} s^2 h^2 \, .
\eea
The potential~\eqref{ScalarPotential_singlet_total} can give rise to a two-step phase transition process, which may then 
result in a very strong electroweak phase transition, 
as follows: for $\mu_s^2 < 0$ in~\eqref{ScalarPotential_singlet}, the singlet field direction is destabilized from the origin 
$(h,s) = (0,0)$ at $T = 0$. In the early Universe, the singlet field would then be destabilized from the origin before the Higgs doublet field if
\begin{equation}
T_s \equiv \sqrt{\frac{-\mu_s^2}{2\,D_s}} > T_h \equiv \sqrt{\frac{\mu_h^2}{2\,D_h}} \, ,
\label{2Step_EWPT}
\end{equation}
with $T_s$ and $T_h$ being the respective temperatures below which the point $(h,s) = (0,0)$ becomes unstable along the singlet and doublet 
field directions. 
If~\eqref{2Step_EWPT} holds, the singlet field develops a vev $x_0$ in the early Universe prior to electroweak symmetry breaking.
The subsequent electroweak phase transition $(0,\, x_0) \to (v,\, 0)$ is then in general strongly first-order, as the two minima will be  
separated by a tree-level potential barrier for some range of temperatures below $T_s$~\cite{Espinosa:2011ax}.

\begin{figure}[t]
\begin{center}
\includegraphics[width=0.8\textwidth]{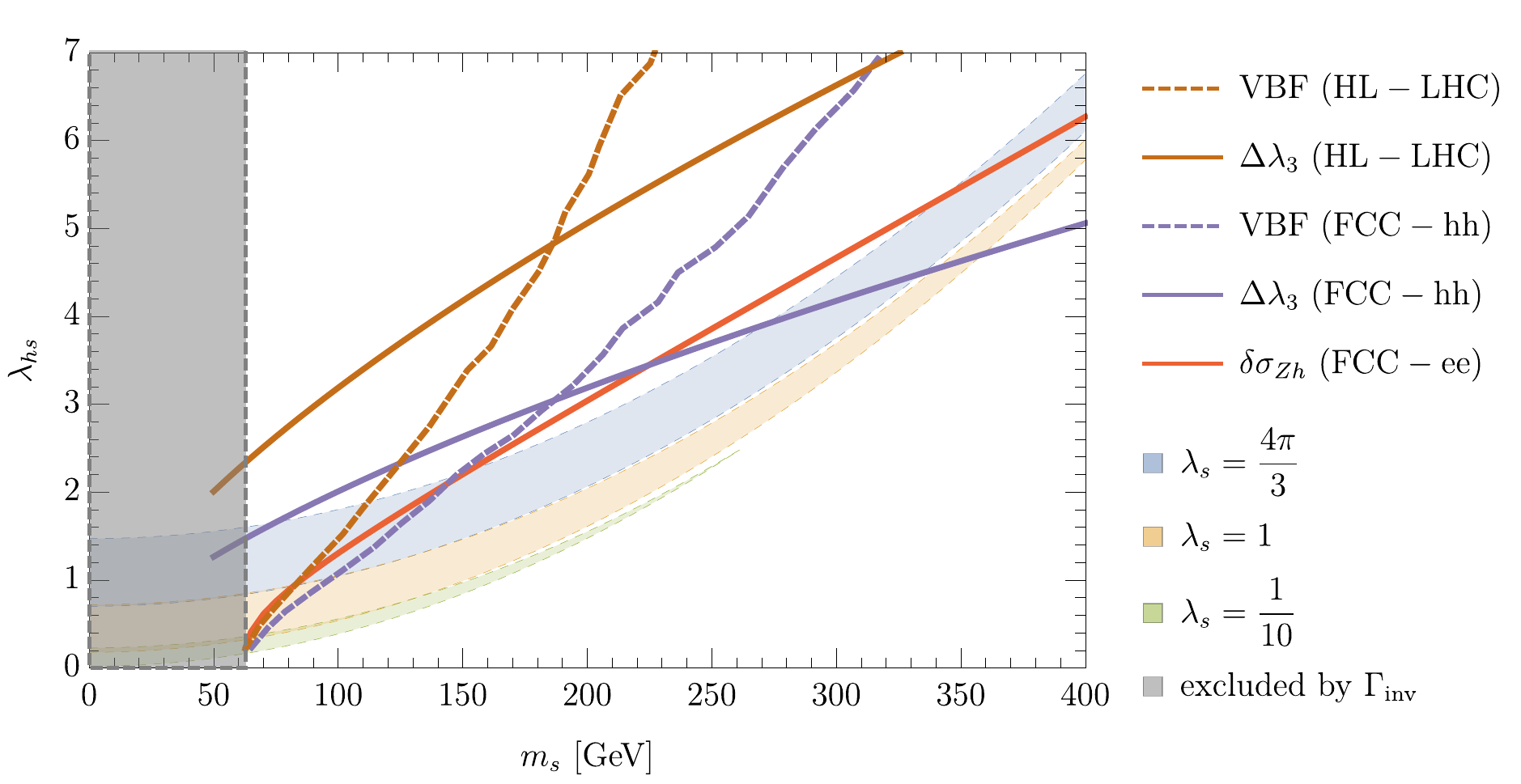}
\end{center}
\vspace{-5mm}
\caption{\it Parameter space region in the ($m_s,\,\lambda_{hs}$) 
plane yielding a viable two-step phase transition (satisfying both~\eqref{2Step_EWPT} and~\eqref{EW_2Step_EWPT}) 
in the real singlet scalar extension of the SM with a $\mathbb{Z}_2$ symmetry, 
for $\lambda_s = 0.1$ (green region), $\lambda_s = 1$ (red region) and $\lambda_s = 4 \pi/3$ (blue region).
The grey region corresponding to $m_h < 63$ GeV is excluded by the present LHC constraint on the Higgs invisible decay width. 
Also shown are the future 95\% C.L. sensitivities from multi-jet + $E^{\rm{miss}}_T$ searches via VBF at HL-LHC (dashed brown) and FCC-$hh$ (dashed blue), 
measurements of the Higgs self-coupling $\lambda_3$ at the HL-LHC (solid brown) and FCC-$hh$ (solid blue), 
and measurements of the Higgs associated production cross section $\sigma_{Zh}$ 
at FCC-$ee$ (solid red), see text for details.}
\label{Fig_Singlet_Strips}
\vspace{-3mm}
\end{figure}

The requirement~\eqref{2Step_EWPT}, together with $-\mu_s^2 = \lambda_{hs} \, v^2/2 - m_s^2 > 0$, 
yields a lower bound on $\lambda_{hs}$ as a function of $m_s$ and $\lambda_s$ 
for a two-step phase transition to be possible.
At the same time, we must require the electroweak minimum to be the absolute minimum of the potential at $T = 0$. 
The condition $V(v,0, T= 0) < V(0,x_0, T=0)$ translates into
\begin{equation}
\sqrt{\lambda_s\, \lambda_h}\,v^2 > \frac{\lambda_{hs}\, v^2}{2} - m_s^2
\label{EW_2Step_EWPT}
\end{equation}
yielding a corresponding upper bound on $\lambda_{hs}$ as a function of $m_s$ and $\lambda_s$. In Figure~\ref{Fig_Singlet_Strips} we show the 
region of parameter space in the ($m_s,\,\lambda_{hs}$)
plane satisfying~\eqref{2Step_EWPT} and~\eqref{EW_2Step_EWPT}
for $\lambda_s = 0.1,\,1,\,4\pi/3$ (the latter being the maximum allowed value by unitarity~\cite{Lewis:2017dme}).
Figure~\ref{Fig_Singlet_Strips} highlights that the two-step phase transition parameter region is very small for 
small $\lambda_s$, becoming larger as $\lambda_s$ increases.
We also note a particular value of $\lambda_{hs}$ for which the extremum $(h, s) = (0, s_0)$ at $T = 0$ turns from 
a saddle point (unstable along the $h$ direction) to a minimum. In the former case, a potential barrier is present when $T > 0$ but disappears at $T = 0$, and the 
phase transition is guaranteed to complete, while in the latter a barrier exists at $T = 0$ and the 
transition may not complete. The corresponding condition on $\lambda_{hs}$ reads
\begin{equation}
\left. \frac{\partial^2 V}{\partial \,h^2}\right|_{(h,s) = (0,x_0)} = 0 \quad \longrightarrow \quad
\frac{\lambda_{hs}}{2\,\lambda_s} \left(\frac{\lambda_{hs}\, v^2}{2} - m_s^2 \right) - \lambda_h\, v^2 = 0 \, .
\end{equation}
Above this value of $\lambda_{hs}$, the transition (if possible at all) is expected to be very strong.

Having illustrated the qualitative behaviour of the model, we now turn 
to a precise description of the electroweak phase transition in the singlet scenario including 
both one-loop corrections to the zero-temperature potential and thermal corrections (the specific details of the potential we use can be found in 
Appendix A of~\cite{Beniwal:2017eik}).
We first show in Figure~\ref{fig:Singlet_Tplot} the relevant temperatures discussed in Section~\ref{sec:FOPT} for 
$\lambda_s = 1$ and $m_s =100$ GeV (left) and $m_s = 200$ GeV (right). The amount of 
supercooling for $m_s =100$ GeV is significant, with both $T_p$ and $T_n$ dropping below 
$T_V$ for the largest allowed values of $\lambda_{hs}$. The grey region signals the range of $\lambda_{hs}$ for which the 
percolation criterion~\eqref{eq:falsevacuumvol} is not satisfied at any temperature. We note that, 
as opposed to the $\left|H\right|^6$ scenario analyzed in Section~\ref{subsec:H6},  
in the present scenario it is not possible for the percolation criterion to fail at $T = T_p$ but be 
satisfied at some lower temperature.

\begin{figure}[h]
\begin{center}
\includegraphics[width=0.495\textwidth]{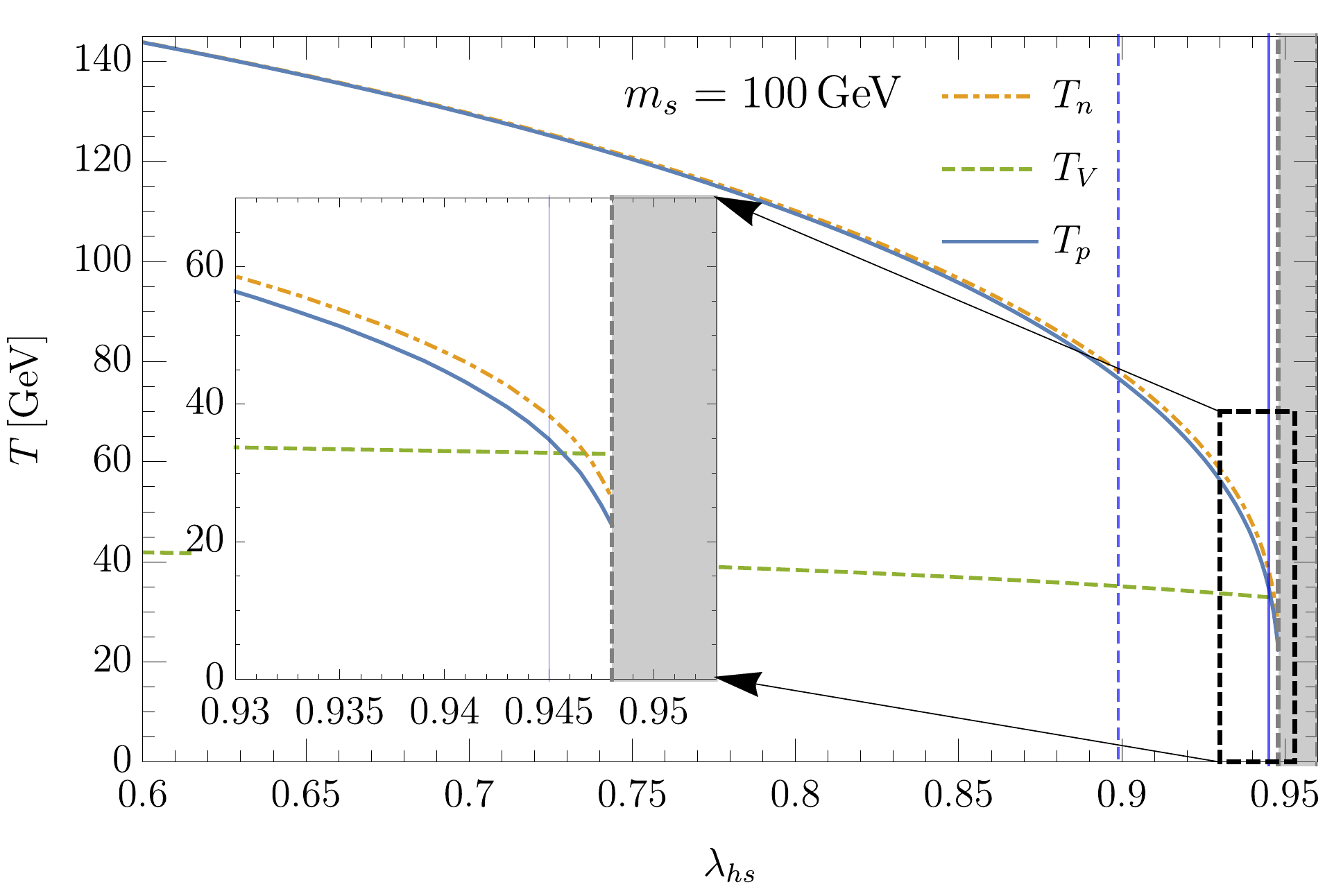}
\includegraphics[width=0.495\textwidth]{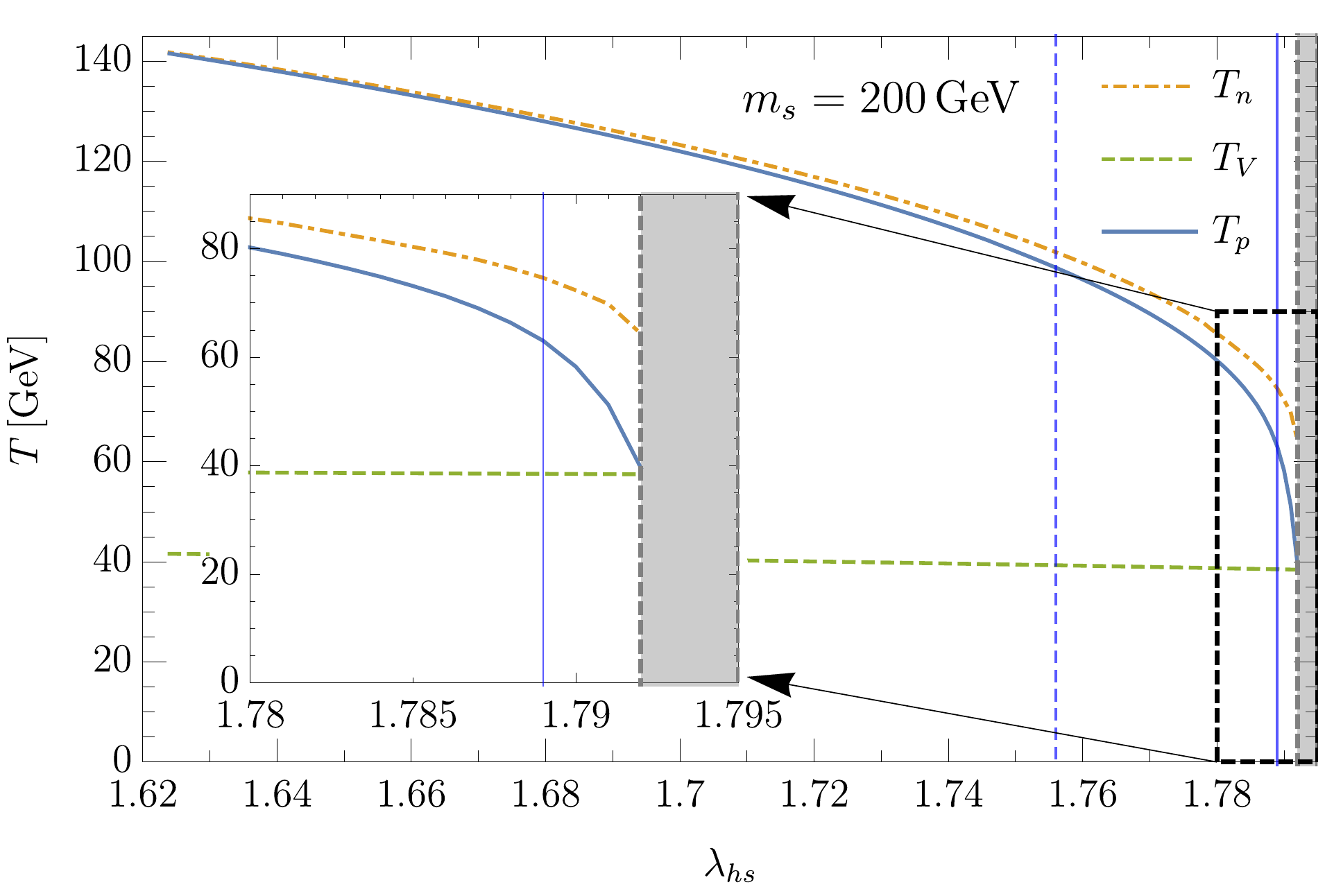}
\end{center}
\vspace{-5mm}
\caption{\it 
Nucleation temperature $T_n$ (orange dashed), percolation temperature $T_p$ (blue solid)
and temperature $T_V$ (green dashed) 
as functions of $\lambda_{hs}$ for $\lambda_s = 1$ and $m_s = 100$ GeV (left panel), $m_s = 200$ GeV (right panel).
The dark grey area is excluded by our percolation criterion~\eqref{eq:falsevacuumvol}.
Vertical lines show the projected reach of LISA for a GW signal
from sound waves (dashed blue line) and turbulence (solid blue line).
}
\label{fig:Singlet_Tplot}
\end{figure}

\begin{figure}[h]
\begin{center}
\includegraphics[width=0.495\textwidth]{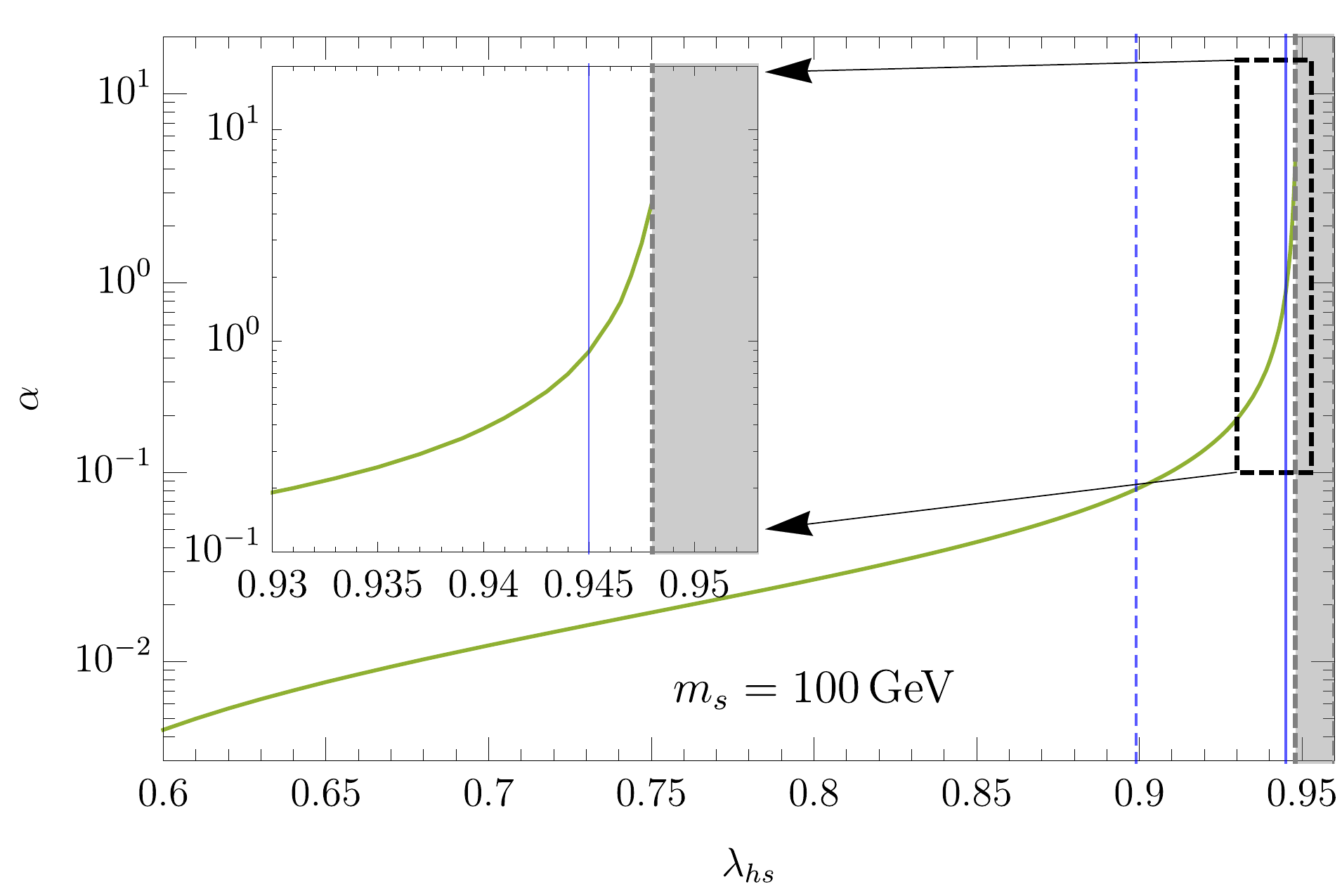}
\includegraphics[width=0.495\textwidth]{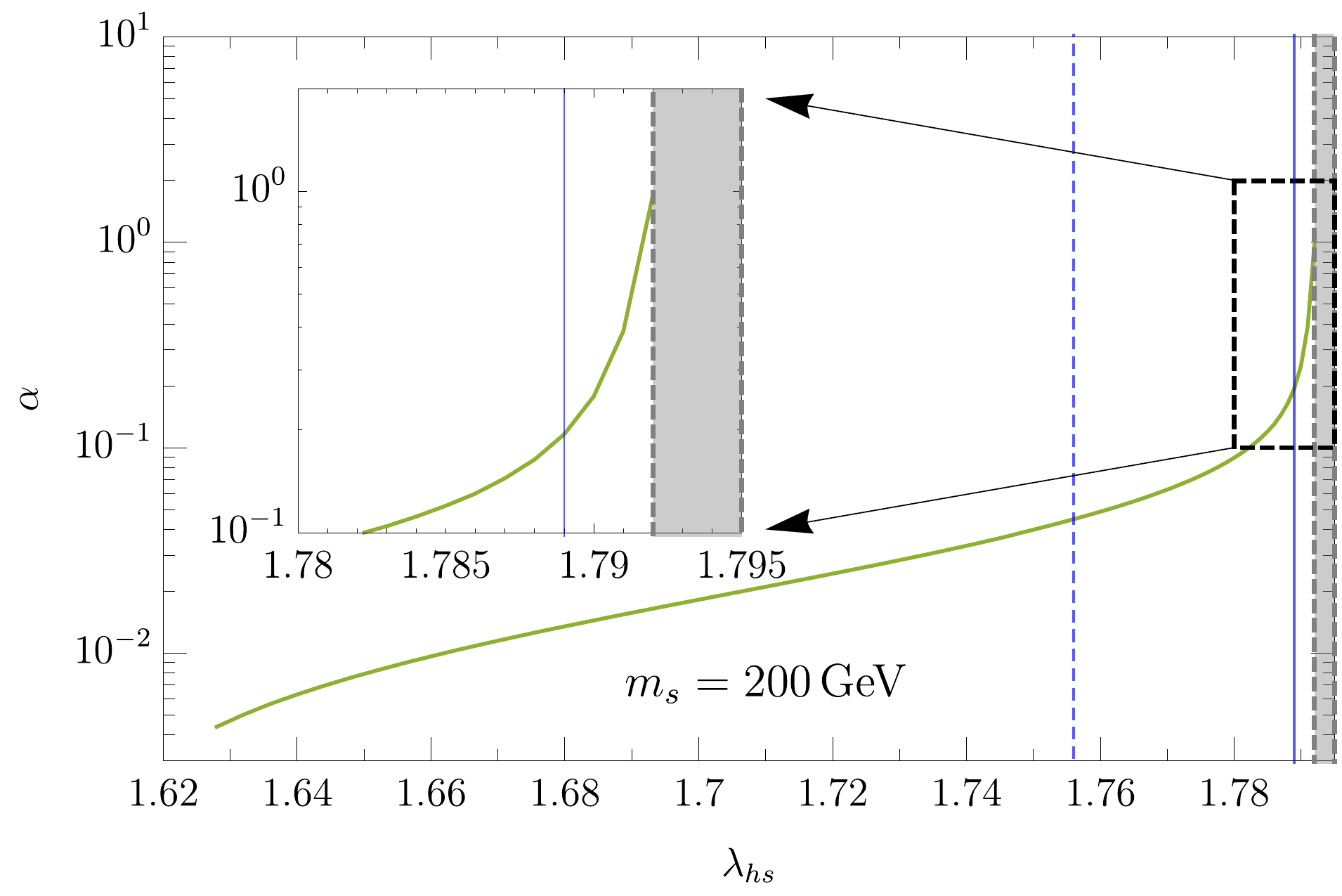}
\end{center}
\vspace{-5mm}
\caption{\it Values of $\alpha$ as a function of $\lambda_{hs}$ 
for $\lambda_s = 1$ and $m_s = 100$ GeV (left panel), $m_s = 200$ GeV (right panel).
The dark grey area is excluded by our percolation criterion \eqref{eq:falsevacuumvol}.
Vertical lines show the projected reach of LISA for a GW signal
from sound waves (dashed blue vertical line) and turbulence (solid blue vertical line).
}
\label{fig:Singlet_alphaplot}
\end{figure}

In Figure~\ref{fig:Singlet_alphaplot} we show the corresponding values of $\alpha$ for $\lambda_s = 1$ and $m_s = 100$ GeV (left) 
and $m_s = 200$ GeV (right). We note in particular that $\alpha \gtrsim 1$ 
is possible for $m_s = 100$ GeV and $\lambda_{hs}$ close to its maximal allowed value, yielding a brief period of 
vacuum domination prior to the successful completion of the phase transition.
Then, in Figure~\ref{fig:Singlet_Rplot} we plot 
$H R$ for both $R_{\rm MAX}$ (solid green) and $R_*$ (dashed orange), which are shown to be very similar in this case.

\begin{figure}[h]
\begin{center}
\includegraphics[width=0.495\textwidth]{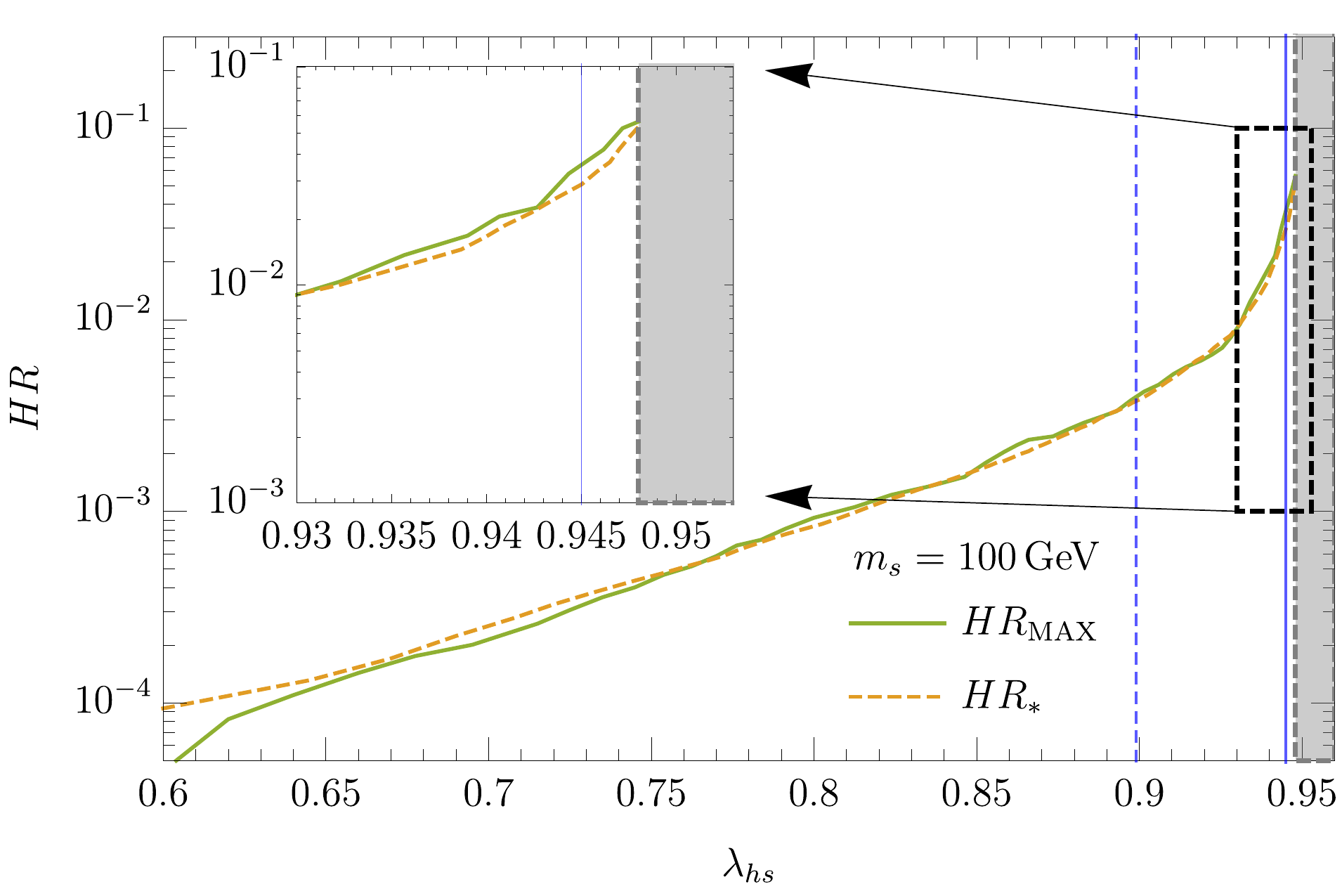}
\includegraphics[width=0.495\textwidth]{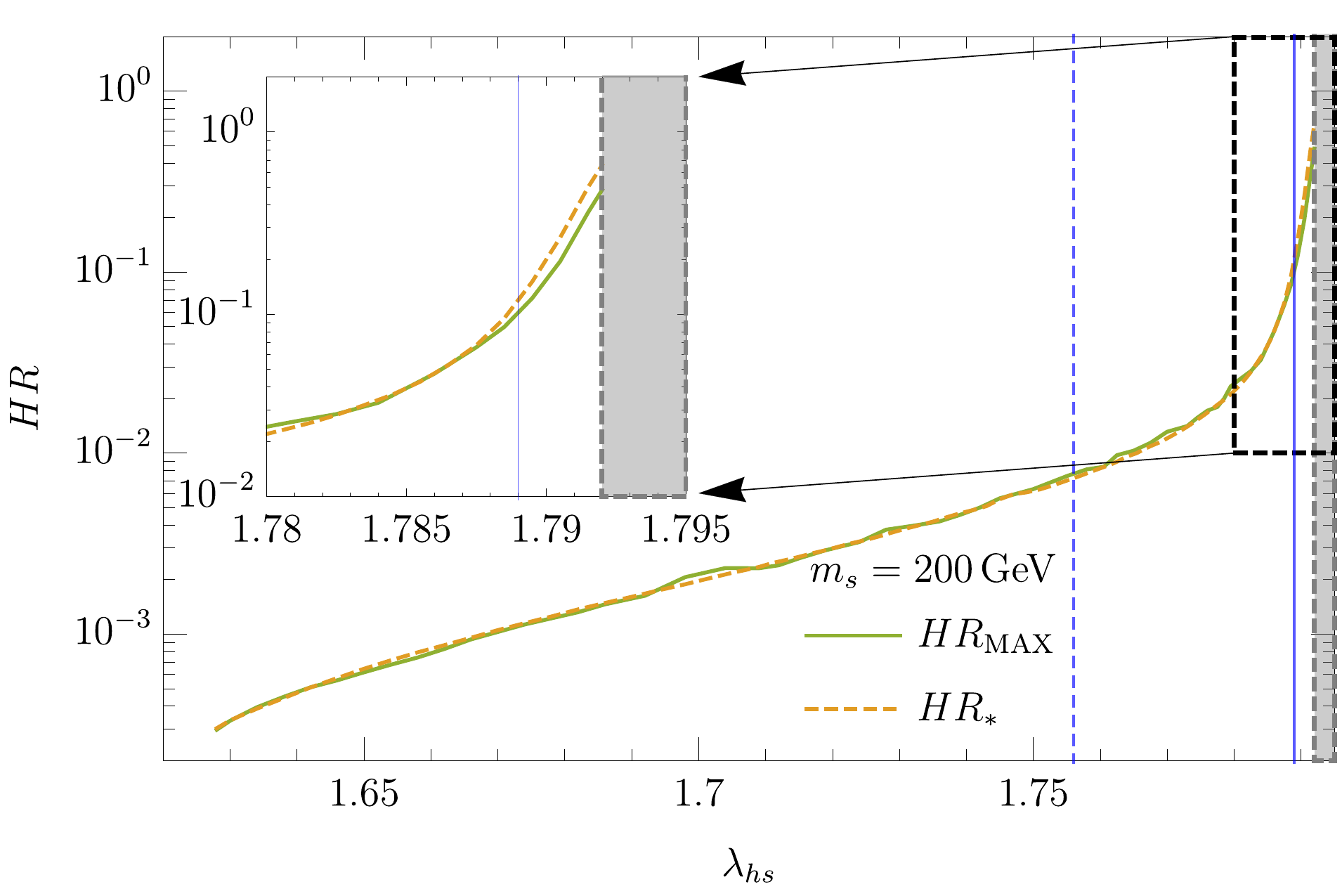}
\end{center}
\vspace{-5mm}
\caption{\it Relevant scales for GW generation: $R_{\rm MAX}$ (solid green) and $R_*$ (dashed red) as a function of $\lambda_{hs}$ 
for $\lambda_s = 1$ and $m_s = 100$ GeV (left panel), $m_s = 200$ GeV (right panel).
Vertical lines show the projected reach of LISA for a GW signal
from sound waves (dashed blue line) and turbulence (solid blue line).
}
\label{fig:Singlet_Rplot}
\end{figure}

\begin{figure}[h]
\begin{center}
\includegraphics[width=0.495\textwidth]{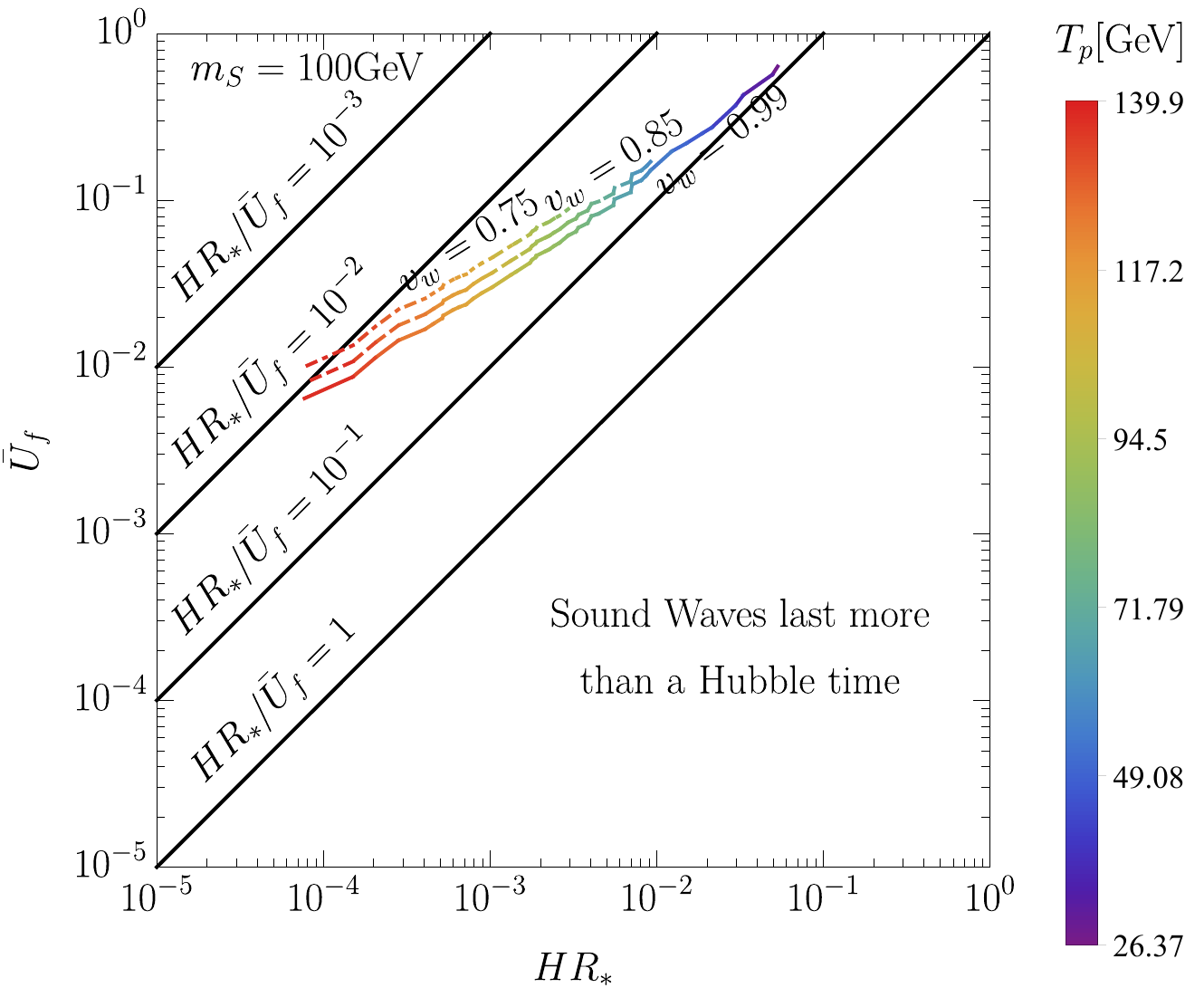}
\includegraphics[width=0.495\textwidth]{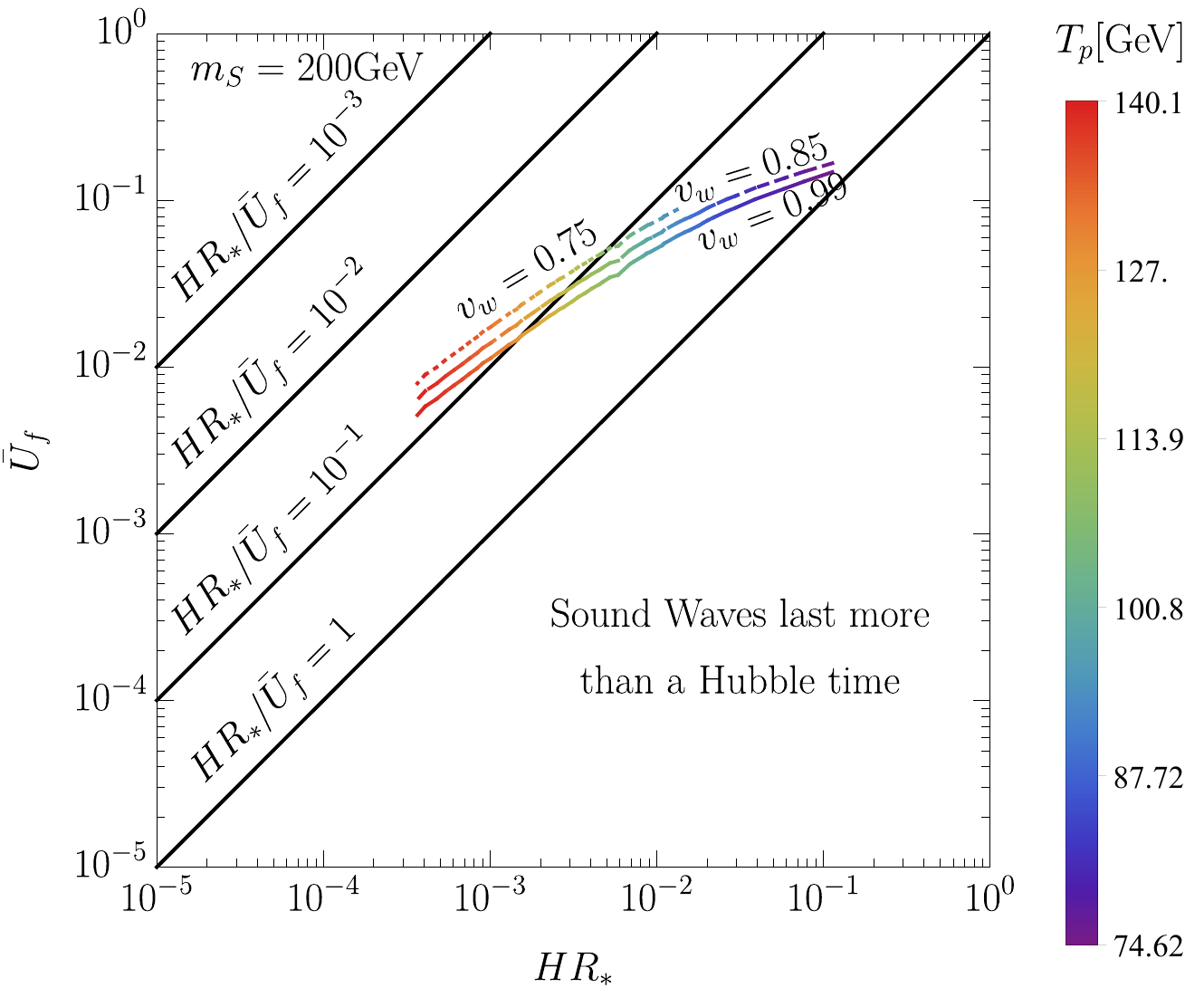}
\end{center}
\vspace{-5mm}
\caption{\it 
Plasma RMS velocity $\bar{U}_f$~\eqref{eq:plasma_rms_velocity} as a function of the relevant scale for GW generation $R_*$ (normalized to 
$H^{-1}$) for $\lambda_s = 1$ and $m_s = 100$ GeV (left panel), $m_s = 200$ GeV (right panel).
Only if $H R_*/\bar{U}_f > 1$ can~\eqref{eq:amplitude_soundwaves} be used to reliably predict the GW spectrum
from sound waves. We show the results for several different bubble wall velocities $v_w$ as indicated in the plot.
}
\label{fig:Singlet_RUplot}
\end{figure}

We can now check the applicability of the sound wave GW result~\eqref{eq:amplitude_soundwaves},  
requiring $H R_*/\bar{U}_f > 1$. The relation between $H R_*$ and 
$\bar{U}_f$ is then shown in Figure~\ref{fig:Singlet_RUplot} for $\lambda_s = 1$ and $m_s =100$ GeV (left) and $m_s = 200$ GeV (right). 
We generically expect for this model $v_w \to 1$, but we show the results for several different bubble wall 
velocities $v_w$ so as to provide a more thorough check.
Similarly to the $\left|H\right|^6$ scenario (see Figure~\ref{fig:SWapplicableplot}), our results 
indicate that the sound wave GW amplitude~\eqref{eq:amplitude_soundwaves} will significantly overestimate the overall amplitude of the GW 
signal in the present scenario. This is also shown in Figure~\ref{fig:Singlet_abplot} in terms of 
the more commonly used variables $\alpha$ and $\beta/H$, with the 
blue regions at the bottom of the figure indicating parameter values for which the sound wave GW spectrum 
prediction should be robust, which are however never reached within this model.

\begin{figure}[t]
\begin{center}
\includegraphics[width=0.495\textwidth]{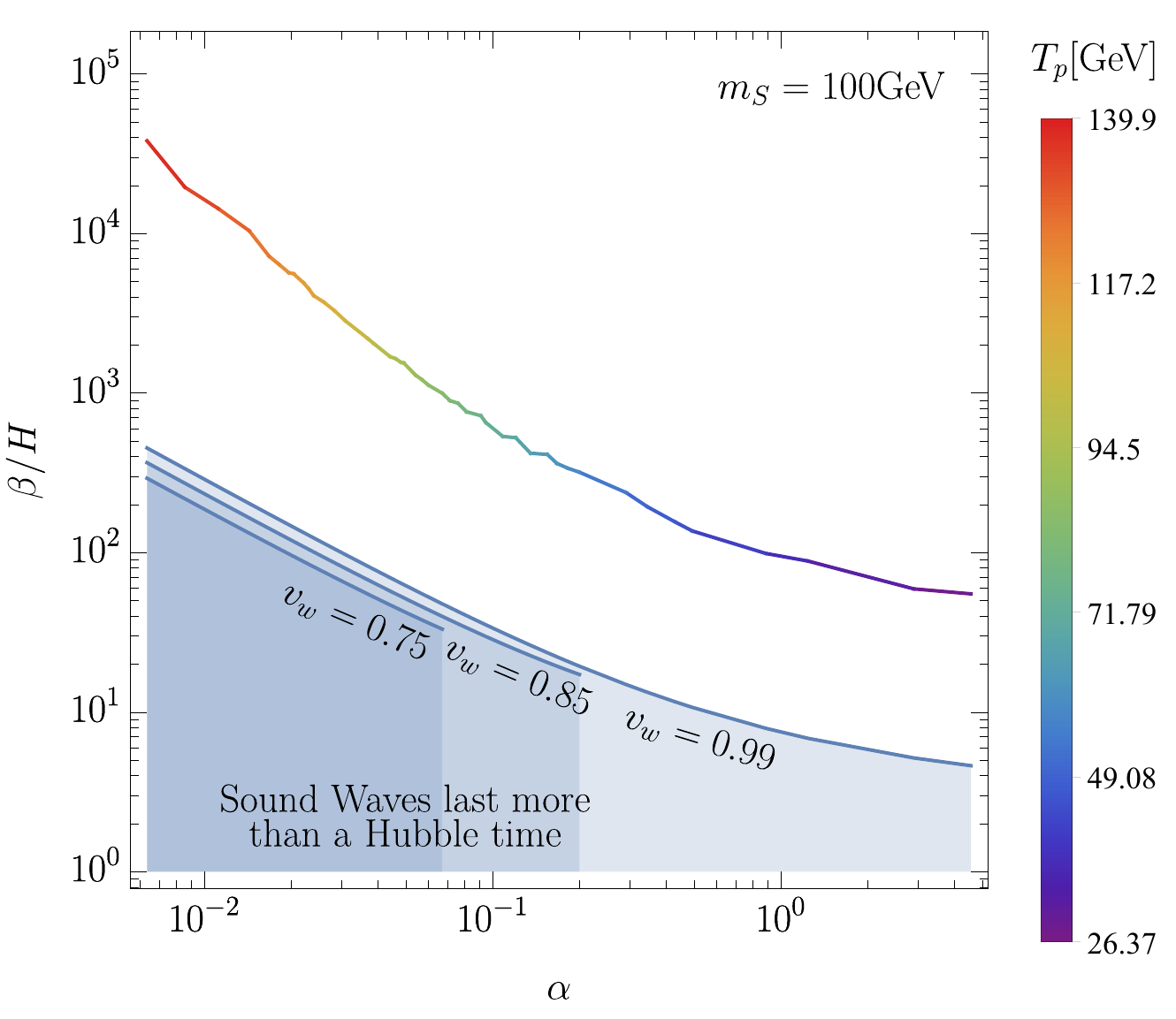}
\includegraphics[width=0.495\textwidth]{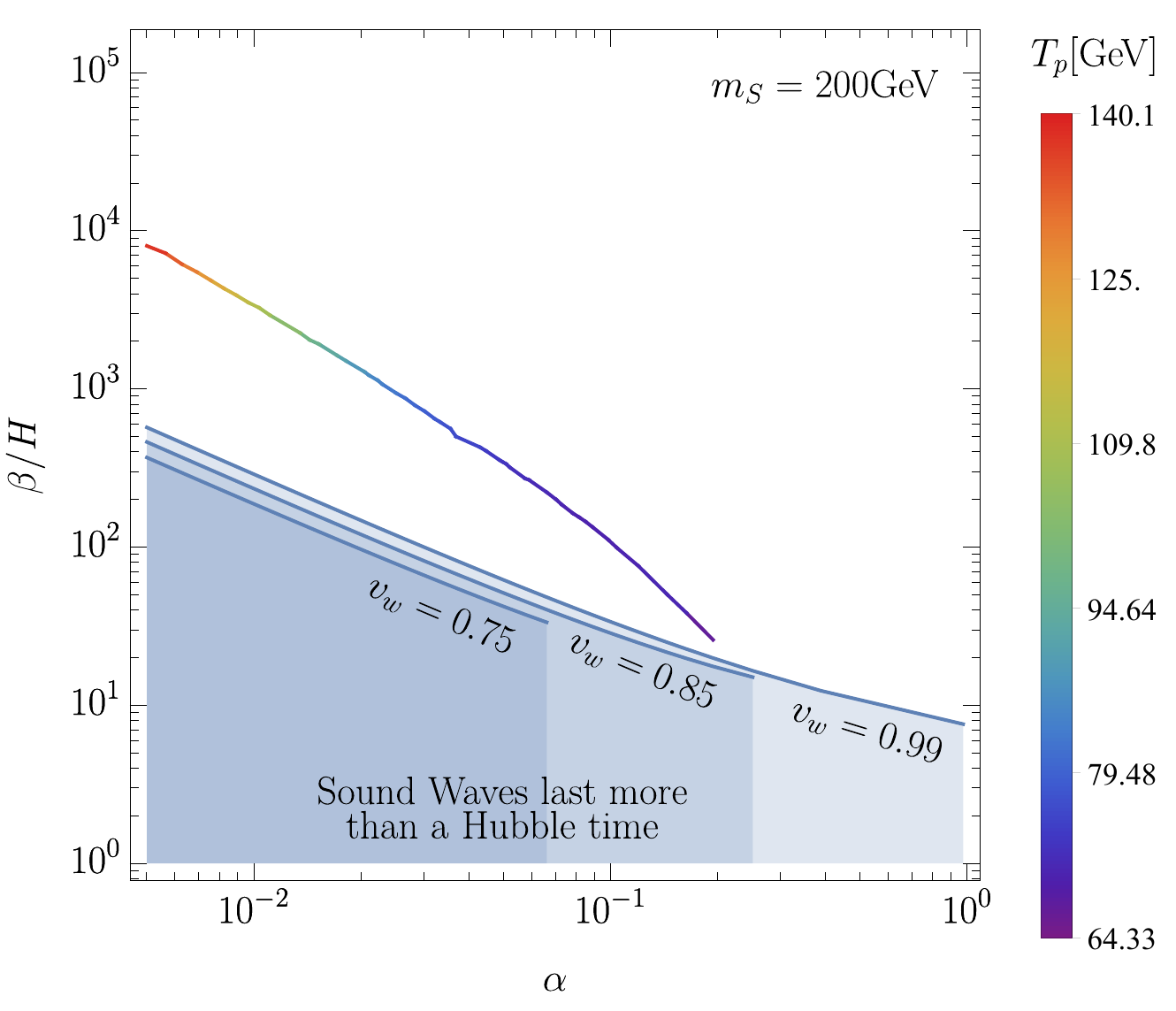}
\end{center}
\vspace{-5mm}
\caption{\it Same result as shown in Figure~\ref{fig:Singlet_RUplot} in terms of the more commonly used variables $\alpha$ and $\beta/H$. 
Blue regions at the bottom of the figure indicate parameter values for which the sound wave GW spectrum prediction should be robust.}
\label{fig:Singlet_abplot}
\end{figure}

\begin{figure}[h]
\begin{center}
\includegraphics[width=0.85\textwidth]{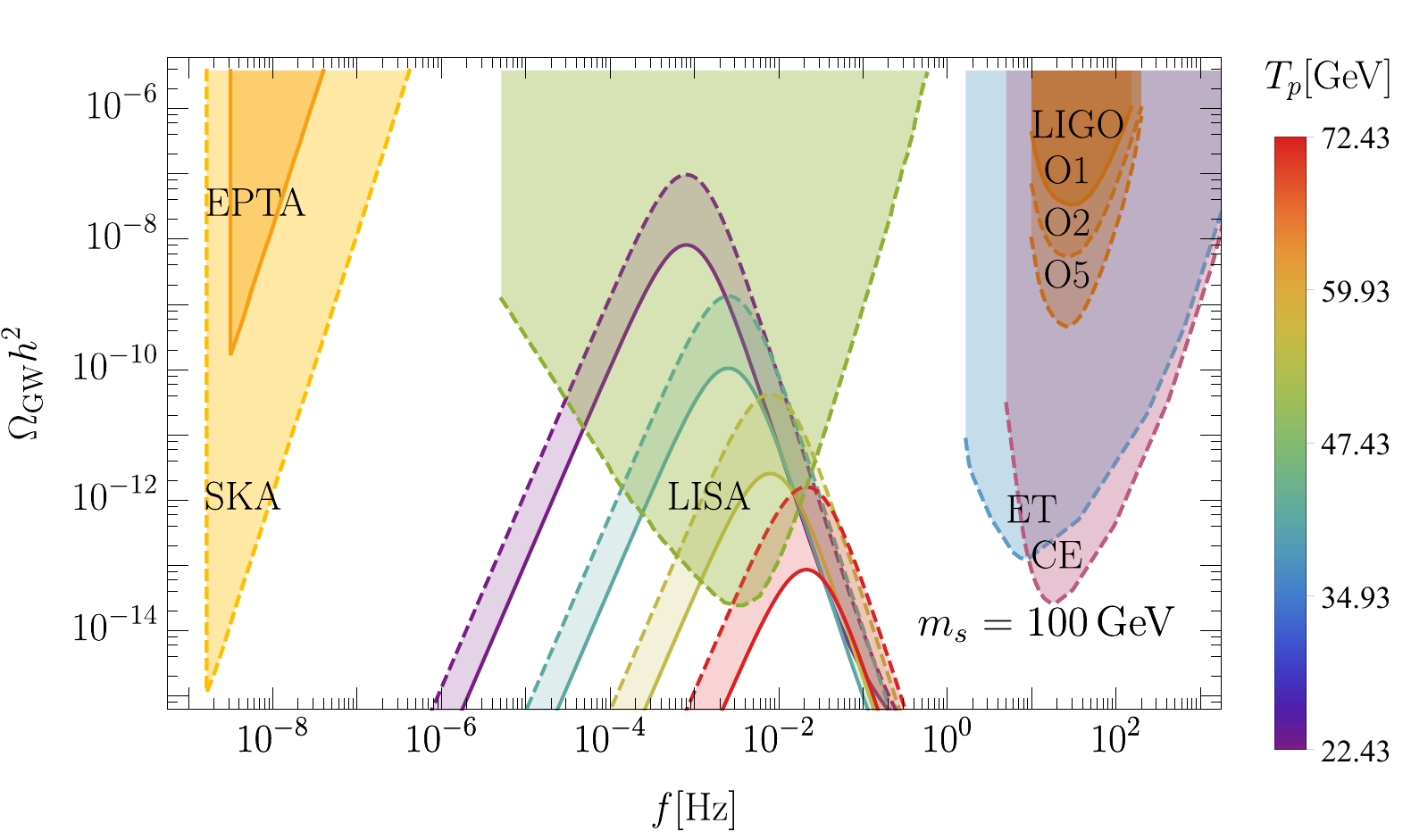}
\end{center}
\vspace{-5mm}
\caption{\it 
Combined GW signal from sound waves and turbulence as a function of frequency for different values of the percolation temperature $T_p$ (being a function 
of $\lambda_{hs}$), for $\lambda_s = 1$ and $m_s = 100$ GeV. 
The uncertainty bands correspond to the GW amplitude range from including / not including a reduction factor in the sound wave GW amplitude by a factor 
$H R_*/\bar{U}_f$ (corresponding to the shortening of the sound wave period as active GW source w.r.t. the {\it long-lasting}  
estimate).}
\label{fig:Singlet_GW100plot}
\end{figure}

We then compute the combined GW signal from sound waves and turbulence using the formulae from Section~\ref{sec:GWsignals},
and show the results in Figures~\ref{fig:Singlet_GW100plot} and~\ref{fig:Singlet_GW200plot} respectively for 
$m_s = 100$ GeV and $m_s = 200$ GeV (for $\lambda_s = 1$ in both cases). In each case, we also show the effect of 
including a reduction in the GW amplitude from sound waves by a factor $H R_*/\bar{U}_f$, which accounts for the shortening of the sound 
wave period as active GW source w.r.t. the naive {\it long-lasting} estimate.
Figures~\ref{fig:Singlet_GW100plot} and~\ref{fig:Singlet_GW200plot} also show the sensitivities of various present and planned 
GW observatories, namely LISA, LIGO, ET, DECIGO and BBO, as well as the PTA sensitivities for EPTA and SKA
(see the discussion in Section~\ref{subsec:H6} for further details).
As depicted in Figures~\ref{fig:Singlet_Tplot},~\ref{fig:Singlet_alphaplot},~\ref{fig:Singlet_Rplot}, the value of $\lambda_{hs}$ that could be probed 
by LISA from the sound wave (turbulence) GW spectrum prediction is 
$\lambda_{hs} > 0.899$ ($\lambda_{hs} > 0.945$) for $m_s = 100$ GeV and $\lambda_{hs} > 1.756$ ($\lambda_{hs} > 1.789$)  for $m_s = 200$ GeV, 
which again highlights that sound waves are in general expected to yield the dominant contribution to GW in this scenario.
Figures~\ref{fig:Singlet_GW100plot} and~\ref{fig:Singlet_GW200plot} also indicate clearly that 
there exists in this scenario a lower bound on the GW peak frequency from successful percolation, corresponding 
to $f \sim 10^{-4}$ Hz.

\begin{figure}[h]
\begin{center}
\includegraphics[width=0.85\textwidth]{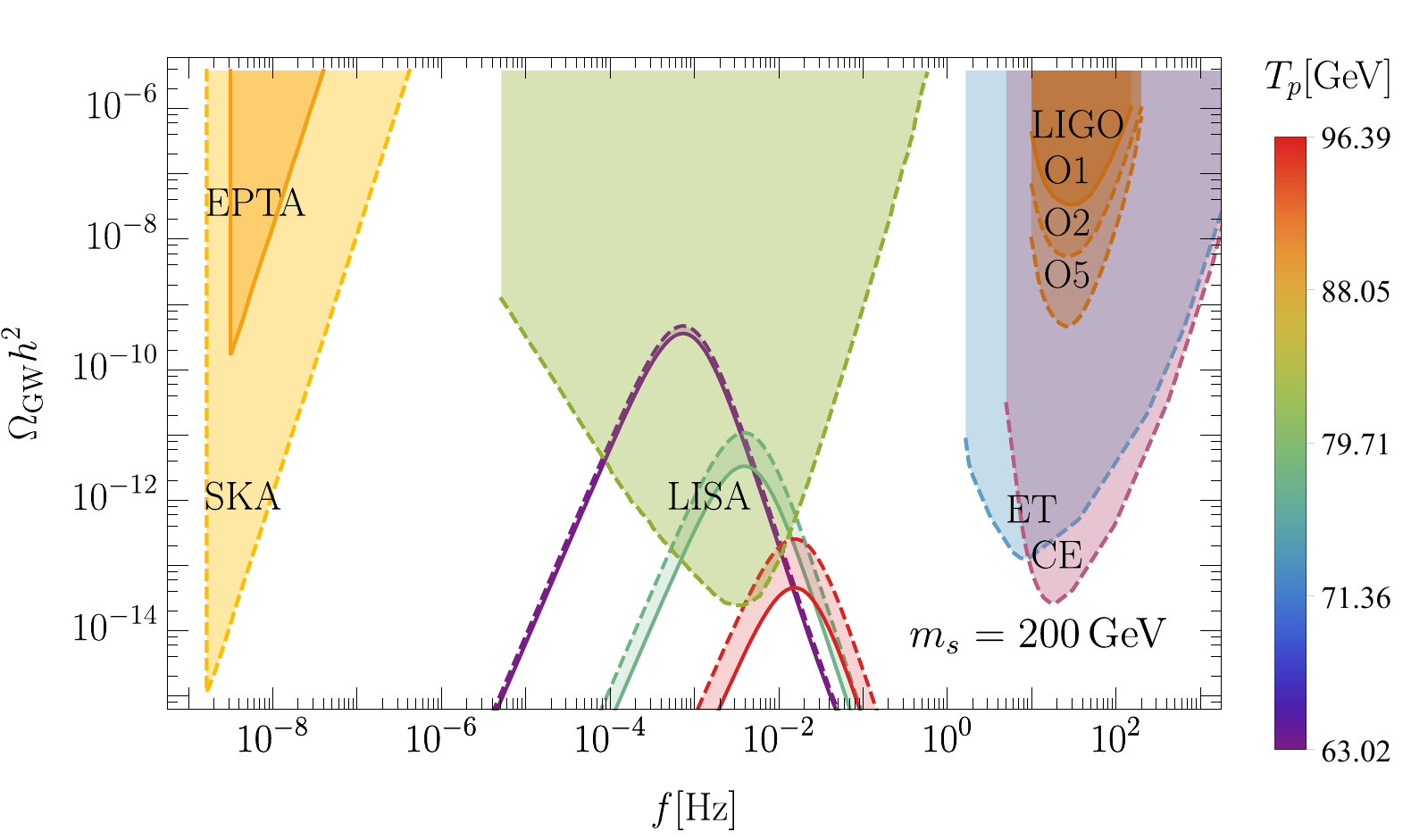}
\end{center}
\vspace{-5mm}
\caption{\it Same as Figure~\ref{fig:Singlet_GW100plot}, but for $m_s = 200$ GeV.}
\label{fig:Singlet_GW200plot}
\end{figure}

Before concluding our analysis of the scalar singlet extension of the SM with a $\mathbb{Z}_2$ symmetry, 
let us remark that this scenario is very challenging to probe at colliders for $m_s > m_h/2 \simeq 63$ 
GeV~\cite{Curtin:2014jma} \footnote{For $m_s < m_h/2$ the constraints on the Higgs invisible decay width from ATLAS/CMS 
searches~\cite{Aad:2015txa,Khachatryan:2016whc} completely exclude the region of parameter space compatible with a first-order phase 
transition (see Figure~\ref{Fig_Singlet_Strips}).}. 
The possibility of exploring this scenario in multi-jet + $E^{\rm{miss}}_T$ signatures at the HL-LHC and a future 
100 TeV hadron collider (FCC-$hh$), through singlet pair-production in association with jets via the vector boson fusion (VBF) 
process $p p \to j j s s$, has been discussed in~\cite{Curtin:2014jma,Craig:2014lda}, with the expected 95\% C.L. 
sensitivities for an integrated luminosity of 3 ab$^{-1}$ 
in each case (taken from~\cite{Craig:2014lda}) shown in Figure~\ref{Fig_Singlet_Strips}.
In addition, there are various indirect probes of this scenario: 
{\it (i)} The precise measurement of the Higgs boson self-coupling $\lambda_{3}$ -
the coupling $\lambda_{hs}$ induces a modification at 1-loop w.r.t. the SM value
\be
\lambda_{3} = \lambda^{\rm{SM}}_{3} + 
\frac{1}{16\pi^2} \frac{\lambda_{hs}^3 \,v^3}{12\, m_s^2} \quad   \longrightarrow \quad \Delta \lambda_{3} =  
\frac{1}{96\pi^2} \frac{\lambda_{hs}^3 \,v^4}{m_s^2\,m_h^2} \, .
\label{g111_loop}
\ee
The 95\% C.L. sensitivities achievable by HL-LHC (corresponding to $\Delta \lambda_3 \sim 0.8$~\cite{Goertz:2013kp,Barger:2013jfa,Barr:2014sga}) 
and by FCC-$hh$ 
(corresponding to $\Delta \lambda_3 \sim 0.2$~\cite{Contino:2016spe}) are shown in Figure~\ref{Fig_Singlet_Strips}. 
{\it (ii)} The very precise measurement of the 
Higgs production cross section in association with a $Z$ boson $\sigma_{Zh}$ at a future circular
$e^+\, e^-$ collider (FCC-$ee$), through the deviation w.r.t. the SM prediction induced
by the coupling $\lambda_{hs}$~\cite{Curtin:2014jma,Huang:2016cjm}
\be
\delta\sigma_{Zh} = \left| \frac{\lambda_{hs}^2\, v^2}{64\,\pi^2\,m_h^2} \left[1 - F(\tau)\right] \right|\, ,
\label{deltaZh}
\ee
with $\tau = m_h^2/(4 m_s^2)$ and $F(\tau)$ given by
\be
F(\tau) = \frac{\mathrm{Arcsin}(\sqrt{\tau})}{\sqrt{\tau (1- \tau)}}\, .
\label{loopZh}
\ee
The projected 95\% C.L. precision of FCC-$ee$ on the Higgs associated production cross section 
is $\delta\sigma_{Zh} \sim 0.4\%$~\cite{Gomez-Ceballos:2013zzn,dEnterria:2016fpc}, and the corresponding sensitivity to the parameter space of the singlet 
scalar extension of the SM is shown on Figure~\ref{Fig_Singlet_Strips}. 
Altogether, these results highlight the challenge of probing the electroweak phase transition with colliders 
in such a scenario, and emphasize the role of LISA as a complementary avenue to explore the electroweak epoch.

\subsection{The case of conformal (dilaton-like) potentials}
\label{sec:dilaton}

The conclusions from the analysis of Sections~\ref{subsec:H6} and~\ref{sec:singlet} 
may in principle be generalizable to other theories beyond the SM, with the notable exception of 
models with an approximate conformal symmetry, which include scenarios 
with a pseudo-Nambu-Goldstone boson associated with a spontaneously-broken 
approximate conformal symmetry~\cite{Creminelli:2001th,Randall:2006py,Nardini:2007me,Konstandin:2010cd,Konstandin:2011dr} 
(see also~\cite{Bruggisser:2018mrt,Megias:2018sxv,Bunk:2017fic,Dillon:2017ctw,vonHarling:2017yew}) (dilaton-like) as well as models
which feature classical scale invariance~\cite{Jaeckel:2016jlh,Baldes:2018emh}. Focusing on the former (though the following discussion does also apply 
to the latter scenarios),
the effective potential then contains a scale-invariant term multiplied by a function that 
varies weakly with the scale~\cite{Goldberger:1999uk}:
\begin{equation}
V(\sigma)=\sigma^4 \times P(\sigma^{\epsilon}) \, ,
\end{equation} 
where $\epsilon \ll 1$. In such potentials it is possible to obtain a large false vacuum decay probability at very low temperatures,
and even a period of inflation ending with the phase transition~\cite{Konstandin:2011dr}.

However, within an EFT context, the lack of deviations from the SM~\cite{Ellis:2018gqa} already pushes the possible 
energy scale of the composite Higgs models, that these dilaton-like scenarios can be recast into~(see e.g.~\cite{Bruggisser:2018mrt}), 
towards the TeV scale. This means that, generally, even though the primordial plasma could be very strongly 
supercooled before the phase transition, the vacuum energy released during the transition would reheat the universe to temperatures 
of the order or above the electroweak scale, up to the TeV.
Since the peak frequency of the GW signal today is directly linked to the reheating temperature through redshifting, 
we expect the generic bound on the peak frequency of GW waves from the electroweak phase transition 
$f \gtrsim 10^{-4}$ Hz to also hold in this case.

There is however a crucial phenomenological difference between the scenarios with an approximate conformal symmetry and those 
discussed in Sections~\ref{subsec:H6} and~\ref{sec:singlet}, coming from the fact that dilaton-like potentials may allow for a 
(generally brief) period of inflation before the phase transition. As discussed in Sections~\ref{subsec:H6} and~\ref{sec:singlet}, 
bubble percolation in models with polynomial potentials allows at most for a very brief period of vacuum domination, 
which does not decrease the temperature of the primordial plasma significantly. In contrast, the possible
inflationary period in models with dilaton-like potentials would still be compatible with successful percolation.
This suggests that dilaton-like scenarios with sufficient supercooling could be the only ones 
in which the GW signal from the electroweak phase transition could be sourced primarily by 
bubble collisions (instead of by sound waves and plasma turbulence). If the plasma is severely diluted by the 
inflationary period, the friction exerted by the plasma on the bubble wall might not prevent the bubble from accelerating 
before the end of the transition. The required amount of supercooling can be obtained approximately
by balancing the NLO ``runaway" friction~\cite{Bodeker:2017cim} (see Section~\ref{sec:hydro}) and the pressure difference
$\Delta {\cal P}_{\mathrm{NLO}} = \gamma\, g^2\, \Delta m \,T^3 \sim (\rm{TeV})^4$ with $\Delta m \sim$ TeV and 
$\gamma \sim M_{\mathrm{pl}}/\Delta m$. 
This yields $T \lesssim 10$ MeV, below which most of the energy of the phase transition can be used 
to accelerate the bubble walls until the end of the phase 
transition~\footnote{In some of these scenarios, e.g., for holographic duals \`a la Randall-Sundrum, it is expected that 
many degrees of freedom become massive during the phase transition, and as such the LO ``runaway" friction 
reads $\Delta {\cal P}_{\mathrm{LO}} \sim N \Delta m^2 \,T^2$ with $N \gg 1$. Nevertheless, for the extreme supercooling considered above ($T\lesssim 10-100$ MeV)
we expect the plasma to be sufficiently diluted for the effect of $N \gg 1$ not to be important. However, we issue a word of caution until this is further investigated.}, 
as would be the case in a transition in pure vacuum, with the 
GW signal being produced via bubble collisions. 
These substantial differences in the calculation of the GW signal in models with dilaton-like potentials
warrant a detailed discussion, which we leave for future work.

\section{Conclusions}
\label{sec:conx}

We have re-examined in this paper the maximum possible strength of a first-order electroweak phase transition and the corresponding GW signal 
that it could generate. We have focused our discussion on two key issues:~{\it (i)} 
the constraint on the duration of near-exponential expansion of the Universe while
dominated by vacuum energy in the supercooled phase that is imposed by the requirement of successful percolation and completion of 
the phase transition.~{\it (ii)} the conditions under which the current estimates for the amplitude of the GW spectrum 
generated from plasma sound waves and MHD turbulence at the end of the transition can be reliably applied to specific scenarios. 
We find that these aspects result in a reduction of the possible strength of the GW signal compared to previous estimates.

Concerning the first issue, it is known that the GW signal from a first-order electroweak phase transition could be enhanced by a period of strong supercooling. However, 
it has also been known since the demise of `old inflation' that most of the Universe might remain trapped in the
exponentially-expanding false vacuum, with bubbles of the true vacuum never percolating to
complete the transition, the `graceful-exit' problem~\cite{Guth:1982pn}. We have provided a general treatment 
of bubble nucleation, growth and percolation, specifically discussing in Section~\ref{sec:percolation} the constraint 
on the amount of supercooling imposed by successful percolation and completion of the electroweak phase 
transition for scenarios with polynomial potentials. 
This was then followed in Section~\ref{sec:GWsignals} by a discussion of the implications of the percolation constraint for the 
possible strength of the GW signal, as well as the possible range for its peak frequency.

We have applied our treatment to specific models that could yield a strongly first-order
electroweak phase transition, namely the SM with a supplementary $|H|^6/\Lambda^2$
interaction and the SM extended by a real singlet scalar field $s$. 
These models capture well the general features expected for a phase transition driven by polynomial potentials.
In the first model we find that the percolation criterion is satisfied only for a cut-off scale $\Lambda > 545$~GeV, with LISA at most being able to 
probe scales $\Lambda \lesssim 580$ GeV.
While current LHC data do not constrain the value of $\Lambda$, 
future measurements with the HL-LHC should be sensitive to $\Lambda \simeq 766$~GeV at the
2-$\sigma$ level via measurements of the Higgs self-coupling.  
This would probe well into the range where this model might yield a GW signal detectable by LISA.
In contrast, for the scenario where the SM is extended by a singlet scalar field $s$, the
prospects for detection at the LHC or future colliders are poor~\cite{Curtin:2014jma}, with LISA providing a 
complementary avenue to probe this scenario.
We emphasize that in both scenarios the requirement of successful percolation yields a lower bound on the value of the GW spectrum peak frequency 
$f \gtrsim 10^{-4}$ Hz, strongly suggesting that detecting a GW signal from a supercooled electroweak phase transition with PTA is not possible in these scenarios.

When discussing specific scenarios for a strongly first-order electroweak phase transition giving rise to a GW signal, 
we have found that the condition $H R_*/\bar{U}_f > 1$  for sound waves to 
be a ``long-lasting" source (active approximately for a Hubble time) 
of GW is generally not satisfied. This implies the period of GW generation 
via plasma sound waves would be cut short, probably resulting in a reduction of the GW amplitude from 
sound waves by approximately a factor $H R_*/\bar{U}_f$ (depending on the transition properties, we can have $H R_*/\bar{U}_f \ll 1$) 
compared to current estimates in the literature. Such shortening would probably be accompanied by an enhancement 
of the GW spectrum from turbulence, but we expect the overall effect to still be a reduction of the GW amplitude as compared to the current predictions 
for BSM scenarios. 

Finally, as briefly discussed in Section~\ref{sec:dilaton}, models with dilaton-like potentials would exhibit different
features from the two models we have studied in detail~\cite{Bruggisser:2018mrt,Megias:2018sxv,Randall:2006py,Nardini:2007me,Konstandin:2010cd,Konstandin:2011dr}, including the possibility of a longer,
inflationary period of vacuum enegry domination, and the possibility that bubble collisions could
dominate GW production. 

\section*{Acknowledgements}

We thank Mark Hindmarsh, Thomas Konstandin and Geraldine Servant for useful discussions and comments. ML and JMN also thank the
LISA Cosmology Working Group for encouragement and for providing a forum to discuss the implications of this work.
The work of JE and ML was supported by the United Kingdom STFC Grant ST/P000258/1. 
JE was also supported by the Estonian Research Council via a Mobilitas Pluss grant and ML by the Polish MNiSW grant IP2015 043174. 
JMN was partially supported by the European Research Council under the European Union's Horizon 2020 program, ERC
Grant Agreement 648680 (DARKHORIZONS) during the early stages of this work, 
and subsequently by the Programa Atracci\'on de Talento de la Comunidad de Madrid via grant 2017-T1/TIC-5202.
JMN also acknowledges support from the Spanish MINECO's ``Centro de Excelencia Severo Ochoa" Programme via grant SEV-2012-0249.

\appendix
\section{Effective potential for SM with $|H|^6/\Lambda^2$}\label{sec:effpotappendixLambda}
We include one-loop corrections to the zero-temperature potential in the on-shell scheme~\cite{Delaunay:2007wb,Curtin:2014jma},
\begin{equation}\label{eqn:V1}
 V_{1-\rm loop}(h)=\sum_{i=h,\chi,W,Z,t}\frac{n_i}{64\pi^2}\left[m_{i}^4 \left( \log\frac{m^2_{i}}{m^2_{0 i}}-\frac{3}{2}\right)+2 m^2_{i} m^2_{0i}\right] ,
\end{equation}
where $n_{\{h,\chi,W,Z,t\}}=\{1,3,6,3,-12\}$, the field-dependent masses read
\begin{equation}
\begin{split} 
m^2_h & = -m^2+3\lambda h^2 +  \frac{15}{4}\frac{h^4}{\Lambda^2}, \quad
m_\chi^2 =-m^2+\lambda h^2 + \frac{3}{4}\frac{h^4}{\Lambda^2} \, ,
\\
m_W^2 & =\frac{g^2}{4}h^2, \quad \quad \ \ m_Z^2=\frac{g^2+g'^2}{4}h^2, \quad \quad \ \ m_t^2=\frac{y_t^2}{2}h^2.
\end{split}
\end{equation}
  and the $m_0$ in (\ref{eqn:V1}) are masses calculated with $\ h=v$.
  
Finally, we also include the finite-temperature corrections given by
\begin{equation}\label{eqn:VT}
 V_T(h,T)=\sum_{i= h,\chi,W,Z,\gamma}\frac{n_iT^4}{2\pi^2} J_b\left(\frac{m^2_i}{T^2}\right)+\sum_{i= t}\frac{n_iT^4}{2\pi^2} J_f\left(\frac{m^2_i}{T^2}\right),
 \end{equation}
where 
\begin{equation}
J_{b/f }\left(\frac{m^2_i}{T^2}\right)=\int_0^\infty dk \, 
k^2\log\left[1\mp {\rm exp}\left(-\sqrt{\frac{k^2+m_i^2}{T^2}} \right) \right].
\end{equation}
It is important to include in this contribution a correction coming from resumming the multi-loop contributions of
longitudinal polarisations of bosons~\cite{Arnold:1992rz,Carrington:1991hz}. 
We achieve this by shifting the masses of the longitudinal polarisations of the gauge bosons and scalars 
by their thermal corrections such that $m_i^2 \rightarrow m_i^2 + \Pi_i$. 
In our model, these shifts read \cite{Carrington:1991hz,Delaunay:2007wb}
\begin{equation}
\begin{split}
\Pi_{h ,{\chi_i}}(T)&=\frac{T^2}{4v^2}\left(m_h^2+2m_W^2+m_Z^2+2m_t^2\right)-\frac{3}{4} T^2 \frac{v^2}{\Lambda^2} \, ,
\\
\Pi_W(T)&=\frac{22}{3}\frac{m^2_W}{v^2}T^2 \, ,\\ 
\end{split}
\end{equation}
while the shifted masses of $Z$ and $\gamma$ ($m^2_{Z/\gamma}+\Pi_{Z/\gamma}(T)$) are eigenvalues of the following mass matrix
\begin{equation}
\begin{pmatrix}
\frac{1}{4} g^2 \phi^2+\frac{11}{6}g^2T^2 & -\frac{1}{4}g'^2 g^2 \phi^2 \\
-\frac{1}{4}g'^2 g^2 \phi^2 & \frac{1}{4} g'^2 \phi^2+\frac{11}{6}g'^2T^2
\end{pmatrix}.
\end{equation}
We can then write the final form of the potential in the form
\begin{equation}\label{eqn:Veff}
V(h,T)=V(h)^{\textrm tree}(h)+ V_{1-\rm loop}(h)+V_T(h,T) \, ,
\end{equation}
with the three contributions given in Eq.~\eqref{eqn:treepot}, Eq.~\eqref{eqn:V1} and Eq.~\eqref{eqn:VT}.


\bibliographystyle{JHEP}
\bibliography{EWPT}

\providecommand{\href}[2]{#2}\begingroup\raggedright\begin{thebibliography}{100}

\bibitem{Kuzmin:1985mm}
V.~A. Kuzmin, V.~A. Rubakov and M.~E. Shaposhnikov, \emph{{On the Anomalous
  Electroweak Baryon Number Nonconservation in the Early Universe}},
  \href{https://doi.org/10.1016/0370-2693(85)91028-7}{\emph{Phys. Lett.}
  {\bfseries B155} (1985) 36}.

\bibitem{Cohen:1993nk}
A.~G. Cohen, D.~B. Kaplan and A.~E. Nelson, \emph{{Progress in electroweak
  baryogenesis}},
  \href{https://doi.org/10.1146/annurev.ns.43.120193.000331}{\emph{Ann. Rev.
  Nucl. Part. Sci.} {\bfseries 43} (1993) 27--70},
  [\href{https://arxiv.org/abs/hep-ph/9302210}{{\ttfamily hep-ph/9302210}}].

\bibitem{Riotto:1999yt}
A.~Riotto and M.~Trodden, \emph{{Recent progress in baryogenesis}},
  \href{https://doi.org/10.1146/annurev.nucl.49.1.35}{\emph{Ann. Rev. Nucl.
  Part. Sci.} {\bfseries 49} (1999) 35--75},
  [\href{https://arxiv.org/abs/hep-ph/9901362}{{\ttfamily hep-ph/9901362}}].

\bibitem{Morrissey:2012db}
D.~E. Morrissey and M.~J. Ramsey-Musolf, \emph{{Electroweak baryogenesis}},
  \href{https://doi.org/10.1088/1367-2630/14/12/125003}{\emph{New J. Phys.}
  {\bfseries 14} (2012) 125003},
  [\href{https://arxiv.org/abs/1206.2942}{{\ttfamily 1206.2942}}].

\bibitem{Kamionkowski:1993fg}
M.~Kamionkowski, A.~Kosowsky and M.~S. Turner, \emph{{Gravitational radiation
  from first order phase transitions}},
  \href{https://doi.org/10.1103/PhysRevD.49.2837}{\emph{Phys. Rev.} {\bfseries
  D49} (1994) 2837--2851},
  [\href{https://arxiv.org/abs/astro-ph/9310044}{{\ttfamily
  astro-ph/9310044}}].

\bibitem{Apreda:2001us}
R.~Apreda, M.~Maggiore, A.~Nicolis and A.~Riotto, \emph{{Gravitational waves
  from electroweak phase transitions}},
  \href{https://doi.org/10.1016/S0550-3213(02)00264-X}{\emph{Nucl. Phys.}
  {\bfseries B631} (2002) 342--368},
  [\href{https://arxiv.org/abs/gr-qc/0107033}{{\ttfamily gr-qc/0107033}}].

\bibitem{Grojean:2006bp}
C.~Grojean and G.~Servant, \emph{{Gravitational Waves from Phase Transitions at
  the Electroweak Scale and Beyond}},
  \href{https://doi.org/10.1103/PhysRevD.75.043507}{\emph{Phys. Rev.}
  {\bfseries D75} (2007) 043507},
  [\href{https://arxiv.org/abs/hep-ph/0607107}{{\ttfamily hep-ph/0607107}}].

\bibitem{Huber:2008hg}
S.~J. Huber and T.~Konstandin, \emph{{Gravitational Wave Production by
  Collisions: More Bubbles}},
  \href{https://doi.org/10.1088/1475-7516/2008/09/022}{\emph{JCAP} {\bfseries
  0809} (2008) 022}, [\href{https://arxiv.org/abs/0806.1828}{{\ttfamily
  0806.1828}}].

\bibitem{Espinosa:2008kw}
J.~R. Espinosa, T.~Konstandin, J.~M. No and M.~Quiros, \emph{{Some Cosmological
  Implications of Hidden Sectors}},
  \href{https://doi.org/10.1103/PhysRevD.78.123528}{\emph{Phys. Rev.}
  {\bfseries D78} (2008) 123528},
  [\href{https://arxiv.org/abs/0809.3215}{{\ttfamily 0809.3215}}].

\bibitem{Ashoorioon:2009nf}
A.~Ashoorioon and T.~Konstandin, \emph{{Strong electroweak phase transitions
  without collider traces}},
  \href{https://doi.org/10.1088/1126-6708/2009/07/086}{\emph{JHEP} {\bfseries
  07} (2009) 086}, [\href{https://arxiv.org/abs/0904.0353}{{\ttfamily
  0904.0353}}].

\bibitem{Dorsch:2014qpa}
G.~C. Dorsch, S.~J. Huber and J.~M. No, \emph{{Cosmological Signatures of a
  UV-Conformal Standard Model}},
  \href{https://doi.org/10.1103/PhysRevLett.113.121801}{\emph{Phys. Rev. Lett.}
  {\bfseries 113} (2014) 121801},
  [\href{https://arxiv.org/abs/1403.5583}{{\ttfamily 1403.5583}}].

\bibitem{Enqvist:2014zqa}
K.~Enqvist, S.~Nurmi, T.~Tenkanen and K.~Tuominen, \emph{{Standard Model with a
  real singlet scalar and inflation}},
  \href{https://doi.org/10.1088/1475-7516/2014/08/035}{\emph{JCAP} {\bfseries
  1408} (2014) 035}, [\href{https://arxiv.org/abs/1407.0659}{{\ttfamily
  1407.0659}}].

\bibitem{Kakizaki:2015wua}
M.~Kakizaki, S.~Kanemura and T.~Matsui, \emph{{Gravitational waves as a probe
  of extended scalar sectors with the first order electroweak phase
  transition}}, \href{https://doi.org/10.1103/PhysRevD.92.115007}{\emph{Phys.
  Rev.} {\bfseries D92} (2015) 115007},
  [\href{https://arxiv.org/abs/1509.08394}{{\ttfamily 1509.08394}}].

\bibitem{Jaeckel:2016jlh}
J.~Jaeckel, V.~V. Khoze and M.~Spannowsky, \emph{{Hearing the signal of dark
  sectors with gravitational wave detectors}},
  \href{https://doi.org/10.1103/PhysRevD.94.103519}{\emph{Phys. Rev.}
  {\bfseries D94} (2016) 103519},
  [\href{https://arxiv.org/abs/1602.03901}{{\ttfamily 1602.03901}}].

\bibitem{Dev:2016feu}
P.~S.~B. Dev and A.~Mazumdar, \emph{{Probing the Scale of New Physics by
  Advanced LIGO/VIRGO}},
  \href{https://doi.org/10.1103/PhysRevD.93.104001}{\emph{Phys. Rev.}
  {\bfseries D93} (2016) 104001},
  [\href{https://arxiv.org/abs/1602.04203}{{\ttfamily 1602.04203}}].

\bibitem{Hashino:2016rvx}
K.~Hashino, M.~Kakizaki, S.~Kanemura and T.~Matsui, \emph{{Synergy between
  measurements of gravitational waves and the triple-Higgs coupling in probing
  the first-order electroweak phase transition}},
  \href{https://doi.org/10.1103/PhysRevD.94.015005}{\emph{Phys. Rev.}
  {\bfseries D94} (2016) 015005},
  [\href{https://arxiv.org/abs/1604.02069}{{\ttfamily 1604.02069}}].

\bibitem{Chala:2016ykx}
M.~Chala, G.~Nardini and I.~Sobolev, \emph{{Unified explanation for dark matter
  and electroweak baryogenesis with direct detection and gravitational wave
  signatures}}, \href{https://doi.org/10.1103/PhysRevD.94.055006}{\emph{Phys.
  Rev.} {\bfseries D94} (2016) 055006},
  [\href{https://arxiv.org/abs/1605.08663}{{\ttfamily 1605.08663}}].

\bibitem{Tenkanen:2016idg}
T.~Tenkanen, K.~Tuominen and V.~Vaskonen, \emph{{A Strong Electroweak Phase
  Transition from the Inflaton Field}},
  \href{https://doi.org/10.1088/1475-7516/2016/09/037}{\emph{JCAP} {\bfseries
  1609} (2016) 037}, [\href{https://arxiv.org/abs/1606.06063}{{\ttfamily
  1606.06063}}].

\bibitem{Kobakhidze:2016mch}
A.~Kobakhidze, A.~Manning and J.~Yue, \emph{{Gravitational Waves from the Phase
  Transition of a Non-linearly Realised Electroweak Gauge Symmetry}},
  \href{https://arxiv.org/abs/1607.00883}{{\ttfamily 1607.00883}}.

\bibitem{Huang:2016cjm}
P.~Huang, A.~J. Long and L.-T. Wang, \emph{{Probing the Electroweak Phase
  Transition with Higgs Factories and Gravitational Waves}},
  \href{https://doi.org/10.1103/PhysRevD.94.075008}{\emph{Phys. Rev.}
  {\bfseries D94} (2016) 075008},
  [\href{https://arxiv.org/abs/1608.06619}{{\ttfamily 1608.06619}}].

\bibitem{Artymowski:2016tme}
M.~Artymowski, M.~Lewicki and J.~D. Wells, \emph{{Gravitational wave and
  collider implications of electroweak baryogenesis aided by non-standard
  cosmology}}, \href{https://doi.org/10.1007/JHEP03(2017)066}{\emph{JHEP}
  {\bfseries 03} (2017) 066},
  [\href{https://arxiv.org/abs/1609.07143}{{\ttfamily 1609.07143}}].

\bibitem{Hashino:2016xoj}
K.~Hashino, M.~Kakizaki, S.~Kanemura, P.~Ko and T.~Matsui, \emph{{Gravitational
  waves and Higgs boson couplings for exploring first order phase transition in
  the model with a singlet scalar field}},
  \href{https://arxiv.org/abs/1609.00297}{{\ttfamily 1609.00297}}.

\bibitem{Vaskonen:2016yiu}
V.~Vaskonen, \emph{{Electroweak baryogenesis and gravitational waves from a
  real scalar singlet}},
  \href{https://doi.org/10.1103/PhysRevD.95.123515}{\emph{Phys. Rev.}
  {\bfseries D95} (2017) 123515},
  [\href{https://arxiv.org/abs/1611.02073}{{\ttfamily 1611.02073}}].

\bibitem{Dorsch:2016nrg}
G.~C. Dorsch, S.~J. Huber, T.~Konstandin and J.~M. No, \emph{{A Second Higgs
  Doublet in the Early Universe: Baryogenesis and Gravitational Waves}},
  \href{https://doi.org/10.1088/1475-7516/2017/05/052}{\emph{JCAP} {\bfseries
  1705} (2017) 052}, [\href{https://arxiv.org/abs/1611.05874}{{\ttfamily
  1611.05874}}].

\bibitem{Beniwal:2017eik}
A.~Beniwal, M.~Lewicki, J.~D. Wells, M.~White and A.~G. Williams,
  \emph{{Gravitational wave, collider and dark matter signals from a scalar
  singlet electroweak baryogenesis}},
  \href{https://doi.org/10.1007/JHEP08(2017)108}{\emph{JHEP} {\bfseries 08}
  (2017) 108}, [\href{https://arxiv.org/abs/1702.06124}{{\ttfamily
  1702.06124}}].

\bibitem{Marzola:2017jzl}
L.~Marzola, A.~Racioppi and V.~Vaskonen, \emph{{Phase transition and
  gravitational wave phenomenology of scalar conformal extensions of the
  Standard Model}},
  \href{https://doi.org/10.1140/epjc/s10052-017-4996-1}{\emph{Eur. Phys. J.}
  {\bfseries C77} (2017) 484},
  [\href{https://arxiv.org/abs/1704.01034}{{\ttfamily 1704.01034}}].

\bibitem{Kang:2017mkl}
Z.~Kang, P.~Ko and T.~Matsui, \emph{{Strong first order EWPT \& strong
  gravitational waves in Z$_{3}$-symmetric singlet scalar extension}},
  \href{https://doi.org/10.1007/JHEP02(2018)115}{\emph{JHEP} {\bfseries 02}
  (2018) 115}, [\href{https://arxiv.org/abs/1706.09721}{{\ttfamily
  1706.09721}}].

\bibitem{Chala:2018ari}
M.~Chala, C.~Krause and G.~Nardini, \emph{{Signals of the electroweak phase
  transition at colliders and gravitational wave observatories}},
  \href{https://doi.org/10.1007/JHEP07(2018)062}{\emph{JHEP} {\bfseries 07}
  (2018) 062}, [\href{https://arxiv.org/abs/1802.02168}{{\ttfamily
  1802.02168}}].

\bibitem{Huang:2015izx}
F.~P. Huang, P.-H. Gu, P.-F. Yin, Z.-H. Yu and X.~Zhang, \emph{{Testing the
  electroweak phase transition and electroweak baryogenesis at the LHC and a
  circular electron-positron collider}},
  \href{https://doi.org/10.1103/PhysRevD.93.103515}{\emph{Phys. Rev.}
  {\bfseries D93} (2016) 103515},
  [\href{https://arxiv.org/abs/1511.03969}{{\ttfamily 1511.03969}}].

\bibitem{Huang:2016odd}
F.~P. Huang, Y.~Wan, D.-G. Wang, Y.-F. Cai and X.~Zhang, \emph{{Hearing the
  echoes of electroweak baryogenesis with gravitational wave detectors}},
  \href{https://doi.org/10.1103/PhysRevD.94.041702}{\emph{Phys. Rev.}
  {\bfseries D94} (2016) 041702},
  [\href{https://arxiv.org/abs/1601.01640}{{\ttfamily 1601.01640}}].

\bibitem{Huang:2017kzu}
F.~P. Huang and C.~S. Li, \emph{{Probing the baryogenesis and dark matter
  relaxed in phase transition by gravitational waves and colliders}},
  \href{https://doi.org/10.1103/PhysRevD.96.095028}{\emph{Phys. Rev.}
  {\bfseries D96} (2017) 095028},
  [\href{https://arxiv.org/abs/1709.09691}{{\ttfamily 1709.09691}}].

\bibitem{Hashino:2018zsi}
K.~Hashino, M.~Kakizaki, S.~Kanemura, P.~Ko and T.~Matsui, \emph{{Gravitational
  waves from first order electroweak phase transition in models with the
  U(1)$_{X}$ gauge symmetry}},
  \href{https://doi.org/10.1007/JHEP06(2018)088}{\emph{JHEP} {\bfseries 06}
  (2018) 088}, [\href{https://arxiv.org/abs/1802.02947}{{\ttfamily
  1802.02947}}].

\bibitem{Vieu:2018zze}
T.~Vieu, A.~P. Morais and R.~Pasechnik, \emph{{Multi-peaked signatures of
  primordial gravitational waves from multi-step electroweak phase
  transition}},  \href{https://arxiv.org/abs/1802.10109}{{\ttfamily
  1802.10109}}.

\bibitem{Huang:2018aja}
F.~P. Huang, Z.~Qian and M.~Zhang, \emph{{Exploring dynamical CP violation
  induced baryogenesis by gravitational waves and colliders}},
  \href{https://doi.org/10.1103/PhysRevD.98.015014}{\emph{Phys. Rev.}
  {\bfseries D98} (2018) 015014},
  [\href{https://arxiv.org/abs/1804.06813}{{\ttfamily 1804.06813}}].

\bibitem{Bruggisser:2018mrt}
S.~Bruggisser, B.~Von~Harling, O.~Matsedonskyi and G.~Servant,
  \emph{{Electroweak Phase Transition and Baryogenesis in Composite Higgs
  Models}},  \href{https://arxiv.org/abs/1804.07314}{{\ttfamily 1804.07314}}.

\bibitem{Megias:2018sxv}
E.~Megias, G.~Nardini and M.~Quiros, \emph{{Cosmological Phase Transitions in
  Warped Space: Gravitational Waves and Collider Signatures}},
  \href{https://arxiv.org/abs/1806.04877}{{\ttfamily 1806.04877}}.

\bibitem{Croon:2018erz}
D.~Croon, V.~Sanz and G.~White, \emph{{Model Discrimination in Gravitational
  Wave spectra from Dark Phase Transitions}},
  \href{https://doi.org/10.1007/JHEP08(2018)203}{\emph{JHEP} {\bfseries 08}
  (2018) 203}, [\href{https://arxiv.org/abs/1806.02332}{{\ttfamily
  1806.02332}}].

\bibitem{Caprini:2015zlo}
C.~Caprini et~al., \emph{{Science with the space-based interferometer eLISA.
  II: Gravitational waves from cosmological phase transitions}},
  \href{https://doi.org/10.1088/1475-7516/2016/04/001}{\emph{JCAP} {\bfseries
  1604} (2016) 001}, [\href{https://arxiv.org/abs/1512.06239}{{\ttfamily
  1512.06239}}].

\bibitem{Audley:2017drz}
{\scshape LISA} collaboration, H.~Audley et~al., \emph{{Laser Interferometer
  Space Antenna}},  \href{https://arxiv.org/abs/1702.00786}{{\ttfamily
  1702.00786}}.

\bibitem{Caprini:2018mtu}
C.~Caprini and D.~G. Figueroa, \emph{{Cosmological Backgrounds of Gravitational
  Waves}}, \href{https://doi.org/10.1088/1361-6382/aac608}{\emph{Class. Quant.
  Grav.} {\bfseries 35} (2018) 163001},
  [\href{https://arxiv.org/abs/1801.04268}{{\ttfamily 1801.04268}}].

\bibitem{Kobakhidze:2017mru}
A.~Kobakhidze, C.~Lagger, A.~Manning and J.~Yue, \emph{{Gravitational waves
  from a supercooled electroweak phase transition and their detection with
  pulsar timing arrays}},
  \href{https://doi.org/10.1140/epjc/s10052-017-5132-y}{\emph{Eur. Phys. J.}
  {\bfseries C77} (2017) 570},
  [\href{https://arxiv.org/abs/1703.06552}{{\ttfamily 1703.06552}}].

\bibitem{Cai:2017tmh}
R.-G. Cai, M.~Sasaki and S.-J. Wang, \emph{{The gravitational waves from the
  first-order phase transition with a dimension-six operator}},
  \href{https://doi.org/10.1088/1475-7516/2017/08/004}{\emph{JCAP} {\bfseries
  1708} (2017) 004}, [\href{https://arxiv.org/abs/1707.03001}{{\ttfamily
  1707.03001}}].

\bibitem{Guth:1982pn}
A.~H. Guth and E.~J. Weinberg, \emph{{Could the Universe Have Recovered from a
  Slow First Order Phase Transition?}},
  \href{https://doi.org/10.1016/0550-3213(83)90307-3}{\emph{Nucl. Phys.}
  {\bfseries B212} (1983) 321--364}.

\bibitem{Turner:1992tz}
M.~S. Turner, E.~J. Weinberg and L.~M. Widrow, \emph{{Bubble nucleation in
  first order inflation and other cosmological phase transitions}},
  \href{https://doi.org/10.1103/PhysRevD.46.2384}{\emph{Phys. Rev.} {\bfseries
  D46} (1992) 2384--2403}.

\bibitem{Megevand:2016lpr}
A.~Megevand and S.~Ramirez, \emph{{Bubble nucleation and growth in very strong
  cosmological phase transitions}},
  \href{https://doi.org/10.1016/j.nuclphysb.2017.03.009}{\emph{Nucl. Phys.}
  {\bfseries B919} (2017) 74--109},
  [\href{https://arxiv.org/abs/1611.05853}{{\ttfamily 1611.05853}}].

\bibitem{Hindmarsh:2013xza}
M.~Hindmarsh, S.~J. Huber, K.~Rummukainen and D.~J. Weir, \emph{{Gravitational
  waves from the sound of a first order phase transition}},
  \href{https://doi.org/10.1103/PhysRevLett.112.041301}{\emph{Phys. Rev. Lett.}
  {\bfseries 112} (2014) 041301},
  [\href{https://arxiv.org/abs/1304.2433}{{\ttfamily 1304.2433}}].

\bibitem{Hindmarsh:2015qta}
M.~Hindmarsh, S.~J. Huber, K.~Rummukainen and D.~J. Weir, \emph{{Numerical
  simulations of acoustically generated gravitational waves at a first order
  phase transition}},
  \href{https://doi.org/10.1103/PhysRevD.92.123009}{\emph{Phys. Rev.}
  {\bfseries D92} (2015) 123009},
  [\href{https://arxiv.org/abs/1504.03291}{{\ttfamily 1504.03291}}].

\bibitem{Hindmarsh:2017gnf}
M.~Hindmarsh, S.~J. Huber, K.~Rummukainen and D.~J. Weir, \emph{{Shape of the
  acoustic gravitational wave power spectrum from a first order phase
  transition}}, \href{https://doi.org/10.1103/PhysRevD.96.103520}{\emph{Phys.
  Rev.} {\bfseries D96} (2017) 103520},
  [\href{https://arxiv.org/abs/1704.05871}{{\ttfamily 1704.05871}}].

\bibitem{Caprini:2006jb}
C.~Caprini and R.~Durrer, \emph{{Gravitational waves from stochastic
  relativistic sources: Primordial turbulence and magnetic fields}},
  \href{https://doi.org/10.1103/PhysRevD.74.063521}{\emph{Phys. Rev.}
  {\bfseries D74} (2006) 063521},
  [\href{https://arxiv.org/abs/astro-ph/0603476}{{\ttfamily
  astro-ph/0603476}}].

\bibitem{Caprini:2009yp}
C.~Caprini, R.~Durrer and G.~Servant, \emph{{The stochastic gravitational wave
  background from turbulence and magnetic fields generated by a first-order
  phase transition}},
  \href{https://doi.org/10.1088/1475-7516/2009/12/024}{\emph{JCAP} {\bfseries
  0912} (2009) 024}, [\href{https://arxiv.org/abs/0909.0622}{{\ttfamily
  0909.0622}}].

\bibitem{Kosowsky:2001xp}
A.~Kosowsky, A.~Mack and T.~Kahniashvili, \emph{{Gravitational radiation from
  cosmological turbulence}},
  \href{https://doi.org/10.1103/PhysRevD.66.024030}{\emph{Phys. Rev.}
  {\bfseries D66} (2002) 024030},
  [\href{https://arxiv.org/abs/astro-ph/0111483}{{\ttfamily
  astro-ph/0111483}}].

\bibitem{Gogoberidze:2007an}
G.~Gogoberidze, T.~Kahniashvili and A.~Kosowsky, \emph{{The Spectrum of
  Gravitational Radiation from Primordial Turbulence}},
  \href{https://doi.org/10.1103/PhysRevD.76.083002}{\emph{Phys. Rev.}
  {\bfseries D76} (2007) 083002},
  [\href{https://arxiv.org/abs/0705.1733}{{\ttfamily 0705.1733}}].

\bibitem{Niksa:2018ofa}
P.~Niksa, M.~Schlederer and G.~Sigl, \emph{{Gravitational Waves produced by
  Compressible MHD Turbulence from Cosmological Phase Transitions}},
  \href{https://doi.org/10.1088/1361-6382/aac89c}{\emph{Class. Quant. Grav.}
  {\bfseries 35} (2018) 144001},
  [\href{https://arxiv.org/abs/1803.02271}{{\ttfamily 1803.02271}}].

\bibitem{Creminelli:2001th}
P.~Creminelli, A.~Nicolis and R.~Rattazzi, \emph{{Holography and the
  electroweak phase transition}},
  \href{https://doi.org/10.1088/1126-6708/2002/03/051}{\emph{JHEP} {\bfseries
  03} (2002) 051}, [\href{https://arxiv.org/abs/hep-th/0107141}{{\ttfamily
  hep-th/0107141}}].

\bibitem{Randall:2006py}
L.~Randall and G.~Servant, \emph{{Gravitational waves from warped spacetime}},
  \href{https://doi.org/10.1088/1126-6708/2007/05/054}{\emph{JHEP} {\bfseries
  05} (2007) 054}, [\href{https://arxiv.org/abs/hep-ph/0607158}{{\ttfamily
  hep-ph/0607158}}].

\bibitem{Nardini:2007me}
G.~Nardini, M.~Quiros and A.~Wulzer, \emph{{A Confining Strong First-Order
  Electroweak Phase Transition}},
  \href{https://doi.org/10.1088/1126-6708/2007/09/077}{\emph{JHEP} {\bfseries
  09} (2007) 077}, [\href{https://arxiv.org/abs/0706.3388}{{\ttfamily
  0706.3388}}].

\bibitem{Konstandin:2010cd}
T.~Konstandin, G.~Nardini and M.~Quiros, \emph{{Gravitational Backreaction
  Effects on the Holographic Phase Transition}},
  \href{https://doi.org/10.1103/PhysRevD.82.083513}{\emph{Phys. Rev.}
  {\bfseries D82} (2010) 083513},
  [\href{https://arxiv.org/abs/1007.1468}{{\ttfamily 1007.1468}}].

\bibitem{Konstandin:2011dr}
T.~Konstandin and G.~Servant, \emph{{Cosmological Consequences of Nearly
  Conformal Dynamics at the TeV scale}},
  \href{https://doi.org/10.1088/1475-7516/2011/12/009}{\emph{JCAP} {\bfseries
  1112} (2011) 009}, [\href{https://arxiv.org/abs/1104.4791}{{\ttfamily
  1104.4791}}].

\bibitem{Bunk:2017fic}
D.~Bunk, J.~Hubisz and B.~Jain, \emph{{A Perturbative RS I Cosmological Phase
  Transition}},
  \href{https://doi.org/10.1140/epjc/s10052-018-5529-2}{\emph{Eur. Phys. J.}
  {\bfseries C78} (2018) 78},
  [\href{https://arxiv.org/abs/1705.00001}{{\ttfamily 1705.00001}}].

\bibitem{Dillon:2017ctw}
B.~M. Dillon, B.~K. El-Menoufi, S.~J. Huber and J.~P. Manuel, \emph{{A rapid
  holographic phase transition with brane-localized curvature}},
  \href{https://arxiv.org/abs/1708.02953}{{\ttfamily 1708.02953}}.

\bibitem{vonHarling:2017yew}
B.~von Harling and G.~Servant, \emph{{QCD-induced Electroweak Phase
  Transition}}, \href{https://doi.org/10.1007/JHEP01(2018)159}{\emph{JHEP}
  {\bfseries 01} (2018) 159},
  [\href{https://arxiv.org/abs/1711.11554}{{\ttfamily 1711.11554}}].

\bibitem{Baldes:2018emh}
I.~Baldes and C.~Garcia-Cely, \emph{{Strong gravitational radiation from a
  simple dark matter model}},
  \href{https://arxiv.org/abs/1809.01198}{{\ttfamily 1809.01198}}.

\bibitem{Coleman:1977py}
S.~R. Coleman, \emph{{The Fate of the False Vacuum. 1. Semiclassical Theory}},
  \href{https://doi.org/10.1103/PhysRevD.15.2929,
  10.1103/PhysRevD.16.1248}{\emph{Phys. Rev.} {\bfseries D15} (1977)
  2929--2936}.

\bibitem{Linde:1980tt}
A.~D. Linde, \emph{{Fate of the False Vacuum at Finite Temperature: Theory and
  Applications}},
  \href{https://doi.org/10.1016/0370-2693(81)90281-1}{\emph{Phys. Lett.}
  {\bfseries B100} (1981) 37}.

\bibitem{Linde:1981zj}
A.~D. Linde, \emph{{Decay of the False Vacuum at Finite Temperature}},
  \href{https://doi.org/10.1016/0550-3213(83)90293-6}{\emph{Nucl. Phys.}
  {\bfseries B216} (1983) 421}.

\bibitem{Guth:1979bh}
A.~H. Guth and S.~H.~H. Tye, \emph{{Phase Transitions and Magnetic Monopole
  Production in the Very Early Universe}},
  \href{https://doi.org/10.1103/PhysRevLett.44.631,
  10.1103/PhysRevLett.44.963.2}{\emph{Phys. Rev. Lett.} {\bfseries 44} (1980)
  631}.

\bibitem{Guth:1981uk}
A.~H. Guth and E.~J. Weinberg, \emph{{Cosmological Consequences of a First
  Order Phase Transition in the SU(5) Grand Unified Model}},
  \href{https://doi.org/10.1103/PhysRevD.23.876}{\emph{Phys. Rev.} {\bfseries
  D23} (1981) 876}.

\bibitem{Bodeker:2009qy}
D.~Bodeker and G.~D. Moore, \emph{{Can electroweak bubble walls run away?}},
  \href{https://doi.org/10.1088/1475-7516/2009/05/009}{\emph{JCAP} {\bfseries
  0905} (2009) 009}, [\href{https://arxiv.org/abs/0903.4099}{{\ttfamily
  0903.4099}}].

\bibitem{Bodeker:2017cim}
D.~Bodeker and G.~D. Moore, \emph{{Electroweak Bubble Wall Speed Limit}},
  \href{https://doi.org/10.1088/1475-7516/2017/05/025}{\emph{JCAP} {\bfseries
  1705} (2017) 025}, [\href{https://arxiv.org/abs/1703.08215}{{\ttfamily
  1703.08215}}].

\bibitem{Steinhardt:1981ct}
P.~J. Steinhardt, \emph{{Relativistic Detonation Waves and Bubble Growth in
  False Vacuum Decay}},
  \href{https://doi.org/10.1103/PhysRevD.25.2074}{\emph{Phys. Rev.} {\bfseries
  D25} (1982) 2074}.

\bibitem{Laine:1993ey}
M.~Laine, \emph{{Bubble growth as a detonation}},
  \href{https://doi.org/10.1103/PhysRevD.49.3847}{\emph{Phys. Rev.} {\bfseries
  D49} (1994) 3847--3853},
  [\href{https://arxiv.org/abs/hep-ph/9309242}{{\ttfamily hep-ph/9309242}}].

\bibitem{Ignatius:1993qn}
J.~Ignatius, K.~Kajantie, H.~Kurki-Suonio and M.~Laine, \emph{{The growth of
  bubbles in cosmological phase transitions}},
  \href{https://doi.org/10.1103/PhysRevD.49.3854}{\emph{Phys. Rev.} {\bfseries
  D49} (1994) 3854--3868},
  [\href{https://arxiv.org/abs/astro-ph/9309059}{{\ttfamily
  astro-ph/9309059}}].

\bibitem{KurkiSuonio:1995pp}
H.~Kurki-Suonio and M.~Laine, \emph{{Supersonic deflagrations in cosmological
  phase transitions}},
  \href{https://doi.org/10.1103/PhysRevD.51.5431}{\emph{Phys. Rev.} {\bfseries
  D51} (1995) 5431--5437},
  [\href{https://arxiv.org/abs/hep-ph/9501216}{{\ttfamily hep-ph/9501216}}].

\bibitem{Espinosa:2010hh}
J.~R. Espinosa, T.~Konstandin, J.~M. No and G.~Servant, \emph{{Energy Budget of
  Cosmological First-order Phase Transitions}},
  \href{https://doi.org/10.1088/1475-7516/2010/06/028}{\emph{JCAP} {\bfseries
  1006} (2010) 028}, [\href{https://arxiv.org/abs/1004.4187}{{\ttfamily
  1004.4187}}].

\bibitem{Konstandin:2010dm}
T.~Konstandin and J.~M. No, \emph{{Hydrodynamic obstruction to bubble
  expansion}}, \href{https://doi.org/10.1088/1475-7516/2011/02/008}{\emph{JCAP}
  {\bfseries 1102} (2011) 008},
  [\href{https://arxiv.org/abs/1011.3735}{{\ttfamily 1011.3735}}].

\bibitem{ThomasK_private}
T.~Konstandin, \emph{{Private communication}}, .

\bibitem{1971Shante}
V.~K.~S. {Shante} and S.~{Kirkpatrick}, \emph{{An introduction to percolation
  theory}}, \href{https://doi.org/10.1080/00018737100101261}{\emph{Advances in
  Physics} {\bfseries 20} (May, 1971) 325--357}.

\bibitem{Enqvist:1991xw}
K.~Enqvist, J.~Ignatius, K.~Kajantie and K.~Rummukainen, \emph{{Nucleation and
  bubble growth in a first order cosmological electroweak phase transition}},
  \href{https://doi.org/10.1103/PhysRevD.45.3415}{\emph{Phys. Rev.} {\bfseries
  D45} (1992) 3415--3428}.

\bibitem{Jinno:2017ixd}
R.~Jinno, S.~Lee, H.~Seong and M.~Takimoto, \emph{{Gravitational waves from
  first-order phase transitions: Towards model separation by bubble nucleation
  rate}}, \href{https://doi.org/10.1088/1475-7516/2017/11/050}{\emph{JCAP}
  {\bfseries 1711} (2017) 050},
  [\href{https://arxiv.org/abs/1708.01253}{{\ttfamily 1708.01253}}].

\bibitem{Cutting:2018tjt}
D.~Cutting, M.~Hindmarsh and D.~J. Weir, \emph{{Gravitational waves from vacuum
  first-order phase transitions: from the envelope to the lattice}},
  \href{https://doi.org/10.1103/PhysRevD.97.123513}{\emph{Phys. Rev.}
  {\bfseries D97} (2018) 123513},
  [\href{https://arxiv.org/abs/1802.05712}{{\ttfamily 1802.05712}}].

\bibitem{Nicolis:2003tg}
A.~Nicolis, \emph{{Relic gravitational waves from colliding bubbles and cosmic
  turbulence}}, \href{https://doi.org/10.1088/0264-9381/21/4/L05}{\emph{Class.
  Quant. Grav.} {\bfseries 21} (2004) L27},
  [\href{https://arxiv.org/abs/gr-qc/0303084}{{\ttfamily gr-qc/0303084}}].

\bibitem{Huber:2007vva}
S.~J. Huber and T.~Konstandin, \emph{{Production of gravitational waves in the
  nMSSM}}, \href{https://doi.org/10.1088/1475-7516/2008/05/017}{\emph{JCAP}
  {\bfseries 0805} (2008) 017},
  [\href{https://arxiv.org/abs/0709.2091}{{\ttfamily 0709.2091}}].

\bibitem{Hindmarsh:2016lnk}
M.~Hindmarsh, \emph{{Sound shell model for acoustic gravitational wave
  production at a first-order phase transition in the early Universe}},
  \href{https://doi.org/10.1103/PhysRevLett.120.071301}{\emph{Phys. Rev. Lett.}
  {\bfseries 120} (2018) 071301},
  [\href{https://arxiv.org/abs/1608.04735}{{\ttfamily 1608.04735}}].

\bibitem{Jinno:2017fby}
R.~Jinno and M.~Takimoto, \emph{{Gravitational waves from bubble dynamics:
  Beyond the Envelope}},  \href{https://arxiv.org/abs/1707.03111}{{\ttfamily
  1707.03111}}.

\bibitem{Konstandin:2017sat}
T.~Konstandin, \emph{{Gravitational radiation from a bulk flow model}},
  \href{https://doi.org/10.1088/1475-7516/2018/03/047}{\emph{JCAP} {\bfseries
  1803} (2018) 047}, [\href{https://arxiv.org/abs/1712.06869}{{\ttfamily
  1712.06869}}].

\bibitem{Bodeker:2004ws}
D.~Bodeker, L.~Fromme, S.~J. Huber and M.~Seniuch, \emph{{The Baryon asymmetry
  in the standard model with a low cut-off}},
  \href{https://doi.org/10.1088/1126-6708/2005/02/026}{\emph{JHEP} {\bfseries
  02} (2005) 026}, [\href{https://arxiv.org/abs/hep-ph/0412366}{{\ttfamily
  hep-ph/0412366}}].

\bibitem{Delaunay:2007wb}
C.~Delaunay, C.~Grojean and J.~D. Wells, \emph{{Dynamics of Non-renormalizable
  Electroweak Symmetry Breaking}},
  \href{https://doi.org/10.1088/1126-6708/2008/04/029}{\emph{JHEP} {\bfseries
  04} (2008) 029}, [\href{https://arxiv.org/abs/0711.2511}{{\ttfamily
  0711.2511}}].

\bibitem{Figueroa:2018xtu}
D.~G. Figueroa, E.~Megias, G.~Nardini, M.~Pieroni, M.~Quiros, A.~Ricciardone
  et~al., \emph{{LISA as a probe for particle physics: electroweak scale tests
  in synergy with ground-based experiments}},
  \href{https://arxiv.org/abs/1806.06463}{{\ttfamily 1806.06463}}.

\bibitem{vanHaasteren:2011ni}
R.~van Haasteren et~al., \emph{{Placing limits on the stochastic
  gravitational-wave background using European Pulsar Timing Array data}},
  \href{https://doi.org/10.1111/j.1365-2966.2011.18613.x,
  10.1111/j.1365-2966.2012.20916.x}{\emph{Mon. Not. Roy. Astron. Soc.}
  {\bfseries 414} (2011) 3117--3128},
  [\href{https://arxiv.org/abs/1103.0576}{{\ttfamily 1103.0576}}].

\bibitem{TheLIGOScientific:2014jea}
{\scshape LIGO Scientific} collaboration, J.~Aasi et~al., \emph{{Advanced
  LIGO}}, \href{https://doi.org/10.1088/0264-9381/32/7/074001}{\emph{Class.
  Quant. Grav.} {\bfseries 32} (2015) 074001},
  [\href{https://arxiv.org/abs/1411.4547}{{\ttfamily 1411.4547}}].

\bibitem{Thrane:2013oya}
E.~Thrane and J.~D. Romano, \emph{{Sensitivity curves for searches for
  gravitational-wave backgrounds}},
  \href{https://doi.org/10.1103/PhysRevD.88.124032}{\emph{Phys. Rev.}
  {\bfseries D88} (2013) 124032},
  [\href{https://arxiv.org/abs/1310.5300}{{\ttfamily 1310.5300}}].

\bibitem{TheLIGOScientific:2016wyq}
{\scshape Virgo, LIGO Scientific} collaboration, B.~P. Abbott et~al.,
  \emph{{GW150914: Implications for the stochastic gravitational wave
  background from binary black holes}},
  \href{https://doi.org/10.1103/PhysRevLett.116.131102}{\emph{Phys. Rev. Lett.}
  {\bfseries 116} (2016) 131102},
  [\href{https://arxiv.org/abs/1602.03847}{{\ttfamily 1602.03847}}].

\bibitem{Bartolo:2016ami}
N.~Bartolo et~al., \emph{{Science with the space-based interferometer LISA. IV:
  Probing inflation with gravitational waves}},
  \href{https://doi.org/10.1088/1475-7516/2016/12/026}{\emph{JCAP} {\bfseries
  1612} (2016) 026}, [\href{https://arxiv.org/abs/1610.06481}{{\ttfamily
  1610.06481}}].

\bibitem{Punturo:2010zz}
M.~Punturo et~al., \emph{{The Einstein Telescope: A third-generation
  gravitational wave observatory}},
  \href{https://doi.org/10.1088/0264-9381/27/19/194002}{\emph{Class. Quant.
  Grav.} {\bfseries 27} (2010) 194002}.

\bibitem{Hild:2010id}
S.~Hild et~al., \emph{{Sensitivity Studies for Third-Generation Gravitational
  Wave Observatories}},
  \href{https://doi.org/10.1088/0264-9381/28/9/094013}{\emph{Class. Quant.
  Grav.} {\bfseries 28} (2011) 094013},
  [\href{https://arxiv.org/abs/1012.0908}{{\ttfamily 1012.0908}}].

\bibitem{Evans:2016mbw}
{\scshape LIGO Scientific} collaboration, B.~P. Abbott et~al., \emph{{Exploring
  the Sensitivity of Next Generation Gravitational Wave Detectors}},
  \href{https://doi.org/10.1088/1361-6382/aa51f4}{\emph{Class. Quant. Grav.}
  {\bfseries 34} (2017) 044001},
  [\href{https://arxiv.org/abs/1607.08697}{{\ttfamily 1607.08697}}].

\bibitem{Janssen:2014dka}
G.~Janssen et~al., \emph{{Gravitational wave astronomy with the SKA}},
  {\emph{PoS} {\bfseries AASKA14} (2015) 037},
  [\href{https://arxiv.org/abs/1501.00127}{{\ttfamily 1501.00127}}].

\bibitem{Kawamura:2006up}
S.~Kawamura et~al., \emph{{The Japanese space gravitational wave antenna
  DECIGO}}, \href{https://doi.org/10.1088/0264-9381/23/8/S17}{\emph{Class.
  Quant. Grav.} {\bfseries 23} (2006) S125--S132}.

\bibitem{Yagi:2011wg}
K.~Yagi and N.~Seto, \emph{{Detector configuration of DECIGO/BBO and
  identification of cosmological neutron-star binaries}},
  \href{https://doi.org/10.1103/PhysRevD.83.044011}{\emph{Phys. Rev.}
  {\bfseries D83} (2011) 044011},
  [\href{https://arxiv.org/abs/1101.3940}{{\ttfamily 1101.3940}}].

\bibitem{Crowder:2005nr}
J.~Crowder and N.~J. Cornish, \emph{{Beyond LISA: Exploring future
  gravitational wave missions}},
  \href{https://doi.org/10.1103/PhysRevD.72.083005}{\emph{Phys. Rev.}
  {\bfseries D72} (2005) 083005},
  [\href{https://arxiv.org/abs/gr-qc/0506015}{{\ttfamily gr-qc/0506015}}].

\bibitem{Graham:2017pmn}
{\scshape MAGIS} collaboration, P.~W. Graham, J.~M. Hogan, M.~A. Kasevich,
  S.~Rajendran and R.~W. Romani, \emph{{Mid-band gravitational wave detection
  with precision atomic sensors}},
  \href{https://arxiv.org/abs/1711.02225}{{\ttfamily 1711.02225}}.

\bibitem{Goertz:2013kp}
F.~Goertz, A.~Papaefstathiou, L.~L. Yang and J.~Zurita, \emph{{Higgs Boson
  self-coupling measurements using ratios of cross sections}},
  \href{https://doi.org/10.1007/JHEP06(2013)016}{\emph{JHEP} {\bfseries 06}
  (2013) 016}, [\href{https://arxiv.org/abs/1301.3492}{{\ttfamily 1301.3492}}].

\bibitem{Barger:2013jfa}
V.~Barger, L.~L. Everett, C.~B. Jackson and G.~Shaughnessy, \emph{{Higgs-Pair
  Production and Measurement of the Triscalar Coupling at LHC(8,14)}},
  \href{https://doi.org/10.1016/j.physletb.2013.12.013}{\emph{Phys. Lett.}
  {\bfseries B728} (2014) 433--436},
  [\href{https://arxiv.org/abs/1311.2931}{{\ttfamily 1311.2931}}].

\bibitem{Barr:2014sga}
A.~J. Barr, M.~J. Dolan, C.~Englert, D.~E. Ferreira~de Lima and M.~Spannowsky,
  \emph{{Higgs Self-Coupling Measurements at a 100 TeV Hadron Collider}},
  \href{https://doi.org/10.1007/JHEP02(2015)016}{\emph{JHEP} {\bfseries 02}
  (2015) 016}, [\href{https://arxiv.org/abs/1412.7154}{{\ttfamily 1412.7154}}].

\bibitem{Anderson:1991zb}
G.~W. Anderson and L.~J. Hall, \emph{{The Electroweak phase transition and
  baryogenesis}}, \href{https://doi.org/10.1103/PhysRevD.45.2685}{\emph{Phys.
  Rev.} {\bfseries D45} (1992) 2685--2698}.

\bibitem{Choi:1993cv}
J.~Choi and R.~R. Volkas, \emph{{Real Higgs singlet and the electroweak phase
  transition in the Standard Model}},
  \href{https://doi.org/10.1016/0370-2693(93)91013-D}{\emph{Phys. Lett.}
  {\bfseries B317} (1993) 385--391},
  [\href{https://arxiv.org/abs/hep-ph/9308234}{{\ttfamily hep-ph/9308234}}].

\bibitem{Espinosa:1993bs}
J.~R. Espinosa and M.~Quiros, \emph{{The Electroweak phase transition with a
  singlet}}, \href{https://doi.org/10.1016/0370-2693(93)91111-Y}{\emph{Phys.
  Lett.} {\bfseries B305} (1993) 98--105},
  [\href{https://arxiv.org/abs/hep-ph/9301285}{{\ttfamily hep-ph/9301285}}].

\bibitem{Profumo:2007wc}
S.~Profumo, M.~J. Ramsey-Musolf and G.~Shaughnessy, \emph{{Singlet Higgs
  phenomenology and the electroweak phase transition}},
  \href{https://doi.org/10.1088/1126-6708/2007/08/010}{\emph{JHEP} {\bfseries
  08} (2007) 010}, [\href{https://arxiv.org/abs/0705.2425}{{\ttfamily
  0705.2425}}].

\bibitem{Espinosa:2011ax}
J.~R. Espinosa, T.~Konstandin and F.~Riva, \emph{{Strong Electroweak Phase
  Transitions in the Standard Model with a Singlet}},
  \href{https://doi.org/10.1016/j.nuclphysb.2011.09.010}{\emph{Nucl. Phys.}
  {\bfseries B854} (2012) 592--630},
  [\href{https://arxiv.org/abs/1107.5441}{{\ttfamily 1107.5441}}].

\bibitem{Chen:2017qcz}
C.-Y. Chen, J.~Kozaczuk and I.~M. Lewis, \emph{{Non-resonant Collider
  Signatures of a Singlet-Driven Electroweak Phase Transition}},
  \href{https://doi.org/10.1007/JHEP08(2017)096}{\emph{JHEP} {\bfseries 08}
  (2017) 096}, [\href{https://arxiv.org/abs/1704.05844}{{\ttfamily
  1704.05844}}].

\bibitem{Lewis:2017dme}
I.~M. Lewis and M.~Sullivan, \emph{{Benchmarks for Double Higgs Production in
  the Singlet Extended Standard Model at the LHC}},
  \href{https://doi.org/10.1103/PhysRevD.96.035037}{\emph{Phys. Rev.}
  {\bfseries D96} (2017) 035037},
  [\href{https://arxiv.org/abs/1701.08774}{{\ttfamily 1701.08774}}].

\bibitem{Curtin:2014jma}
D.~Curtin, P.~Meade and C.-T. Yu, \emph{{Testing Electroweak Baryogenesis with
  Future Colliders}},
  \href{https://doi.org/10.1007/JHEP11(2014)127}{\emph{JHEP} {\bfseries 11}
  (2014) 127}, [\href{https://arxiv.org/abs/1409.0005}{{\ttfamily 1409.0005}}].

\bibitem{Aad:2015txa}
{\scshape ATLAS} collaboration, G.~Aad et~al., \emph{{Search for invisible
  decays of a Higgs boson using vector-boson fusion in $pp$ collisions at
  $\sqrt{s}=8$ TeV with the ATLAS detector}},
  \href{https://doi.org/10.1007/JHEP01(2016)172}{\emph{JHEP} {\bfseries 01}
  (2016) 172}, [\href{https://arxiv.org/abs/1508.07869}{{\ttfamily
  1508.07869}}].

\bibitem{Khachatryan:2016whc}
{\scshape CMS} collaboration, V.~Khachatryan et~al., \emph{{Searches for
  invisible decays of the Higgs boson in pp collisions at $\sqrt{s}$ = 7, 8,
  and 13 TeV}}, \href{https://doi.org/10.1007/JHEP02(2017)135}{\emph{JHEP}
  {\bfseries 02} (2017) 135},
  [\href{https://arxiv.org/abs/1610.09218}{{\ttfamily 1610.09218}}].

\bibitem{Craig:2014lda}
N.~Craig, H.~K. Lou, M.~McCullough and A.~Thalapillil, \emph{{The Higgs Portal
  Above Threshold}}, \href{https://doi.org/10.1007/JHEP02(2016)127}{\emph{JHEP}
  {\bfseries 02} (2016) 127},
  [\href{https://arxiv.org/abs/1412.0258}{{\ttfamily 1412.0258}}].

\bibitem{Contino:2016spe}
R.~Contino et~al., \emph{{Physics at a 100 TeV pp collider: Higgs and EW
  symmetry breaking studies}},
  \href{https://doi.org/10.23731/CYRM-2017-003.255}{\emph{CERN Yellow Report}
  (2017) 255--440}, [\href{https://arxiv.org/abs/1606.09408}{{\ttfamily
  1606.09408}}].

\bibitem{Gomez-Ceballos:2013zzn}
{\scshape TLEP Design Study Working Group} collaboration, M.~Bicer et~al.,
  \emph{{First Look at the Physics Case of TLEP}},
  \href{https://doi.org/10.1007/JHEP01(2014)164}{\emph{JHEP} {\bfseries 01}
  (2014) 164}, [\href{https://arxiv.org/abs/1308.6176}{{\ttfamily 1308.6176}}].

\bibitem{dEnterria:2016fpc}
D.~d'Enterria, \emph{{Physics case of FCC-ee}}, {\emph{Frascati Phys. Ser.}
  {\bfseries 61} (2016) 17},
  [\href{https://arxiv.org/abs/1601.06640}{{\ttfamily 1601.06640}}].

\bibitem{Goldberger:1999uk}
W.~D. Goldberger and M.~B. Wise, \emph{{Modulus stabilization with bulk
  fields}}, \href{https://doi.org/10.1103/PhysRevLett.83.4922}{\emph{Phys. Rev.
  Lett.} {\bfseries 83} (1999) 4922--4925},
  [\href{https://arxiv.org/abs/hep-ph/9907447}{{\ttfamily hep-ph/9907447}}].

\bibitem{Ellis:2018gqa}
J.~Ellis, C.~W. Murphy, V.~Sanz and T.~You, \emph{{Updated Global SMEFT Fit to
  Higgs, Diboson and Electroweak Data}},
  \href{https://doi.org/10.1007/JHEP06(2018)146}{\emph{JHEP} {\bfseries 06}
  (2018) 146}, [\href{https://arxiv.org/abs/1803.03252}{{\ttfamily
  1803.03252}}].

\bibitem{Arnold:1992rz}
P.~B. Arnold and O.~Espinosa, \emph{{The Effective potential and first order
  phase transitions: Beyond leading-order}},
  \href{https://doi.org/10.1103/PhysRevD.50.6662,
  10.1103/PhysRevD.47.3546}{\emph{Phys. Rev.} {\bfseries D47} (1993) 3546},
  [\href{https://arxiv.org/abs/hep-ph/9212235}{{\ttfamily hep-ph/9212235}}].

\bibitem{Carrington:1991hz}
M.~E. Carrington, \emph{{The Effective potential at finite temperature in the
  Standard Model}}, \href{https://doi.org/10.1103/PhysRevD.45.2933}{\emph{Phys.
  Rev.} {\bfseries D45} (1992) 2933--2944}.

\end{thebibliography}\endgroup
\end{document}